\documentclass[a4paper,12pt,bold]{scrartcl}

\usepackage{color,colortbl}			
\usepackage[dvipsnames]{xcolor}
\usepackage{bookmark}
\usepackage{apacite}
\usepackage{footnote}
\makesavenoteenv{tabular}
\usepackage{enumitem}
\usepackage{adjustbox}

\normalsize

\newcommand{\E}{\mathrm{E}}		

\newcommand{\ER}{\mathrm{ER}\,}
\newcommand{\OV}{\mathrm{OV}\,}

\usepackage{bbm}

\usepackage{algpseudocode,tabularx,ragged2e}
\usepackage{siunitx}

\newcolumntype{C}{>{\centering\arraybackslash}X} 
\newcolumntype{L}{>{\arraybackslash}X} 

\usepackage{algorithmicx}

\usepackage{algorithm}

\let\Algorithm\algorithm
\renewcommand\algorithm[1][]{\Algorithm[#1]\setstretch{1.5}}

\definecolor{lightgrey}{gray}{0.90}	
\definecolor{grey}{gray}{0.85}
\definecolor{darkgrey}{gray}{0.65}
\definecolor{lightblue}{rgb}{0.8,0.85,1}

\newcolumntype{g}{>{\columncolor{gray}}c}
\usepackage{bibentry}
\usepackage{booktabs}
\usepackage{epigraph}
\usepackage[sans]{dsfont}
\usepackage[round,longnamesfirst]{natbib}
\usepackage{bm}																									
\usepackage{setspace}																						
\usepackage{threeparttable}
\usepackage{lscape}																							
\usepackage[latin1]{inputenc}																		
\usepackage{graphicx}
\graphicspath{{../material/}{../material/structural-model/}{../material/results/}{../material/data/}}

\usepackage{amsmath, placeins}
\usepackage{amssymb}
\usepackage{fancybox}																						
\usepackage{appendix}
\usepackage{listings}
\usepackage{xr}

\usepackage{enumerate}

\usepackage{tabularx}
\usepackage{longtable,tabu}

\usepackage[position=top]{subfig}
\usepackage{float}																				
\usepackage{color,colortbl}																			
\usepackage[left=2cm, right=2cm, top=2cm, bottom=2.5cm]{geometry} 
\definecolor{lightgrey}{gray}{0.95}	
\definecolor{grey}{gray}{0.85}
\definecolor{darkgrey}{gray}{0.80}

\usepackage{tikz}
\usetikzlibrary{positioning}

\usepackage[labelfont=bf]{caption}
\usepackage{url}  

\deffootnote[1em]{1.1em}{0em}{\textsuperscript{\thefootnotemark}}

\usepackage{siunitx}
\DeclareMathOperator*{\argmin}{arg\,min}



\newcommand{\ind}{{\textbf{I}}}

\definecolor{tabblue}{RGB}{31,119,180}  
\definecolor{taborange}{RGB}{255, 127, 14}  
\definecolor{tabgreen}{RGB}{44, 160, 44}  
\definecolor{tabred}{RGB}{214, 39, 40}  
\definecolor{tabpurple}{RGB}{148, 103, 189}  
\definecolor{tabbrown}{RGB}{140, 86, 75}
\definecolor{tabrose}{RGB}{227, 119, 194}
\definecolor{tabgrey}{RGB}{127, 127, 127}
\definecolor{tablime}{RGB}{188, 189, 34}
\definecolor{tabcyan}{RGB}{23, 190, 207}

\definecolor{bwtabblue}{RGB}{77, 77, 77}
\definecolor{bwtabred}{RGB}{42, 42, 42}
\definecolor{bwtabpurple}{RGB}{130, 130, 130}
\definecolor{bwtaborange}{RGB}{209, 209, 209}
\definecolor{bwtabgreen}{RGB}{187 ,187 ,187}

\begin{document}

\title{\textbf{Sequential Choices, Option Values, and the Returns to Education}%
\thanks{\protect\linespread{1}\protect\selectfont Acknowledgements: We thank James Heckman for numerous helpful discussions, and seminar participants at University of Chicago (Lifecycle Working Group), Arizona State University, Royal Holloway-University of London, Aarhus University and University of Copenhagen for helpful comments. We thank Annica Gehlen, Emily Schwab, and Leiqiong Wan for their outstanding research assistance. We are grateful to the Social Sciences Computing Service (SSCS) at the University of Chicago for the permission to use their computational resources and Statistics Norway for providing access to micro data and computational resources. Manudeep Bhuller received financial support from the Research Council of Norway through the Young Research Talents Grant 275123. Philipp Eisenhauer is funded by the Deutsche Forschungsgemeinschaft (DFG, German Research Foundation) under Germany's Excellence Strategy - EXC 2126/1- 390838866, the TRA Modelling (University of Bonn) as part of the Excellence Strategy of the federal and state governments. Moritz Mendel gratefully acknowledges funding by the German Research Foundation (DFG)  through CRC TR 224 (Project C01)  and the Research Training Group ``The Macroeconomics of Inequality''.}}

\author{Manudeep Bhuller\thanks{ University of Oslo; Statistics Norway; CEPR; CESifo; IZA. 
 \href{mailto:manudeep.bhuller@econ.uio.no}{manudeep.bhuller@econ.uio.no}} \qquad
 Philipp Eisenhauer\thanks{ Amazon.
 \href{mailto:peisen@amazon.com }{peisen@amazon.com}}  \qquad 
 Moritz Mendel\thanks{ University of Bonn. \href{mailto:moritz.mendel12@gmail.com}{moritz.mendel12@gmail.com}}
}

\date{\today}

\maketitle
\begin{abstract}
\noindent \textbf{Abstract}: Using detailed Norwegian data on earnings, education and work histories, we estimate a dynamic structural model of education and sector choices that captures rich life-cycle patterns by ability. We validate the model against variation in education choices induced by a compulsory schooling reform. Our approach allows us to estimate the ex-ante returns to different education tracks across the life-cycle by individual ability and quantify the contribution of option values. We find substantial heterogeneity in returns and establish crucial roles for option values and re-enrollment in determining education choices and the impact of schooling policies.\\

\end{abstract}

\newpage
\noindent 

\section{Introduction}
Standard models of human capital \citep{Becker.1964, Mincer.1958} assume that individuals compare the potential future earnings streams at the beginning of their schooling career, choose the alternative with the highest net benefit, and subsequently complete their desired level of schooling.\footnote{\cite{Heckman.2006a} and \cite{Card.1999} provide extensive reviews of this literature on returns to schooling based on the Becker-Mincer models of human capital investments.} This view ignores both the sequential nature of human capital investments and the uncertainties embedded in this decision-making. The decision to take an additional year of schooling may open up further schooling opportunities as, for instance, a high school diploma is a stepping stone for a college education. And, individuals make such important decisions often facing considerable uncertainty about the associated costs and gains (see, e.g., \citet{Wiswall.2015a}, \citet{Attanasio.2017} and \citet{Wiswall.2021}).\\

\noindent Our paper is motivated by two stylized facts about educational choices related to their sequential and uncertain nature. Firstly, across a wide range of settings, educational researchers have documented the prevalence of drop-out, re-enrollment, and track switching in educational histories. These patterns are present not only in countries with highly subsidized educational system like the Scandinavian countries and Germany, but also in the U.S., where students face high monetary costs of higher education.\footnote{For instance, according to data from the National Student Clearinghouse (NSC), thirty-six million Americans held some postsecondary schooling in 2019 without completing a college degree or being currently enrolled \citep{Shapiro.2019}.} Secondly, a recurrent finding in the literature on compulsory schooling policies is that such policies tend to have so-called ``inframarginal'' impacts, i.e., educational choices beyond the minimum schooling requirements are impacted. In a seminal study of the compulsory attendance laws in the U.S., \cite{Lang.1986} documented the prevalence of such impacts, while similar findings have echoed in later studies.\footnote{For the U.S., \cite{Acemoglu.2000} provide further evidence on how the compulsory attendance laws affected the distribution of schooling, while \cite{Bedard.2001} provides evidence on how better university access increased high school drop-out rates. Relatedly, \cite{Meghir.2005} found evidence on ``inframarginal'' responses to a compulsory schooling reform in Sweden, while similar evidence for Norway is reported (though not emphasized) in \cite{Black.2005}. While these responses can reflect equilibrium adjustments to policy reforms in line with models of educational signaling \citep{Spence.1973}, these may also reflect the sequential and uncertain nature of educational decisions, as argued by \cite{Altonji.1993} and \cite{Heckman.2018}.}\\

\noindent In this paper, we develop and estimate a dynamic structural model of educational choices in a life-cycle context that can accommodate and explain both of the above-mentioned features of educational careers. Agents in our model are forward-looking and make sequential decisions every period from age 15 onwards on whether to attend school, and if so, the type of educational track to attend, to work in one of three sectors, or to stay at home, while they face uncertainty in terms of their work productivity in each sector and tastes for schooling tracks, work and leisure, and also differ in terms of their ability and a vector of latent types. Our model is able to generate a rich set of educational and work histories that feature (i) interruptions and re-enrollment in educational careers, (ii) persistence in choices across the life-cycle, (iii) costly switching between educational tracks, (vi) sector-specific human capital accumulation, and (v) persistent individual heterogeneity. The model is also able to re-produce ``inframarginal'' responses to actual and hypothetical compulsory schooling reforms, provide evidence on who is affected and inform about the potential economic mechanisms driving these patterns.\\

\noindent Using our model, we define and quantify two key objects related to educational choices. First, we consider the \textit{ex-ante return} to each schooling track choice for a given state in our model. This object takes into account the sequential and uncertain nature of schooling choices and captures both the immediate wage and non-monetary rewards as well as the discounted lifetime rewards associated with a choice as compared to the best alternative choice. Unlike the monetary wage rewards that have been the focus of much of the returns to schooling literature (see, e.g., \cite{Card.1999} for a review), the ex-ante returns are the objects that drive the educational choices of agents in our model. To illustrate the role of uncertainty, we further contrast \textit{ex-ante} and \textit{ex-post} returns, where the latter depend on the realizations of productivity and taste shocks, while the former reflect agents' expectations. Second, we consider the \textit{option value} to each schooling track choice for a given state in our model. The sequential nature of schooling decisions generates this value, for instance, as completing a high school diploma generates the option to enroll in college, and college enrollment generates the option to obtain a college degree \citep{Comay.1973,Weisbrod.1962}. Non-linearities in the wage returns to schooling choices \citep{Hungerford.1987} and the sequential resolution of uncertainties embedded in these choices can further exacerbate the importance of option values \citep{ Trachter.2015, Stange.2012, Altonji.1993}. As we will show, these objects are crucial determinants of schooling decisions, essential in understanding the impacts of policy reforms, and under-appreciated in the existing literature.\\

\noindent We implement our modeling approach and provide evidence on the ex-ante returns and the option values of education using Norwegian administrative data with career-long earnings information and education histories. We combine these data with detailed demographic information, including measures on individual ability for males collected as part of the compulsory military recruitment testing. This dataset has several advantages, as this provides (i) complete annual information on educational track choices and earnings histories for selected cohorts across four decades, (ii) allows us to capture several sources of persistent heterogeneity, and (iii) only suffers from natural sample attrition due to either death or out-migration. Our dataset further covers a compulsory schooling reform that increased the minimum requirement schooling from seven to nine years across different municipalities at different points in time \citep{Black.2005}. The latter feature of our setting allows us to estimate our model on individuals that were not exposed to the reform and rely on the reform-induced variation to evaluate our model in an out-of-sample validation and to shed light on  ``inframarginal'' responses to educational policy reforms. Specifically, we compare the predictions about policy impacts based on the model estimated on pre-reform data to the observed impacts post-reform \citep{Todd.2021}.\footnote{The recent review by \citet{Galiani.2021} also emphasizes the need for model validation and provides a structured review of the small literature.}\\


\noindent Our analysis presents several insights. We find substantial heterogeneity in the ex-ante returns by the level of schooling, track choice and individual ability. For low-ability individuals, our estimates of average ex-ante returns range between $-3\%$ at the 8th grade in the academic track and $3\%$ at the 9th year in the vocational track.\footnote{A negative return to a choice alternative in our model reflects that an individual expects to receive a higher reward from another choice alternative; i.e., the respective choice alternative is thus not chosen.} By comparison, both medium- and high-ability individuals have sizeable and positive returns to academic schooling, while high-ability individual have negative return at the 8th grade of vocational schooling. Underlying this heterogeneity is a strong pattern of \textit{ability-related sorting} into academic and vocational tracks. Indeed, the structure of the returns reflects the separation of the ability groups in the different schooling tracks at an early stage of the educational pathways. Meanwhile, the relatively low average ex-ante returns for an extra schooling year reflect the presence of \textit{re-enrollment opportunities}, as many individuals who drop out early do re-enroll and complete further education later in their life-cycle. Moving beyond the averages, we document substantial heterogeneity in the \textit{distributions} of ex-ante returns, with more dispersion in returns at the earlier stages of the educational careers, in the academic track and among high-ability individuals.\\

\noindent Our evidence further shows that the option values make up a sizeable fraction of the overall values of educational choices, however, their contribution varies considerably by the stage of educational career, track and individual ability. Option values depend on the likelihood that an individual will continue to pursue schooling further and the rewards they accrue if they do so. A recurring finding is that the \textit{option value contribution} is highest for the year of schooling right before the completion of an academic degree that entails considerable ``diploma'' effects. Intuitively, completing the schooling year right before the degree awarding year makes it more likely that individual will indeed pursue a degree and thus this choice holds a high option value. We also find important heterogeneity by ability and track. The option value contributions in the academic track tend to be highest for high-ability individuals, as they are also likely to benefit the most from the additional schooling opportunities. Indeed, we find more than 90\% of the high-ability individuals facing the choice of attending the 11th year in academic track would not have completed this education had it not triggered the option of continuing further to attain a high school diploma. Our evidence also provides insights on how \textit{sector-specific returns} to academic and vocational schooling influence educational choices for different ability groups. For instance, we find that fewer public sector job opportunities would induce some medium-ability individuals to drop academic schooling and pursue vocational high school. \\

\noindent Finally, we use our model to analyze the impacts of compulsory schooling reforms. Our model predicts that a compulsory schooling reform similar to the one actually implemented in Norway during the 1960s would increase the share of high school graduates by about 3\%, a prediction that resembles the actual reform-induced change observed in our data. Option values provide an economic rationale for such \textit{inframarginal responses}. By forcing more schooling on individuals, who prior to the reform would have taken less schooling than the new minimum requirement, we also bring them closer to the margins of schooling choices that hold stronger rewards through diplomas or degrees, and as a result some of these individuals do indeed pursue further education. Another important mechanism that our model brings forth is that of re-enrollment opportunities. Even prior to the reform, a sizeable fractions of individuals would have attained the new minimum schooling requirement, but only after first dropping-out and re-enrolling at a large stage in their careers. Since the reform forces these individuals to take the new minimum schooling requirement in an uninterrupted manner, their educational trajectories are also impacted. Interestingly, some of these individuals now also go on to pursue further education, since they no longer face high re-enrollment costs. \\

\noindent Our paper provides several contributions. We extend the empirical literature that acknowledges the sequential nature of schooling investments and emphasizes the roles of uncertainties and non-linearities. \citet{Lee.2017}, \citet{Eisenhauer.2015b}, \citet{Trachter.2015} and \citet{Stange.2012} all study the role of uncertainty and option values in shaping schooling decisions in deliberately simplified settings. However, none of these studies analyze life-cycle decisions, or allow for heterogeneity by ability, re-enrollment, sector and track switching at the same time. Our work is closely related to \citet{Heckman.2018}, who develop a sequential educational choice model. However, they restrict their attention to ex-post returns of education, and avoid making specific assumptions about individuals' expectations about costs and benefits of schooling. We impose additional structure on the decision process and are able to quantify ex-ante returns and option values. Our paper also relates to a large literature on compulsory schooling reforms \citep{Brunello.2009,Oreopoulos.2006}, providing evidence on the impacts of such reforms along the distribution of schooling attainment and the potential mechanisms driving these patterns. \\

\noindent The structure of our paper is as follows. We outline our structural model in Section \ref{sec:Model}. Section \ref{sec:Data-and-Setting} describes our data and institutional setting, and discusses model implementation and provide evidence on model fit and validation. Section \ref{sec:Results} presents our main findings. Section \ref{sec:Conclusion}  concludes.

\section{Model}\label{sec:Model} 
We now present a model that takes the sequential and uncertain nature of schooling investments into account, besides allowing for nonlinearities in the rewards to such investments. Our model is an example of the Eckstein-Keane-Wolpin (EKW) class of models \citep{Aguirregabiria.2010}, which are frequently used to study the mechanisms determining human capital investment decisions and to predict the effects of human capital policies \citep{Blundell.2017,Keane.2011d,Low.2017}. The model exploits the richness of our data and captures essential features of the Norwegian school system. We start by describing the model setup, and then define our main objects of interest -- the ex-ante returns and the option values of schooling.

\subsection{Setup}\label{subsec:Model-setup}
\noindent We follow individuals over most of their working life from young adulthood at age $15$ to the final period $T$ at age $50$.\footnote{In an earlier version of our model, we followed individuals up to age $58$. The decision to follow individuals up to age $50$ was motivated by computational ease, however, this choice does not comprise our main findings.} When entering the model, all individuals have seven years of basic compulsory schooling. Each individual is characterized by one of three levels of measured ability, and within each ability group, by one of several $j \in \mathcal{J}$ latent types \citep{Heckman.1984a}.\footnote{To classify individuals in low, medium and high ability groups, we rely on an IQ test score available in our data. See further details in Section \ref{subsec:Data-Sample} below. We allow three latent types among medium and high ability groups, and four latent types in the low ability group, so that the total number of distinct ability-by-latent types equals ten. We choose the number of latent types in each ability group based on model diagnostics.} Our model treats an individual's ability group and latent type as predetermined and individuals are assumed to know their ability and latent type. We allow all model parameters to vary across the three ability groups, effectively estimating our model separately for each ability group. In the following, we suppress subscripts for ability groups to ease notation. \\

\noindent The decision period $t = 15, \dots, 50$  is a school year. Each period individuals observe the state of their choice environment $s_t$ and decide to take action $a_t \in \mathcal{A}$. Individuals can decide whether to work in the private sector ($a_t = PR$), to work in the public sector ($a_t = PU$), to work as self-employed ($a_t = S$), to attend an academic $(a_t = A)$ or a vocational $(a_t = V)$ schooling track, or to stay at home $(a_t = H)$. The decision has two consequences: an individual receives an immediate utility $u(s_t, a_t)$ and the environment is updated to a new state $s_{t + 1}$. The transition from $s_t$ to $s_{t + 1}$ is affected by the action but remains partly uncertain. Individuals are forward-looking. Thus, they do not simply choose the alternative with the highest immediate utility. Instead, they take the future consequences of their current action into account.\\

\begin{figure}[h!]\centering
\caption{Timing of Events.}\label{Timing}
\begin{tikzpicture}[node distance=2cm]
\tikzstyle{startstop} = [circle, rounded corners, minimum width=0.6cm, minimum height=0.3cm,text centered, draw=black]
[
->,
>=stealth',
auto,node distance=3cm,
thick,
main node/.style={circle, draw, font=\sffamily\Large\bfseries}
]
\tikzstyle{arrow} = [thick,->,>=stealth]]
\tikzstyle{darrow} = [dotted,->,>=stealth]]

\node (r0) [startstop, xshift = -3cm, draw = none] {};
\node (r999) [startstop, xshift = 13cm, draw = none] {};

\node (r1) [startstop, xshift = -1cm] {\footnotesize $~\,s_t\,~$};
\node (r2) [startstop, xshift = 5cm] {\footnotesize $s_{t+1}$};  
\node (r3) [startstop, xshift = 11cm] {\footnotesize $s_{t+2}$}; 

\draw [arrow, dashed] (r0) -- node[anchor=south] {} (r1) ;
\draw [arrow] (r1) -- node[anchor=south] {} (r2) ;
\draw [arrow] (r2) -- node[anchor=south] {} (r3) ;
\draw [arrow, dashed] (r3) -- node[anchor=south] {} (r999) ;
t
\node (r4) [startstop, xshift = 0 cm, yshift = -2.5cm, inner sep = 0.08cm] {\footnotesize $~\,a_t\,~$ };
\node (r5) [startstop, xshift = 3.5
 cm, yshift = -2.5cm, inner sep = 0.08cm] {\footnotesize $~\,u_t\,~$ };
\node (r6) [startstop, xshift = 6 cm, yshift = -2.5cm, inner sep = 0.08cm] {\footnotesize $a_{t+1}$ };
\node (r7) [startstop, xshift = 9.5 cm, yshift = -2.5cm, inner sep = 0.08cm] {\footnotesize $u_{t+1}$ };

\draw [arrow] (r1) -- node[anchor=south] {} (r4) ;
\draw [arrow] (r2) -- node[anchor=south] {} (r6) ;

\draw[ thick](0,-2.05).. controls (0.75, -0.2) and (0.8,0)..(1.5, 0);
\draw [arrow] (r4) -- node[anchor=south] {} (r5) ;
\draw[ thick](6,-2.05).. controls (6.75, -0.2) and (6.85,0)..(7.5, 0);
\draw [arrow] (r6) -- node[anchor=south] {} (r7) ;

\node(r8)[startstop, xshift = - 1.3cm, yshift = -1.3cm, draw =none, align=center] {\footnotesize decide \\ \footnotesize $d_{t}$ };
\node(r9)[startstop, xshift= 4.6cm, yshift = -1.3cm, draw =none, align = center] {\footnotesize decide \\ \footnotesize $d_{t + 1}$ };
\node(r10)[startstop, yshift = 0.5cm, xshift = 1.7cm, draw =none, align=center ] {\footnotesize transition \\ \footnotesize $p(s_t, a_t)$};
\node(r11)[startstop, yshift = 0.5cm, xshift = 8cm, draw =none, align=center ] {\footnotesize transition \\ \footnotesize $p(s_{t+1}, a_{t+1})$};
\node(r12)[startstop, yshift = -2cm, xshift = 1.7cm, draw =none, align=center ] {\footnotesize receive\\ \footnotesize $u(s_t, a_t)$};
\node(r13)[startstop, yshift = -2cm, xshift = 7.7cm, draw =none, align=center ] {\footnotesize receive\\ \footnotesize $u(s_{t+1}, a_{t+1})$};

\end{tikzpicture}
\end{figure}

\noindent A policy $\pi =(d_1, \hdots, d_T)$ provides the individual with instructions for choosing an action in any possible future state. It is a sequence of decision rules $d_t$ that specify the planned action at a particular time $t$ for any possible state $s_t$. The implementation of a policy generates a sequence of utilities that depends on the transition probability distribution $p(s_t, a_t)$ for the evolution of state $s_t$ to $s_{t + 1}$ induced by the model. To fix ideas, Figure \ref{Timing} illustrates the timing of events in the model for two generic periods. At the beginning of period $t$, an individual fully learns about each alternative's immediate utility, chooses one of the alternatives, and receives its immediate utility. Then, the state evolves from $s_t$ to $s_{t + 1}$ and the process is repeated in $t + 1$. \\

\noindent Individuals make their decisions facing uncertainty about the future and seek to maximize their expected total discounted utilities over all remaining decision periods. They have rational expectations \citep{Muth.1961}, so their subjective beliefs about the future agree with the objective probabilities for all possible future events determined by the model. Immediate utilities are separable between periods \citep{Kahneman.1997}, and individuals discount future over immediate utilities by a discount factor $\delta$  \citep{Samuelson.1937}. Equation (\ref{Objective Risk}) provides the formal representation of an individual's objective function. Given an initial state $s_1$, they implement a policy $\pi^*$ from the set of all possible policies $\Pi$ that maximizes the expected total discounted utilities over all decision periods given the information available at the time.
\begin{align}\label{Objective Risk}
\max_{\pi \in \Pi} \E_{s_1}^\pi\left[\,\sum^{T}_{t = 16}  \delta^{t - 16} u(s_t, d_t(s_t))\,\right]
\end{align}


\noindent The immediate utility $u(\cdot)$ of each alternative consists of a non-pecuniary utility $\zeta_a(\cdot)$ and, for the working alternatives, an additional monetary wage component $w_{a}(\cdot)$. Both depend on an individual's level of human capital as measured by sector-specific work experience $\bm{k_t} = (k_{a, t})_{a\in\{PU, PR, S\}}$, years of completed schooling in each track $\bm{h_t} = (h_{a, t})_{a\in\{A, V\}}$, and alternative-specific skill endowment $\bm{e} = \left(e_{j,a}\right)_{\mathcal{J} \times \mathcal{A}}$ that depends on the individual's latent type. The immediate utilities are also influenced by the decision $a_{t -1}$ in the previous period, a time trend $t$, and alternative-specific shocks $\bm{\epsilon_t} = \left(\epsilon_{a,t}\right)_{a\in\mathcal{A}}$. Their general form is given by:
\begin{align*}
u(\cdot) =
\begin{cases}
    \zeta_a(\bm{k_t}, \bm{h_t}, t, a_{t -1}, e_{j, a})  + w_a(\bm{k_t}, \bm{h_t}, t, a_{t -1}, e_{j, a}, \epsilon_{a,t})                 &   \text{if}\, a \in\{PR, PU ,S\}  \\
    \zeta_a(\bm{k_t}, \bm{h_t}, t, a_{t-1}, e_{j,a}, \epsilon_{a,t})                                                  &   \text{if}\, a \in\{A, V ,H\} .
\end{cases}
\end{align*}

\noindent Sector-specific work experience $\bm{k_t}$ and years of completed schooling in each track $\bm{h_t}$ evolve deterministically. There is no uncertainty about grade completion \citep{Altonji.1993} and no part-time enrollment. Schooling is defined by time spent in school, not by formal credentials acquired. Once individuals reach a certain amount of schooling, they acquire a degree.
\begin{align*}
\begin{array}{lll}
k_{a,t + 1} & = k_{a,t\phantom{,t}}   + \ind[a_t = a] & \quad  \text{if}\, a \in \{PU, PR, S\}\\
h_{a,t+1} & = h_{a,t} + \ind[a_t = a\,] &\quad  \text{if}\, a \in \{A, V\}
\end{array}
\end{align*}
\\
\noindent The productivity and preference shocks $\bm{\epsilon_t}$ are unknown to the individual in advance, and capture uncertainty about the returns and cost of future schooling. In our model setup, we specify these shocks $\bm{\epsilon_t}$ to be uncorrelated across time and follow a multivariate normal distribution with mean $\bm{0}$ and covariance $\bm{\Sigma}$. Given the structure of the utility functions and the distribution of the shocks, the state at time $t$ is $s_t = \{\bm{k_t}, \bm{h_t}, t, a_{t -1}, \bm{e},\bm{\epsilon_t}\}$.\footnote{While in principle, one could also allow persistence in these shocks over time, the estimation problem becomes computationally more cumbersome as this would increase the state space dramatically. However, as discussed below, since the model includes persistent heterogeneity \textit{and} adjustment costs in moving across states, adding persistence in transitory shocks can also pose challenges for identification \citep{Heckman.1984a}.} \\

\noindent Individuals' ability and latent type are the two sources of persistent heterogeneity in this model. As noted above, we allow all model parameters to vary across three ability groups and further let individuals' alternative-specific skill endowments $\bm{e}$ to vary by their latent type. All remaining differences in life-cycle decisions result from differences in the transitory shocks $\bm{\epsilon}_t$ over time. In each period, individuals can increase their stock of human capital $(\bm{k_t}, \bm{h_t})$ by either working in the labor market or enrolling in school. From the beginning, individuals are aware of their level of ability and alternative-specific skill endowments $\bm{e}$. They only learn about the realizations of shocks $\bm{\epsilon_t}$ at the beginning of each period. As the shocks are distributed independently over time, individuals do not update their prior beliefs about their productivity or alternative-specific tastes (see, e.g, \cite{Arcidiacono.2004}; \cite{Arcidiacono.2016}).\\

\noindent Previous research on the determinants of life-cycle wages and schooling decisions \citep{Keane.1997,Meghir.2011} informs our specification of the immediate utility functions. We specify the wage component $w_a(\cdot) = r\, x_a(\cdot)$ in the immediate utility from working as the product of the market-equilibrium rental price $r$ and a sector-specific skill level $x_a(\cdot)$ for $a \in \{PU, PR, S\}$. The skill level is determined by a skill production function, which includes a deterministic component $\Gamma_a(\cdot)$ and a multiplicative stochastic productivity shock $\epsilon_{a,t}$, as follows:
\begin{align}
    x_a(\bm{k_t}, \bm{h_t}, t, a_{t-1}, e_{j, a}, \epsilon_{a,t}) & = \exp \big( \Gamma_a(\bm{k_t},  \bm{h_t}, t, a_{t-1}, e_{j,a}) \cdot \epsilon_{a,t} \big) \nonumber
\end{align}
\noindent The specification above leads to a standard logarithmic wage equation in which the constant term is the skill rental price $\ln(r)$ and wages follow a log-normal distribution. Equation (\ref{Skill production function}) shows the parametrization of the deterministic component $\Gamma_a(\cdot)$ of the skill production function:
\begin{align}\label{Skill production function}
\Gamma_a(\bm{k_t}, \bm{h_t}, t, a_{t-1}, e_{j, a}) & = e_{j,a} \\ \nonumber
 & + \underbrace{\delta_{1,a} \cdot  h^A_{t} + \delta_{2,a} \cdot  h^V_{t} + \delta_{3,a} \cdot  k^{W}_{t} + \delta_{4,a} \cdot  (k^{W}_{t})^2+ \delta_{5,a} \cdot  k^{a}_{t} + \delta_{6,a} \cdot  (k^{a}_{t})^2}_{\text{Mincer-inspired returns to schooling and experience}} \\\nonumber
 & + \underbrace{\sum_{d\in \{9, 12, 16\}} \gamma^A_{d,a} \cdot \ind[h^A_t \geq d] + \sum_{d\in \{9, 12\}} \gamma^V_{d,a} \cdot \ind[h^V_t \geq d]}_{\text{Non-linear rewards to diplomas or degrees}} \\ \nonumber
& + \underbrace{\eta_{1,a} \cdot \ind[a_{t-1} = W]}_{\text{Skill depreciation}} + \nu_{1,a} \cdot  (t - 15) + \nu_{2,a} \cdot \ind[t < 17]
\end{align}
\noindent The first part of the sectoral skill production function is motivated by \cite{Mincer.1974} and hence linear in years of completed schooling by track $(\delta_{1,a}, \delta_{2,a})$, quadratic in past work experience ($\delta_{3,a}, \delta_{4,a},\delta_{5,a},\delta_{6,a}$), and separable between the two of them. Both sector-specific experience $k_{t}^{a}$ and general work experience $k_{t}^{W} = k_{t}^{PR} + k_{t}^{PU} + k_{t}^{S} $ matter,  where we use $W \in \{PU, PR,  S\}$ as a generic work choice. We include diploma effects $(\gamma^A_{d,a}, \gamma^V_{d,a})$  that capture the non-linear rewards associated with a degree $d$ beyond the years of schooling \citep{Hungerford.1987, Jaeger.1996}. Skills depreciate by $\eta_{1,a}$ if the individual didn't work in the previous period. Finally, there is an age effect $\nu_{1,a}$ and a penalty when working as a minor $\nu_{2,a}$.\\

\noindent In the following, we discuss how we parametrize the immediate non-pecuniary reward of working in each sector and the immediate utilities of attending academic and vocational schooling, and staying at home, respectively.\footnote{Note that to enhance model performance, we varied some of these specifications slightly across the three ability groups. We include a more detailed exposition of reward functions by ability groups in the Appendix.} Equation (\ref{Nonpec Work}) shows the non-pecuniary rewards to working in each sector, i.e., for $a \in \{PU, PR, S\}$. We allow the immediate non-pecuniary reward (i.e., disutility) of work to depend on accumulated work experience in the particular sector $k^a_t$, accumulated general work experience $k^{W}_t$ and years of completed schooling in each track $\bm{h_t}$, and allow for age and diploma effects as in equation (\ref{Skill production function}). Further, we include indicators for previous period sector-specific work $\ind[a_{t-1} = a]$ and previous period private sector work $\ind[a_{t-1} = PR]$, which capture the fixed costs of market entry and sector switching.

\begin{align}\label{Nonpec Work}
\zeta_{a}(k_t, \bm{h_t}, t, a_{t-1})  & = e_{j,a}  + \underbrace{\beta_{1,a} \cdot \ind[t < 17]}_{\text{Age Effects}} + \underbrace{\beta_{2,a} \cdot k^{W}_t+ \beta_{3,a} \cdot k^{a}_t + \beta_{4,a} \cdot h^A_t + \beta_{5,a} \cdot h^V_t}_{\text{Experience Effects}} \\ \nonumber 
& \underbrace{+ \sum_{d\in \{9, 12, 16\}} \vartheta^A_{d,a} \cdot \ind[h^A_t \geq d]
 + \sum_{d\in \{9, 12\}} \vartheta^V_{d,a} \cdot \ind[h^V_t \geq d]}_{\text{Degree Effects}} \\ \nonumber
 & +  \underbrace{ \beta_{6,a} \cdot \ind[a_{t-1} = PR] + \beta_{7,a} \cdot \ind[k^{a}_t > 0]+ \beta_{8,a} \cdot \ind[k^{W}_t > 0] + \beta_{9,a} \cdot \ind[a_{t-1} = a]}_{\text{Labor Market Entry and Sector Switching Costs}} \nonumber 
\end{align}

\noindent Next, in Equation (\ref{Nonpec School}), we show the immediate rewards of academic and vocational education, i.e., $a \in \{A, V\}$. They include age $(t-15)$, an indicator for whether they have been enrolled in the track in the past period $\ind[a_{t-1} = a]$, an indicator for whether they have nonzero experience in the other schooling track $\ind[h^{a^{C}}_{t} > 0]$ to capture the costs of track switching. Furthermore they include diploma effects to capture changes in the programs as people progress over time.

\begin{align}\label{Nonpec School}
\zeta_{a}(k_t, \bm{h_t}, t, a_{t-1})  & = e_{j,a} + \underbrace{\beta_{1,a} \cdot (t-15)  + \beta_{2,a} \cdot \ind[t < 17 ]}_{\text{Age Effects}} + \underbrace{\beta_{3,a} \cdot \ind[h^{a^{C}}_{t} > 0]}_{\text{Track Switching Cost}}\\ \nonumber
&  + \underbrace{\beta_{4,a} \cdot \ind[a_{t-1} = a] + \beta_{5,a} \cdot \ind[a_{t-1} \in \{PR,PU,S,H\}] + \beta_{6,A} \cdot \ind[a_{t-1} \neq A] \cdot \ind[h^A_t \geq 12]}_{\text{Reenrolment Costs}}\\ \nonumber
& +\underbrace{\vartheta^A_{9,a} \cdot \ind[h^V_t \geq 9]+ \vartheta^A_{9,a} \cdot \ind[h^A_t \geq 9]}_{\text{Degree Effects}} + \epsilon_{t,a}
\end{align}

\noindent Finally we discuss the non-pecuniary reward to home production contained in Equation (\ref{Nonpec Home}), i.e., $a=H$. The utility of staying at home is allowed to depend on whether one is below age 17, which captures different opportunity costs of home production for minors. Furthermore it depends on indicators for having completed a high school or a college degree which captured differences in labor market transitions across education levels.

\begin{align}\label{Nonpec Home}
\zeta_{H}(k_t, \bm{h_t}, t, a_{t-1})  & = e_{j,H} + \underbrace{\beta_{1,H} \cdot \ind[t < 17]}_{\text{Age Effects}}  +  \underbrace{\sum_{d\in \{12, 16\}} \vartheta^A_{d,H} \cdot \ind[h^A_t \geq d]  +  \vartheta^V_{12,H}  \cdot \ind[h^V_t \geq 12]}_{\text{Degree Effects}}  +  \epsilon_{t,H}
\end{align}

\subsection{Objects of Interest}\label{subsec:Model-objects}
We now define two primary objects of interest in our analysis within the framework of the above model, namely the ex-ante return to schooling and the option value of schooling. While in the empirical analysis, we will present evidence on these objects separately by academic and vocational schooling, we refer to a generic schooling choice $G \in \{A, V\}$ here to ease the exposition.  \\

\noindent We define these objects in terms of value functions $v(s_t, a)$. The value functions are alternative- and state-specific and summarize the total value that individuals receive of choosing alternative $a$ for a given state $s_t$, including the immediate reward and the discounted future rewards, assuming that the optimal policy $\pi^*$ is followed in the future:
\begin{align*}
v(s_t, a)  & = u(s_t, a) + \delta\, \E_{s_t, a} \left[ v^{\pi^*}(s_{t + 1})\right]\qquad\forall\quad a \in \mathcal{A}
\end{align*}

\noindent Accordingly, the total value of schooling $v(s_t, G)$ in state $s_t$ captures the immediate and expected future benefits from continuing one's education in another period, subject to optimal policy $\pi^*$.

\noindent \paragraph{Ex-ante Return}  Next, we describe the elements that are needed to isolate the ex-ante return to an additional year of schooling. The thought-experiment we perform is to compare the total value of schooling $v(s_t, G)$ against the value of choosing the best alternative choice, which can be another schooling track. Notably, the alternative choice can also contain the option of taking more schooling at a later stage in the life-cycle through re-enrollment. Quantifying the ex-ante return requires making comparisons of counterfactuals as a given individual in state $s_t$ in the estimated model only chooses one of the available alternatives. We construct such counterfactuals through model simulations, where we require each individual to make alternative choices in state $s_t$, but restrict the realizations of shocks in each period to be held fixed across comparisons. \\





\noindent We denote by $ER(s_t)$ the ex-ante return capturing the value of an additional year of schooling against the best alternative choice in state $s_t$. Formally, we can express this object as follows:

\begin{align}\label{Calculation ex-ante return}
\ER(s_t) = \frac{v(s_t, G)  - \tilde{v}(s_t) }{ \tilde{v}(s_t)},\quad \text{where}\quad   \tilde{v}(s_t) = \max_{a \neq G} \{v(s_t, a)\}.
\end{align}

\noindent In our model, the ex-ante return of an additional year of schooling in a particular educational track is positive for all individuals who appear in state $s_t$ making this choice and negative for those that decide to attend another educational track, work or stay at home instead.\\

\noindent We also consider the ex-ante lifetime earnings returns associated with each choice. The expected lifetime earnings value associated with a choice $a$ at state $s_t$ is the expected lifetime discounted earnings conditional on choosing the optimal policy $\pi^*$ from the next period onward.

\begin{align} \label{lt_wages}
W(s_t, a) = \sum_{\tau \geq t} \delta^{\tau -t} E^{\pi^*}_{s_t,a}[w_{\tau}]
\end{align}

\noindent The expected ex-ante earnings return is then defined as follows:
\begin{align} \label{lt_wage_return}
\text{ERW}(s_t) = \frac{W(s_t, G)  - \tilde{W}(s_t) }{ \tilde{W}(s_t)},\quad \text{where}\quad \tilde{W}(s_t) = \max_{a \neq G} \{W(s_t, a)\}
\end{align}

\noindent \paragraph{Option Value}  We are further interested in the option value of schooling $OV(s_t)$. This object captures the part of the value of another year of schooling that can be attributed to having an opportunity to pursue further schooling in the future. This component arises due to the sequential nature of schooling investments. To compute the option value component, we perform another counterfactual comparison. We now compare the total value of schooling $v(s_t, G)$ in state $s_t$ to the value of the same alternative but with an optimal policy $\hat{\pi}$ that does not allow one to increase schooling further beyond the next period. We use superscripts $\pi^*$ and $\hat{\pi}$ to differentiate between the two total values. The latter is by construction a counterfactual scenario, unless one has already reached the maximum level of schooling.
\begin{align*}
\OV(s_t) = v^{\pi^*}(s_t, G) - v^{\hat{\pi}}(s_t, G)
\end{align*}
The option value of schooling is non-negative at all states and zero once an individual attains the maximum schooling level. The option value increases with the future benefits of pursuing higher education and the probability of doing so. To calculate this value, we compare the total value of continued schooling under the scenario that the individual \textit{may} continue to increase their schooling level in the future periods and the counterfactual where it is impossible to do so. \\
%




\noindent As a measure for the importance of the option value, we compute its contribution to the overall value of a state by taking the following ratio:
\begin{align}\label{Calculation option value contribution}
\text{OVC}(s_t) & = \frac{\OV(s_t)}{ v^{\pi^*}(s_t)}.
\end{align}

\noindent In the empirical analysis, we will report estimates of the option value contributions based on the above measure, which provides a decomposition of the total value of schooling in a state.\\

\section{Data and Implementation\label{sec:Data-and-Setting}}
In this section, we first describe our data sources, then briefly describe the Norwegian education system and the compulsory schooling reform, before we discuss the implementation and estimation of our model on these data, and finally, provide evidence on model fit and validation.

\subsection{Data Sources\label{subsec:Data-Sample}}
Our empirical analysis uses several registry databases maintained by Statistics Norway. First, we use the Norwegian National Education Database, a comprehensive population-wide event-history dataset with information on the dates of enrollment, termination and completion of 6-digit educational courses for all residents since 1970. Second, we use a longitudinal dataset containing annual earnings and tax records for all Norwegians for every year from 1967 onwards. Third, we use demographic information (e.g., cohort of birth, gender and childhood municipality of residence) for all individuals ever registered in the Norwegian Central Population Register, established in 1964. Fourth, we are also able to access supplementary demographic data from the Decennial Population Censuses held in 1960 and 1970, and link information on the primary sector of work based on a register of public sector employees established in 1974, and more detailed matched employer-employee data available from 1983. Finally, we received information on IQ test scores from the Norwegian Armed Forces for male conscripts born in 1950 and later. Importantly, each of these datasets include unique personal identifiers which allow us to follow individuals' educational choices and earnings across time, and add a proxy of ability for males.

\paragraph{Sample construction} We restrict our sample to Norwegian males born between 1955 and 1960. We can follow each of these individuals' educational choices and earnings from age 15 and up to age 50.\footnote{We observe educational choices annually for birth cohort 1955 at age 15 and onwards since the National Education Database was established in 1970. Educational histories are partially observed for earlier cohorts.} Our initial sample consists of 176,804 Norwegian-born males. Dropping individuals with missing information on childhood municipality of residence, exposure to compulsory schooling reform and education enrollment or attainment, we retain 165,171 individuals. Further dropping individuals with missing information on IQ test scores in the Norwegian military records, we retain 136,292 individuals (i.e., around 77\% of the initial sample). Using annual information on individuals' educational choices and earnings, we create a weakly-balanced panel of individuals entering our sample across 36 annual observations, which an individual can exit only due to natural attrition (i.e., death or out-migration). Our panel dataset thus consists of 4,820,863 person-year observations.\\

\noindent We further split the analytical sample in two parts, depending on the type of compulsory schooling system each individual was subject to, exploiting variation in the timing of a compulsory schooling reform across different municipalities in Norway (see details in Section \ref{subsec:Model-Validation}). Specifically, there are 9,156 individuals in our sample (i.e., around 7\%) who grew up in one of the 200 (out of 732) municipalities which hadn't implemented the compulsory schooling reform by the year they turned age 14 (threshold age for completely compulsory schooling) and as such were subject to the pre-reform education system. We will utilize of this sample of 9,156 individuals and 324,157 person-year observations to estimate our structural model, accounting for key features of the pre-reform education system, and refer to this as the \textit{estimation sample}. We will use the remaining sample of 127,136 post-reform individuals and 4,496,706 person-year observations to validate the structural model, and refer to this as the \textit{validation sample}.

\paragraph{Education} Our education information is primarily based on the Norwegian National Education Database (NUDB), which is many respects is an ideal dataset to study the enrollment, drop-out and completion behavior of individuals across time. The NUDB is an event-history dataset providing population-wide information on the dates of each enrollment and exit within a 6-digit educational course code across all lower secondary educations to tertiary educations. The detailed classification of educational course codes allows distinguishing educations by the level of attainment, the standard length of each course/degree, and the type of field for secondary (vocational/academic) and teritary educations. Each entry in this dataset further has information on the outcome of each enrollment, e.g., allowing the researcher to distinguish drop-out/early terminations and successful completions of a course, and whether the individual was enrolled as a part-time or as a full-time attendant in a specific course.\\

\noindent Combining information obtained from the NUDB and Statistics Norway's Education Register, where the latter also comprises information on compulsory education attainment, we can classify individuals' educational choices in a detailed manner across all education levels.\footnote{Earlier studies on the returns to education in Norway have relied on Statistics Norway's Education Register and used information on an individual's highest completed education level or the years of schooling corresponding to the highest level of education, see, e.g., \cite{Aryal.2021}, \cite{Mogstad.2017} and \cite{Aakvik.2010}. Neither of these studies consider the ex-ante returns or the option value of educational choices.} Noteworthy, information in both the NUDB and the Education Register is based on the annual reports submitted by educational establishments for each of their attendants directly to Statistics Norway, which minimizes the chance of misreporting. In comparison, survey-based data that are often used to study enrollment, drop-out and completion behavior may suffer from non-response or other biases due to misreporting.

\paragraph{Earnings and Work} Our earnings data are based on annual tax records. Our earnings measure is the sum of labor income (from wages and self-employment) and work-related cash transfers (such as unemployment benefits and short-term sickness benefits). This dataset have several advantages over those available in most countries. First, there is no attrition from the original sample other than natural attrition due to either death or out-migration. Second, our earnings data pertain to all individuals, and are not limited to some sectors or occupations. Third, we can construct long earnings histories that allow us estimate the returns to education across the life-cycle. Finally, using the annual tax records, we are able to distinguish between individuals who work as salaried workers and self-employed, respectively. Further, we add information on the primary sector of employment for salaried workers by relying on information recorded in the register of public sector employees from 1974 and matched employer-employee data from 1983 and onwards. This allows us to track individuals' sectoral choices over their careers.\footnote{Unfortunately, Norwegian register data contain limited information on occupations prior to 2003.}

\paragraph{Ability} An important measure we exploit in our analysis to capture observational heterogeneity is an IQ test score accessed from the Norwegian Armed Forces. In Norway, military service was compulsory for all able males in the birth cohorts we study. Before each male entered the service, his medical and psychological suitability was assessed. Most eligible Norwegian males in our sample took this test around their 18$^{th}$ birthday. The IQ test score is a composite unweighted mean from three speeded tests-{}-arithmetics, word similarities, and figures (\cite{Sundet.2004}). The score is reported in stanine (standard nine) units, a method of standardizing raw scores into a 9-point standard scale with a normal distribution, a mean of 5.8, and a standard deviation of 1.7. This score is strongly related to individuals' actual completed education, with a correlation of around 0.5 with years of schooling in our analytical sample.

\begin{figure}[H]
\begin{centering}
\caption{Descriptive Statistics.\label{fig:Desc-Stats}}
\subfloat[Distribution of Final Years of Education]{
\centering{}\includegraphics[width=0.4\columnwidth]{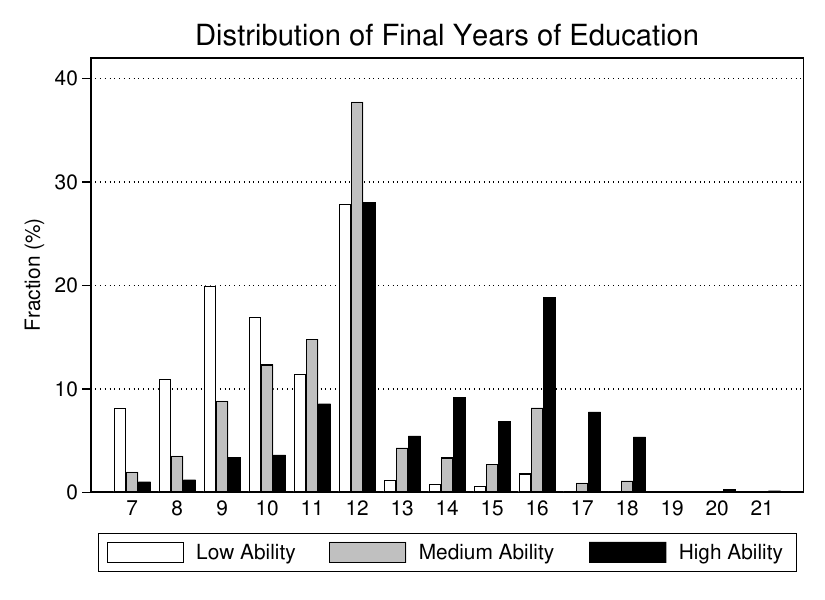}}\subfloat[Fraction Enrolled in Education by Age]{
\centering{}\includegraphics[width=0.4\columnwidth]{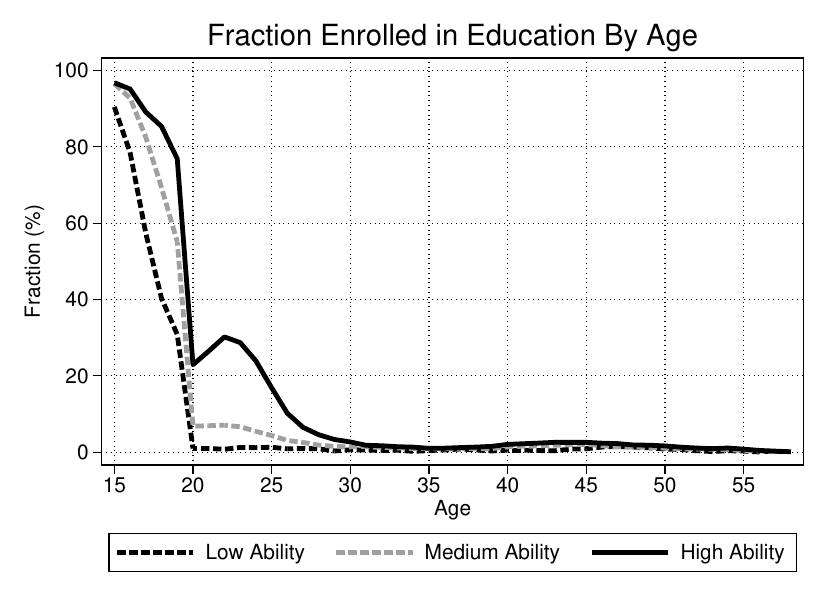}}
\par\end{centering}
\begin{centering}
\subfloat[Exit Rate By Age]{
\centering{}\includegraphics[width=0.4\columnwidth]{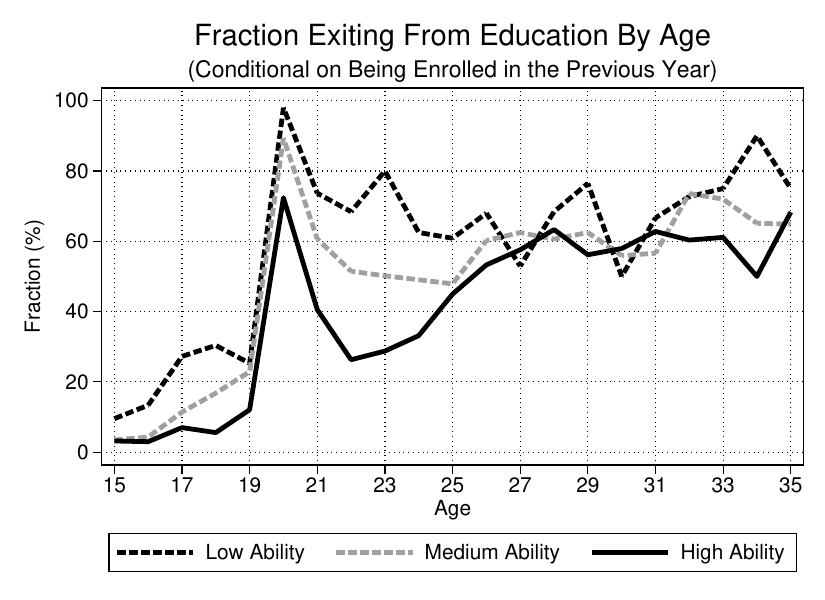}}\subfloat[Re-enrollment Rate by Age]{
\centering{}\includegraphics[width=0.4\columnwidth]{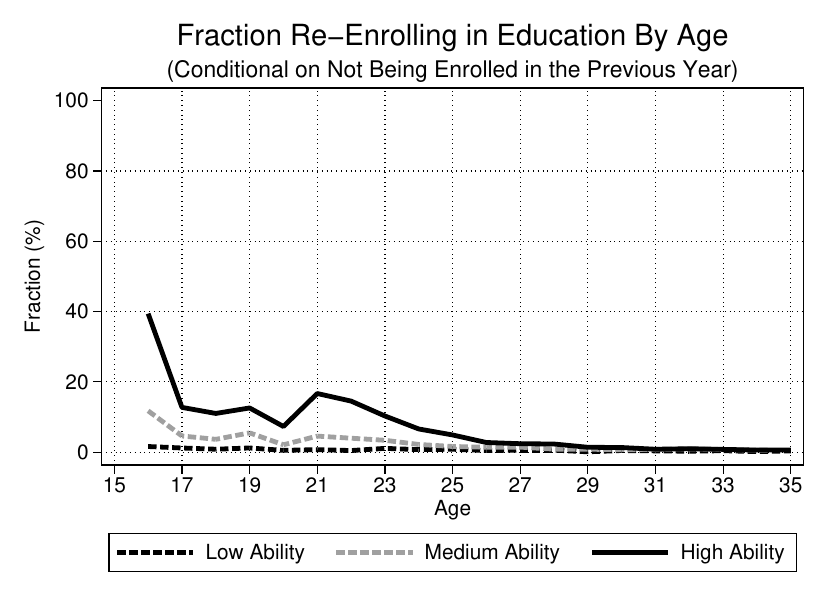}}
\par\end{centering}
\begin{centering}
\subfloat[Track Choice in Middle/High School]{
\centering{}\includegraphics[width=0.4\columnwidth]{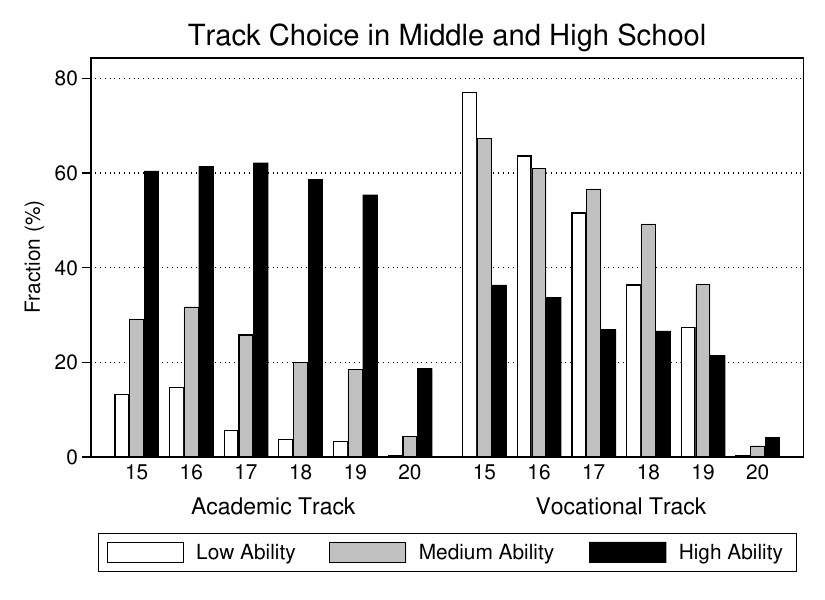}}\subfloat[Fraction working by age]{
\centering{}\includegraphics[width=0.4\columnwidth]{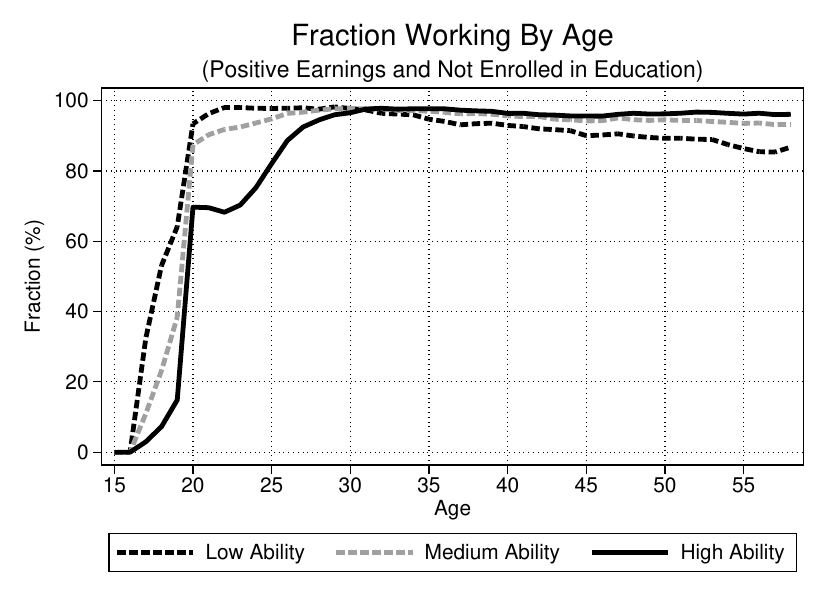}}
\par\end{centering}
\begin{centering}
\subfloat[Mean Earnings by Age]{
\centering{}\includegraphics[width=0.4\columnwidth]{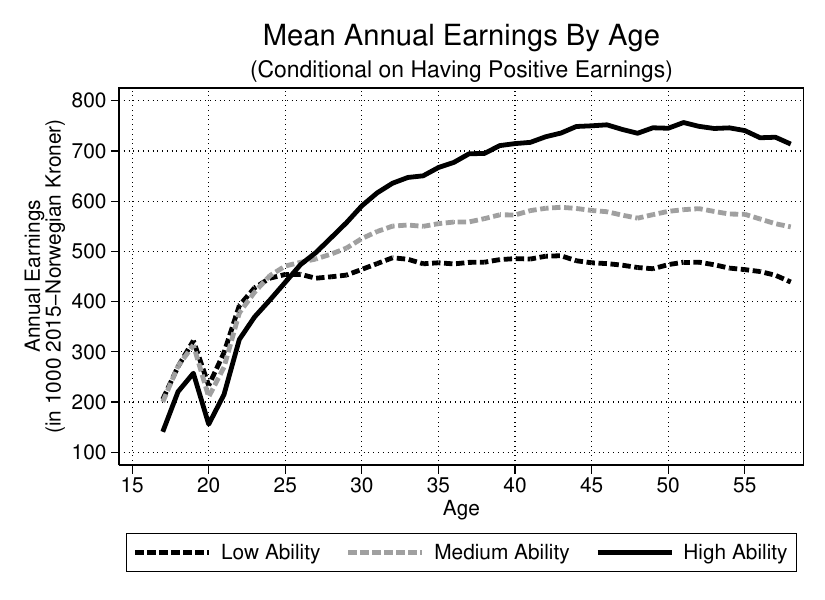}}\subfloat[Standard Deviation of Earnings by Age]{
\centering{}\includegraphics[width=0.4\columnwidth]{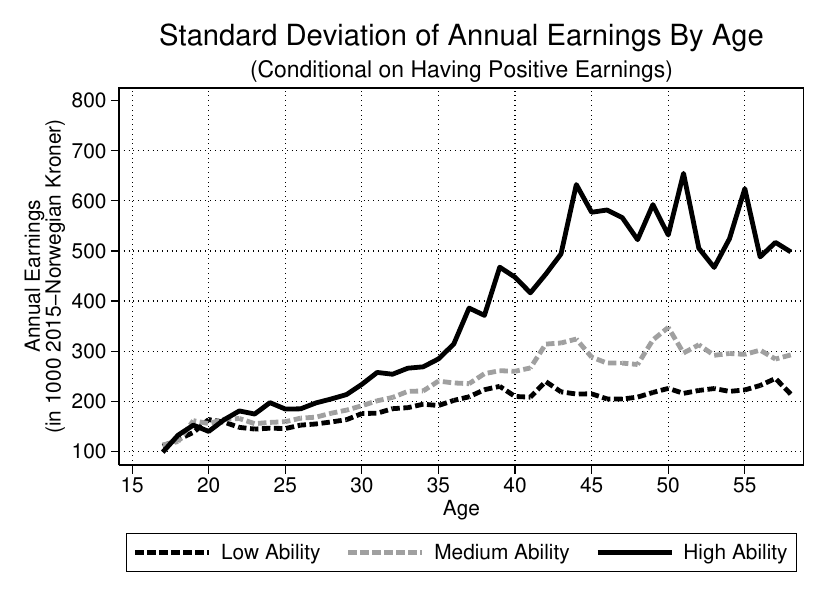}}
\par\end{centering}

\emph{\scriptsize{}Note:}{\scriptsize{} The sample consists of Norwegian
males born 1955-1960, who grew up up in a municipality which hadn't
implemented the compulsory schooling reform by the year they turned
age 14 (see details in Section \ref{subsec:Education-System-Reform}).
Individual are followed over ages 15--58, corresponding to
calendar years 1970--2018, unless there is natural attrition due
to death or out-migration. Individuals' ability is split in three
discrete categories, constructed as low (stanine IQ scores 1-4), medium
(scores 5-6) and high (scores 7-9). N=9,143.}{\scriptsize\par}
\end{figure}

\paragraph{Descriptives} Figure \ref{fig:Desc-Stats} presents key variables in our dataset. In each panel, we categorize individuals as either low (scores 1-4), medium (scores 5-6) or high (scores 7-9) ability. Panel (a) shows the distributions of final years of education, where we find clear associations between education and ability. Panel (b) shows the fraction of individuals enrolled in education by age. Individuals are more likely to attend education during the early part of their life-cycle, with gradual declines in the enrollment rate up to age 30, and virtually no enrollment beyond age 33. Next, in panels (c)-(d), we consider the conditional exit and re-enrollment rates, focusing on the earlier part of the life-cycle. The exit rate is substantially higher among low- and medium-ability individuals, while high-ability individuals have consistently higher re-enollment. Panel (e) illustrates the choices of academic and vocational tracks in middle and high school, reflecting clear ability-related differences in track choices.  Next, panel (f) shows employment rates by ability. While low-ability individuals reach an employment rate of 96 percent already at age 20, high- and medium-ability individuals gradually increase their employment rates until age 30. Beyond age 30, the employment rates remain relatively stable across all groups, though low-ability have earlier labor market exits. Panels (g)-(h) show age-specific means and standard deviations of annual earnings (in 1000s, 2015-Norwegian Kroner) conditional on working, respectively, by ability. The age-earnings profiles are increasing for all ability groups up to age 45, aside from a temporary drop in earnings at ages 19/20 due to military service. Earnings dispersion is stable in the earlier parts of the life-cycle and relatively similar across ability groups, but increases sharply after age 35.

\begin{figure}[h!]\centering
\caption{Illustration of the Decision Tree and Transitions.}\label{Decision tree and data}

\tikzset{
	treenode/.style = {shape=rectangle, rounded corners, draw, align=center, bottom color=blue!20},
	root/.style     = {treenode, font=\small, draw=none},
	env/.style      = {treenode, font=\small, draw=none},
	dummy/.style    = {circle,draw}
}

\subfloat[Academic Track at Age 17]{
\begin{tikzpicture}
[
	x=30pt,
	y=26pt,
	yscale=-1,
	xscale=1,
	baseline=-120pt,
	grow                    = right,
	edge from parent/.style = {draw, -latex},
	every node/.style       = {font=\footnotesize, minimum width={2cm}},
	sloped
]

\node [root, top color = white, bottom color=white, draw] (0) at (-10,0) {\textbf{Academic}};

\node [env, top color = bwtabblue, bottom color = bwtabblue, color = bwtaborange, scale = 0.8] (1) at (-5,-3.2) {\textbf{Work}};
\draw[->, thick] (0) edge node[left of = 0, rotate = 67, node distance = 0.3cm]{\textbf{7 \%}} (1);
\node [env, top color = bwtabblue, bottom color = bwtabblue, color = bwtaborange, scale = 0.55] (11) at (0,-4.2) {Work};
\draw[->] (1) edge node[right of = 4, yshift=0.5cm, rotate=20, node distance=1cm]{\tiny 91\%}(11);
\node [env, top color = bwtabred, bottom color = bwtabred, color = bwtaborange, scale = 0.55] (12) at (0,-3.7) {Academic};
\draw[->] (1) edge node[right of = 4, yshift=0.3cm, rotate=10, node distance=1cm]{\tiny 2\%}(12);
\node [env, top color = bwtabpurple, bottom color = bwtabpurple, scale = 0.55] (13) at (0,-3.2) {Vocational};
\draw[->] (1) edge node[right of = 4, yshift=0.13cm, rotate=-2, node distance=1cm]{\tiny 1\%}(13);
\node [env, top color = bwtaborange, bottom color = bwtaborange, scale = 0.55] (14)  at (0,-2.7) {Home};
\draw[->] (1) edge node[right of = 4, yshift=-0.05cm, rotate=-10, node distance=1cm]{\tiny 6\%}(14);
\coordinate (a1) at (3,-4.2);
\draw [->, dashed, color=tabgrey] (11) to[right] node[auto] {} (a1);
\coordinate (a2) at (3,-3.7);
\draw [->, dashed, color=tabgrey] (12) to[right] node[auto] {} (a2);
\coordinate (a3) at (3,-3.2);
\draw [->, dashed, color=tabgrey] (13) to[right] node[auto] {} (a3);
\coordinate (a4) at (3,-2.7);
\draw [->, dashed, color=tabgrey] (14) to[right] node[auto] {} (a4);

\node [env, top color = bwtabred, bottom color = bwtabred, color = bwtaborange, scale=0.8] (2) at (-5,-1.1) {\textbf{Academic}};
\draw[->, thick] (0) edge node[right of = 0, yshift=0.35cm,  rotate = 20.0, node distance = 0cm]{\textbf{74 \%}} (2);
\node [env, top color = bwtabblue, bottom color = bwtabblue, color = bwtaborange, scale = 0.55] (21) at (0,-2.1) {Work};
\draw[->] (2) edge node[right of = 4, yshift=0.5cm, rotate=20, node distance=1cm]{\tiny 4\%}(21);
\node [env, top color = bwtabred, bottom color = bwtabred, color = bwtaborange, scale = 0.55] (22) at (0,-1.6) {Academic};
\draw[->] (2) edge node[right of = 4, yshift=0.3cm, rotate=10, node distance=1cm]{\tiny 83\%}(22);
\node [env, top color = bwtabpurple, bottom color = bwtabpurple, scale = 0.55] (23) at (0,-1.1) {Vocational};
\draw[->] (2) edge node[right of = 4, yshift=0.13cm, rotate=-2, node distance=1cm]{\tiny 9\%}(23);
\node [env, top color = bwtaborange, bottom color = bwtaborange, scale = 0.55] (24)  at (0,-0.6) {Home};
\draw[->] (2) edge node[right of = 4, yshift=-0.05cm, rotate=-10, node distance=1cm]{\tiny 4\%}(24);
\coordinate (e1) at (3,-2.1);
\draw [->, dashed, color = tabgrey] (21) to[right] node[auto] {} (e1);
\coordinate (e2) at (3,-1.6);
\draw [->, dashed, color = tabgrey] (22) to[right] node[auto] {} (e2);
\coordinate (e3) at (3,-1.1);
\draw [->, dashed, color = tabgrey] (23) to[right] node[auto] {} (e3);
\coordinate (e4) at (3,-0.6);
\draw [->, dashed, color = tabgrey] (24) to[right] node[auto] {} (e4);

\node [env, top color = bwtabpurple, bottom color = bwtabpurple, scale = 0.8] (3) at (-5,1.1) {\textbf{Vocational}};
\draw[->, thick] (0) edge node[right of = 0, rotate=-20, yshift = 0.25cm, node distance = 0.2cm]{\textbf{ 10 \%}} (3);
\node [env, top color = bwtabblue, bottom color = bwtabblue, color = bwtaborange, scale = 0.55] (31) at (0,0.1) {Work};
\draw[->] (3) edge node[right of = 4, yshift=0.5cm, rotate=20, node distance=1cm]{\tiny 5\%}(31);
\node [env, top color = bwtabred, bottom color = bwtabred, color = bwtaborange, scale = 0.55] (32) at (0,0.6) {Academic};
\draw[->] (3) edge node[right of = 4, yshift=0.3cm, rotate=10, node distance=1cm]{\tiny 17\%}(32);
\node [env, top color = bwtabpurple, bottom color = bwtabpurple, scale = 0.55] (33) at (0,1.1) {Vocational};
\draw[->] (3) edge node[right of = 4, yshift=0.13cm, rotate=-2, node distance=1cm]{\tiny 74\%}(33);
\node [env, top color = bwtaborange, bottom color = bwtaborange, scale = 0.55] (34)  at (0,1.6) {Home};
\draw[->] (3) edge node[right of = 4, yshift=-0.05cm, rotate=-10, node distance=1cm]{\tiny 4\%}(34);
\coordinate (c1) at (3,0.1);
\draw [->, dashed, color = tabgrey] (31) to[right] node[auto] {} (c1);
\coordinate (c2) at (3,0.6);
\draw [->, dashed, color = tabgrey] (32) to[right] node[auto] {} (c2);
\coordinate (c3) at (3,1.1);
\draw [->, dashed, color = tabgrey] (33) to[right] node[auto] {} (c3);
\coordinate (c4) at (3,1.6);
\draw [->, dashed, color = tabgrey] (34) to[right] node[auto] {} (c4);

\node [env, top color = bwtaborange, bottom color = bwtaborange, scale = 0.8] (4)  at (-5,3.2) {\textbf{Home}};
\draw[->, thick] (0) edge node[right of = 0, yshift=0.1cm, rotate = -45, node distance = 0.25cm]{\textbf{9 \%}} (4);
\node [env, top color = bwtabblue, bottom color=bwtabblue, color = bwtaborange, scale = 0.55] (41) at (0,2.2) {Work};
\draw[->] (4) edge node[right of = 4, yshift=0.5cm, rotate=20, node distance=1cm]{\tiny 48\%}(41);
\node [env, top color = bwtabred, bottom color=bwtabred, color = bwtaborange, scale = 0.55] (42) at (0,2.7) {Academic};
\draw[->] (4) edge node[right of = 4, yshift=0.3cm, rotate=10, node distance=1cm]{\tiny 4\%}(42);
\node [env, top color = bwtabpurple, bottom color=bwtabpurple, scale = 0.55] (43) at (0,3.2) {Vocational};
\draw[->] (4) edge node[right of = 4, yshift=0.13cm, rotate=-2, node distance=1cm]{\tiny 2\%}(43);
\node [env, top color = bwtaborange, bottom color=bwtaborange, scale = 0.55] (44)  at (0,3.7)
{Home};
\draw[->] (4) edge node[right of = 4, yshift=-0.05cm, rotate=-10, node distance=1cm]{\tiny 46\%}(44);
\coordinate (b1) at (3,2.2);
\draw [->, dashed, color = tabgrey] (41) to[right] node[auto] {} (b1);
\coordinate (b2) at (3,2.7);
\draw [->, dashed, color = tabgrey] (42) to[right] node[auto] {} (b2);
\coordinate (b3) at (3,3.2);
\draw [->, dashed, color = tabgrey] (43) to[right] node[auto] {} (b3);
\coordinate (b4) at (3,3.7);
\draw [->, dashed, color = tabgrey] (44) to[right] node[auto] {} (b4);
\end{tikzpicture}}

\subfloat[Vocational Track at Age 17]{
\begin{tikzpicture}
[
	x=30pt,
	y=26pt,
	yscale=-1,
	xscale=1,
	baseline=-120pt,
	grow                    = right,
	edge from parent/.style = {draw, -latex},
	every node/.style       = {font=\footnotesize, minimum width={2cm}},
	sloped
]

\node [root, top color = white, bottom color=white, draw] (0) at (-10,0) {\textbf{Vocational}};

\node [env, top color = bwtabblue, bottom color = bwtabblue, color = bwtaborange, scale = 0.8] (1) at (-5,-3.2) {\textbf{Work}};
\draw[->, thick] (0) edge node[left of = 0, rotate = 67, node distance = 0.3cm]{\textbf{8 \%}} (1);
\node [env, top color = bwtabblue, bottom color = bwtabblue, color = bwtaborange, scale = 0.55] (11) at (0,-4.2) {Work};
\draw[->] (1) edge node[right of = 4, yshift=0.5cm, rotate=20, node distance=1cm]{\tiny 95\%}(11);
\node [env, top color = bwtabred, bottom color = bwtabred, color = bwtaborange, scale = 0.55] (12) at (0,-3.7) {Academic};
\draw[->] (1) edge node[right of = 4, yshift=0.3cm, rotate=10, node distance=1cm]{\tiny 0\%}(12);
\node [env, top color = bwtabpurple, bottom color = bwtabpurple, scale = 0.55] (13) at (0,-3.2) {Vocational};
\draw[->] (1) edge node[right of = 4, yshift=0.13cm, rotate=-2, node distance=1cm]{\tiny 1\%}(13);
\node [env, top color = bwtaborange, bottom color = bwtaborange, scale = 0.55] (14)  at (0,-2.7) {Home};
\draw[->] (1) edge node[right of = 4, yshift=-0.05cm, rotate=-10, node distance=1cm]{\tiny 4\%}(14);
\coordinate (a1) at (3,-4.2);
\draw [->, dashed, color=tabgrey] (11) to[right] node[auto] {} (a1);
\coordinate (a2) at (3,-3.7);
\draw [->, dashed, color=tabgrey] (12) to[right] node[auto] {} (a2);
\coordinate (a3) at (3,-3.2);
\draw [->, dashed, color=tabgrey] (13) to[right] node[auto] {} (a3);
\coordinate (a4) at (3,-2.7);
\draw [->, dashed, color=tabgrey] (14) to[right] node[auto] {} (a4);

\node [env, top color = bwtabred, bottom color = bwtabred, color = bwtaborange, scale=0.8] (2) at (-5,-1.1) {\textbf{Academic}};
\draw[->, thick] (0) edge node[right of = 0, yshift=0.35cm,  rotate = 20.0, node distance = 0cm]{\textbf{10 \%}} (2);
\node [env, top color = bwtabblue, bottom color = bwtabblue, color = bwtaborange, scale = 0.55] (21) at (0,-2.1) {Work};
\draw[->] (2) edge node[right of = 4, yshift=0.5cm, rotate=20, node distance=1cm]{\tiny 2\%}(21);
\node [env, top color = bwtabred, bottom color = bwtabred, color = bwtaborange, scale = 0.55] (22) at (0,-1.6) {Academic};
\draw[->] (2) edge node[right of = 4, yshift=0.3cm, rotate=10, node distance=1cm]{\tiny 86\%}(22);
\node [env, top color = bwtabpurple, bottom color = bwtabpurple, scale = 0.55] (23) at (0,-1.1) {Vocational};
\draw[->] (2) edge node[right of = 4, yshift=0.13cm, rotate=-2, node distance=1cm]{\tiny 9\%}(23);
\node [env, top color = bwtaborange, bottom color = bwtaborange, scale = 0.55] (24)  at (0,-0.6) {Home};
\draw[->] (2) edge node[right of = 4, yshift=-0.05cm, rotate=-10, node distance=1cm]{\tiny 3\%}(24);
\coordinate (e1) at (3,-2.1);
\draw [->, dashed, color = tabgrey] (21) to[right] node[auto] {} (e1);
\coordinate (e2) at (3,-1.6);
\draw [->, dashed, color = tabgrey] (22) to[right] node[auto] {} (e2);
\coordinate (e3) at (3,-1.1);
\draw [->, dashed, color = tabgrey] (23) to[right] node[auto] {} (e3);
\coordinate (e4) at (3,-0.6);
\draw [->, dashed, color = tabgrey] (24) to[right] node[auto] {} (e4);

\node [env, top color = bwtabpurple, bottom color = bwtabpurple, scale = 0.8] (3) at (-5,1.1) {\textbf{Vocational}};
\draw[->, thick] (0) edge node[right of = 0, rotate=-20, yshift = 0.25cm, node distance = 0.2cm]{\textbf{ 79 \%}} (3);
\node [env, top color = bwtabblue, bottom color = bwtabblue, color = bwtaborange, scale = 0.55] (31) at (0,0.1) {Work};
\draw[->] (3) edge node[right of = 4, yshift=0.5cm, rotate=20, node distance=1cm]{\tiny 16\%}(31);
\node [env, top color = bwtabred, bottom color = bwtabred, color = bwtaborange, scale = 0.55] (32) at (0,0.6) {Academic};
\draw[->] (3) edge node[right of = 4, yshift=0.3cm, rotate=10, node distance=1cm]{\tiny 1\%}(32);
\node [env, top color = bwtabpurple, bottom color = bwtabpurple, scale = 0.55] (33) at (0,1.1) {Vocational};
\draw[->] (3) edge node[right of = 4, yshift=0.13cm, rotate=-2, node distance=1cm]{\tiny 78\%}(33);
\node [env, top color = bwtaborange, bottom color = bwtaborange, scale = 0.55] (34)  at (0,1.6) {Home};
\draw[->] (3) edge node[right of = 4, yshift=-0.05cm, rotate=-10, node distance=1cm]{\tiny 4\%}(34);
\coordinate (c1) at (3,0.1);
\draw [->, dashed, color = tabgrey] (31) to[right] node[auto] {} (c1);
\coordinate (c2) at (3,0.6);
\draw [->, dashed, color = tabgrey] (32) to[right] node[auto] {} (c2);
\coordinate (c3) at (3,1.1);
\draw [->, dashed, color = tabgrey] (33) to[right] node[auto] {} (c3);
\coordinate (c4) at (3,1.6);
\draw [->, dashed, color = tabgrey] (34) to[right] node[auto] {} (c4);

\node [env, top color = bwtaborange, bottom color = bwtaborange, scale = 0.8] (4)  at (-5,3.2) {\textbf{Home}};
\draw[->, thick] (0) edge node[right of = 0, yshift=0.1cm, rotate = -45, node distance = 0.25cm]{\textbf{3 \%}} (4);
\node [env, top color = bwtabblue, bottom color=bwtabblue, color = bwtaborange, scale = 0.55] (41) at (0,2.2) {Work};
\draw[->] (4) edge node[right of = 4, yshift=0.5cm, rotate=20, node distance=1cm]{\tiny 56\%}(41);
\node [env, top color = bwtabred, bottom color=bwtabred, color = bwtaborange, scale = 0.55] (42) at (0,2.7) {Academic};
\draw[->] (4) edge node[right of = 4, yshift=0.3cm, rotate=10, node distance=1cm]{\tiny 2\%}(42);
\node [env, top color = bwtabpurple, bottom color=bwtabpurple, scale = 0.55] (43) at (0,3.2) {Vocational};
\draw[->] (4) edge node[right of = 4, yshift=0.13cm, rotate=-2, node distance=1cm]{\tiny 6\%}(43);
\node [env, top color = bwtaborange, bottom color=bwtaborange, scale = 0.55] (44)  at (0,3.7)
{Home};
\draw[->] (4) edge node[right of = 4, yshift=-0.05cm, rotate=-10, node distance=1cm]{\tiny 37\%}(44);
\coordinate (b1) at (3,2.2);
\draw [->, dashed, color = tabgrey] (41) to[right] node[auto] {} (b1);
\coordinate (b2) at (3,2.7);
\draw [->, dashed, color = tabgrey] (42) to[right] node[auto] {} (b2);
\coordinate (b3) at (3,3.2);
\draw [->, dashed, color = tabgrey] (43) to[right] node[auto] {} (b3);
\coordinate (b4) at (3,3.7);
\draw [->, dashed, color = tabgrey] (44) to[right] node[auto] {} (b4);
\end{tikzpicture}}

\vspace{-.5cm}
\begin{center}
\begin{minipage}[t]{\columnwidth}
\emph{\scriptsize{}Note:}{\scriptsize{} This figure shows the decision tree between ages 17 to 19 for those attending the academic (a) or vocational (b) tracks at ages 15--17, after completing compulsory schooling at age 14. This also shows the fraction transiting from one state to another at each age. The corresponding decision tree and transitions between ages 14 to 17 are illustrated in the Appendix Figure \ref{Decision tree and data app}.}{\scriptsize\par}
\end{minipage}
\end{center}
\end{figure}

\noindent Besides some of the key moments of our dataset that are illustrated in Figure \ref{fig:Desc-Stats}, our model implementation will also rely crucially on the fractions of individuals who transit between different choices over their life-cycle. To illustrate the rich transition patterns present in our data, in Figure \ref{Decision tree and data}, panels (a) and (b), we consider individuals who were enrolled in the academic and vocational tracks, respectively, at ages 15 to 17 after having compulsory schooling at age 14, and follow their transition histories over the following three years in our data. Around 74\% (79\%) of these individuals continue in the academic (vocational) track for a third year while 10\% switch to the other educational track. Less than 10\% enter the labor market, while the remaining stay at home. Among those who decide to stay at home for one period, roughly 6-8\% re-enroll in an academic or vocational school. These rich patterns of (i) persistence in choices over time, (ii) presence of track switching, and (iii) re-enrollment after spells of work or home stay provide a motivation for the flexible modelling of schooling choices in Section \ref{sec:Model}. \\

\subsection{The Norwegian Education System\label{subsec:Education-System-Reform}}
We describe here the structure of the Norwegian education system that existed in the 1960s, which our model is specified to fit. This system had four main stages; see Appendix Figure \ref{fig:Education-System}. The first stage consisted of seven years of compulsory elementary education. The second stage involved tracking, where pupils could attend either a vocational middle school (\emph{framhaldsskole}) or an academic middle school (\emph{realskole}). The vocational middle school could be either one, two or three years, with most attending two years, while the academic middle school could be either two or three years, where the final year was only required for those who wanted to later pursue further academic education. The third stage corresponded to a high school education, which again was track-specific. After attending the academic middle school, students could move on to attend a academic high school (\emph{gymnas}). In contrast, pupils attending the vocational middle school normally didn't quality for the academic high school, but could rather attend a vocational high school (\emph{yrkesskole}). The academic high school was required to be three years, while the vocational high school could be of varying lengths, depending on the particular vocational field. Finally, the fourth stage involved higher education, leading up to an academic degree at a college or a university, enrollment to which was typically contingent on having completed the academic high school. There existed two main degrees in tertiary education; a 4-year (\emph{cand.mag}) and a 6-year (\emph{hovedfag}), but degrees of other durations also existed.

\subsection{Model Implementation and Estimation}\label{subsec:Model-Implementation}

\noindent The model we described in Section \ref{sec:Model} is set up as a standard Markov decision process (MDP), which can be solved by a simple backward induction procedure \citep{Puterman.1994,White.1993,Rust.1994}. In the final period $T$, there is no future to consider, and the optimal action is choosing the alternative with the highest immediate utility in each state. With the decision rule for the final period, we can determine all other optimal decisions recursively.\footnote{We use an open-source research code \textit{respy} \citep{Gabler.2020b} that allows for the flexible specification, simulation, and estimation of EKW models. Detailed documentation of this software and its numerical components is available at \url{http://respy.readthedocs.io}.}

\begin{table}[H]\centering
  \caption{Summary of Moments Used in the Estimation.}\label{Moments summary}\par

\begin{tabular*}{1\textwidth}{@{\extracolsep{\fill}}lr}\toprule\addlinespace
    \textbf{Type of Moment}                                 & \textbf{Number} \\ \addlinespace\toprule\addlinespace
   I. Average of Annual Earnings Per Period and Sector        & 288                         \\\addlinespace
    II. Standard Deviation of Annual Earnings Per Period and Sector & 288                         \\\addlinespace
    III. Fraction in Academic Schooling Per Period   & 108                         \\\addlinespace
    IV. Fraction in Vocational Schooling Per Period  & 108                         \\\addlinespace
    V. Fraction Working in Public Sector Per Period    & 108                         \\\addlinespace
    VI. Fraction Working in Private Sector Per Period    & 108                        \\\addlinespace    
    VII. Fraction Working as Self-Employed Per Period    & 108                         \\\addlinespace
    VIII. Fraction Staying at Home Per Period          & 108                         \\\addlinespace
    IX. Distribution of Final Schooling Levels Per Track  & 21                         \\ \addlinespace
    X. Distribution of Final Schooling Levels Per Sector & 54                         \\ \addlinespace
    \bottomrule
\end{tabular*}
\begin{center}
\begin{minipage}[t]{\columnwidth}
\emph{\scriptsize{}Note:}{\scriptsize{} This table provides an overview of the 1,083 moments used in the estimation by the type of moment. Moments of type I--II are each calculated between ages 18 and 50 in the private sector and ages 21 and 50 in the public sector and in self employment. Thus there are 96 moments in each ability group and $96 \times 3 = 288$ moments in total.
Moments of type III--VII are calculated for each of the 36 periods across all ability groups $36 \times 3 = 108$. 
Moments of type IX capture the distribution of schooling degrees in each of the two schooling types, which can take four values in academic schooling and three values in vocational schooling, calculated for each of the three ability types, i.e., $7 \times 3 = 21$ unique moments.
Moments of type X capture the distribution of final schooling degrees which can take six values in each sector and ability group, i.e., $6 \times 3 \times 3 = 54$ unique moments.
}{\scriptsize\par}
\end{minipage}
\end{center}
\end{table}

\paragraph{Implementation} We use the method of simulated moments \citep{Pakes.1989, Duffie.1993} to estimate the parameters $\bm{\hat{\theta}}$ of the model for each ability group. Equation \eqref{Criterion function} shows our criterion function. We select the parameterization for our analysis that minimizes the weighted squared distance between our specified set of moments computed on the observed $\bm{M_D}$ and the simulated data $\bm{M_S(\theta)}$.  We weigh the moments with a diagonal matrix $\bm{W}$ that contains the variances of the observed moments \citep{Altonji.1996} and use the \textit{estimagic} package \citep{Gabler.2022} for the optimization of the criterion function.\footnote{We use a global version of the BOBYQA algorithm within the package \citep{Powell.2009}.} We simulate a sample of $100,000$ individuals based on the candidate parameterizations of the model.
\begin{align}\label{Criterion function}
\bm{\hat{\theta}} = \argmin_{\bm{\theta} \in \bm{\Theta}}\, (\bm{M_D} - \bm{M_S(\theta)}) \bm{W^{-1}}  (\bm{M_D} - \bm{M_S}(\bm{\theta}))^\prime
\end{align}
Table \ref{Moments summary} provides an overview of the 1,083 empirical moments used in our estimation. These consist of aggregate moments of annual earnings (type I--II), aggregate annual choice proportions in each alternative (type III--VIII), and the distribution of final schooling (type XI--X). Moments of type I--II are each calculated between ages 18 to 50 in the private sector and ages 21 to 50 in the public sector and in self-employment. Thus, there are 96 such moments in each ability group and a total of $96 \times 3 = 288$  unique moments.
Moments of type III--VIII are calculated for each of the 36 periods across all ability groups, i.e., there are $36 \times 3 = 108$  unique moments.
Moments of type IX capture the distribution of schooling levels in each of the two educational tracks, which can take four values in academic schooling and three values in vocational schooling, calculated for each of the three ability types, i.e., $7 \times 3 = 21$ unique moments.
Moments of type X capture the distribution of final schooling levels which can take six values in each sector and ability group, i.e., $6 \times 3 \times 3 = 54$ unique moments.

\paragraph{Identification} The moments are selected to determine the various components of our model.
In terms of identification, it is useful to distinguish five distinct groups of parameters, for each ability group.
These include (i) wage parameters, (ii) non-pecuniary parameters, (iii) latent types, (iv) distribution of shocks, and (v) discount factor. 
Wage parameters and non-pecuniary parameters are fixed across individuals and vary across states. 
Latent types are fixed across states but vary across individuals.
Shocks vary both across states and individuals but are independent of each other.
The discount factor is fixed across individuals and across states.  \\
 
\noindent While all parameter are identified jointly by all the moments, we provide some heuristic arguments for their identification.
Wage parameters determine average wage returns to work choices across different states in the model. Changes in average wage moments as the composition of the work force changes over time is an important factor in the identification of wage returns.
Intuitively, the movements in aggregate wages across periods where people with more schooling enter the labor market help to identify the wage rewards to additional years of schooling. And, aggregate wages across periods where more high school graduates enter the labor market help to identify the wage rewards to a high school degree. Including the distribution of final schooling level as an additional set of moments further allows recovering non-linearities in the wage-schooling relationship and associated bunching at specific degrees.\footnote{Note that we do not rely on the cross-sectional wage-schooling relationship directly in our estimation, as these moments usually suffer from problems related to endogeneity of schooling and sample selection bias. Using the distribution of final schooling levels and average earnings and choice shares by period, we can nonetheless recover the parameters of the wage-schooling relationship in a relatively flexible manner.} Similar arguments apply for the identification of wage rewards to additional years of labor market experience. \\

\noindent Non-pecuniary parameters refer to the non-monetary value of choices in the model. 
Generally they can be identified by differences in choices across states keeping wages and the composition of individuals fixed.
As an example take the non-pecuniary reward to working in the private sector with a high school degree.
Given wage returns and non-pecuniary schooling returns, the fraction of private sector workers with a high school degree and the variation in sector choices across individuals with different schooling levels are important for identifying these returns.
\\

\noindent Latent types and shock parameters introduce heterogeneity across individuals.
Latent types introduce persistent heterogeneity across individuals whereas shocks introduce temporary and serially uncorrelated heterogeneity across individuals.
The model picks the combination of persistent and temporary heterogeneity that leads to the best model fit conditional on wage parameters and non-pecuniary parameters.
Consider as an example the case where half of individuals in our data leave school at age 16 and the other half at age 24.
This pattern would be much easier to reconcile with persistent differences than with serially uncorrelated taste shocks.
In this case the model would load most of the heterogeneity across individuals into two different latent types.
On the other hand, if the shares of individuals leaving school varies more continuously across periods, then this is more suggestive of a larger role of independent taste shocks. The key assumption that this representation places on the data generating process is that persistent heterogeneity does not systematically vary across states. If individuals differ in a systematically different way before and after they have obtained a high school degree for example then the approach would be invalid. Note, however, that both latent types and the distribution of shocks can be distinguished from non-pecuniary returns because they do not systematically vary across states. If choices or outcomes vary across states while the composition of individuals across these states does not change, then non-pecuniary parameters are likely to drive these differences. A similar argument holds for identifying wage parameters versus latent types and the distribution of shocks. If the change in average wage after individuals receive high school degree implies very large returns for labor market entrants, then this is not line with the choices of individuals who have not obtained a high school degree. This would rather suggest that some part of the increase is due to selection as opposed to returns. \\

\noindent Finally, conditional on the structure of wages, non-pecuniary rewards, and latent types, the average changes in choices over time provide variation that help identify the discount factor. 


%

\subsection{Model Validation}\label{subsec:Model-Validation}
We now demonstrate our model's credibility by discussing some selected parameter estimates and comparing them to the existing literature. We then report the estimated model's in-sample model fit and discuss the results from an out-of-sample validation based on a schooling reform.
\paragraph{Parameter Estimates}
All parameter estimates are reported in Appendix, Section \ref{subsec:App-A-estimation-results}, along with the associated standard errors based on simulation-based inference. In the following, we discuss some of these estimates. Most of our parameter estimates are standard and in line with the previous literature \citep{Eisenhauer.2015b,Keane.1997}. The annual discount rate is about 4\% for all ability levels. Returns to sector-specific experience are concave. Latent type heterogeneity plays an important role in shaping schooling decisions even within ability groups. The cost of re-enrollment in school after dropping out is very high. There are interesting differences in the wage returns to academic and vocational schooling by ability and sector. For the academic choice, we find the highest wage returns for high ability individuals, for whom an additional year of schooling increases wages by $16-17\%$ in all sectors. By comparison, we find only around $7\%$ return to an additional year of academic schooling for low ability individuals in the public sector. For the vocational choice, however, we find the largest wage returns for low ability individual in all sectors. It should be noted that these returns abstract from the additional degree premiums associated with completing particular years, which are also important elements of our model specification. Further, the associated non-pecuniary rewards reinforce the sorting of high-ability individuals into academic and low-ability individuals into a vocational school. For example, the non-pecuniary benefits from an academic education are negative for low ability individuals but positive for high ability. These patterns likely reflect the heterogeneous drop-out risks across ability groups found in our data.

\begin{figure}\centering
\caption{Model Fit.}\label{Model fit for choices}
\subfloat[Academic Schooling]{\scalebox{0.4}{\includegraphics{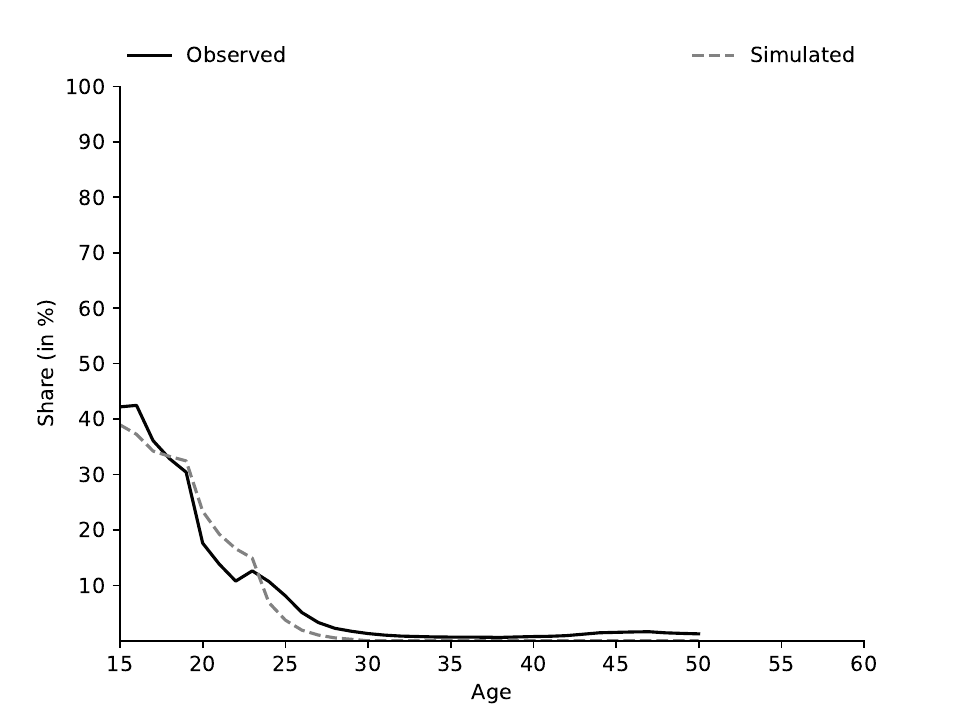}}}\hspace{0.3cm}
\subfloat[Vocational Schooling]{\scalebox{0.4}{\includegraphics{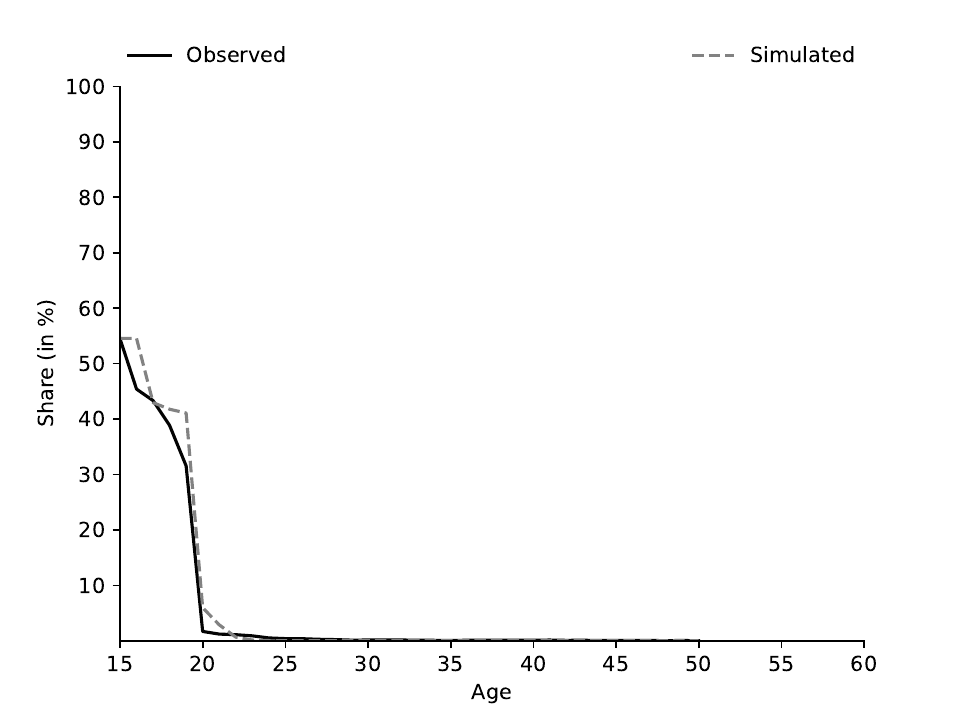}}}
\hspace{0.3cm}
\subfloat[Private Sector]{\scalebox{0.4}{\includegraphics{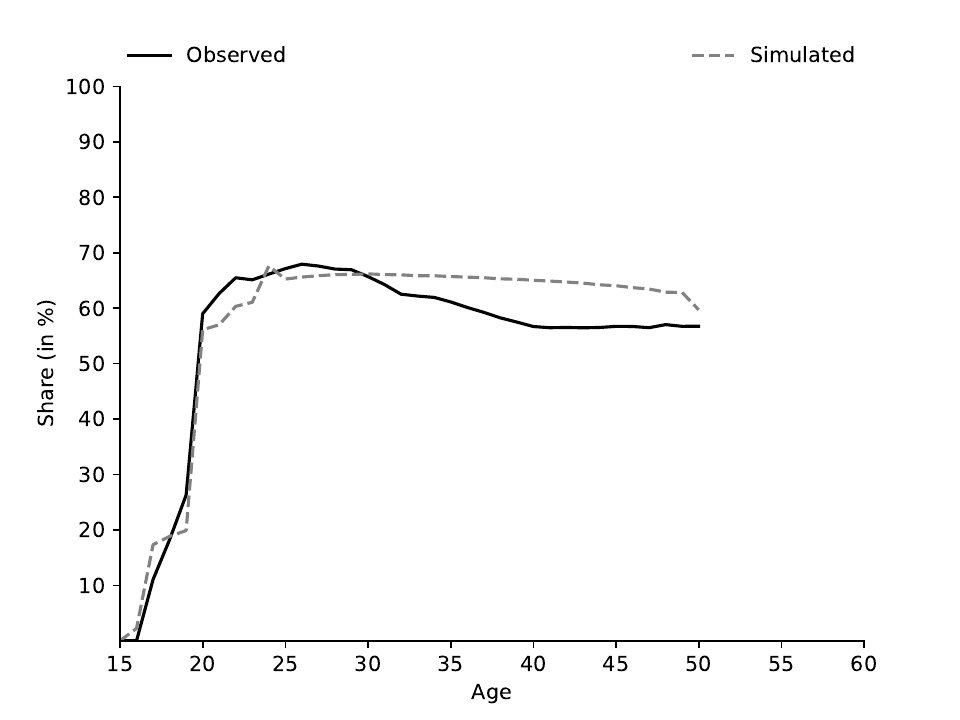}}}\hspace{0.3cm}
\subfloat[Public Sector]{\scalebox{0.4}{\includegraphics{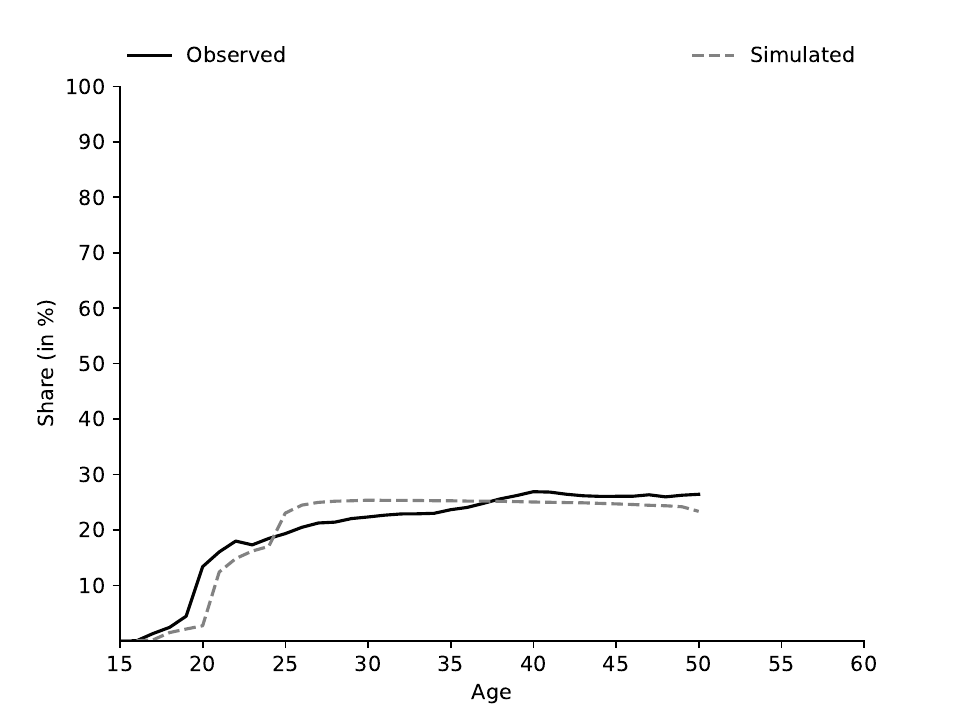}}}
\hspace{0.3cm}
\subfloat[Self-Employment]{\scalebox{0.4}{\includegraphics{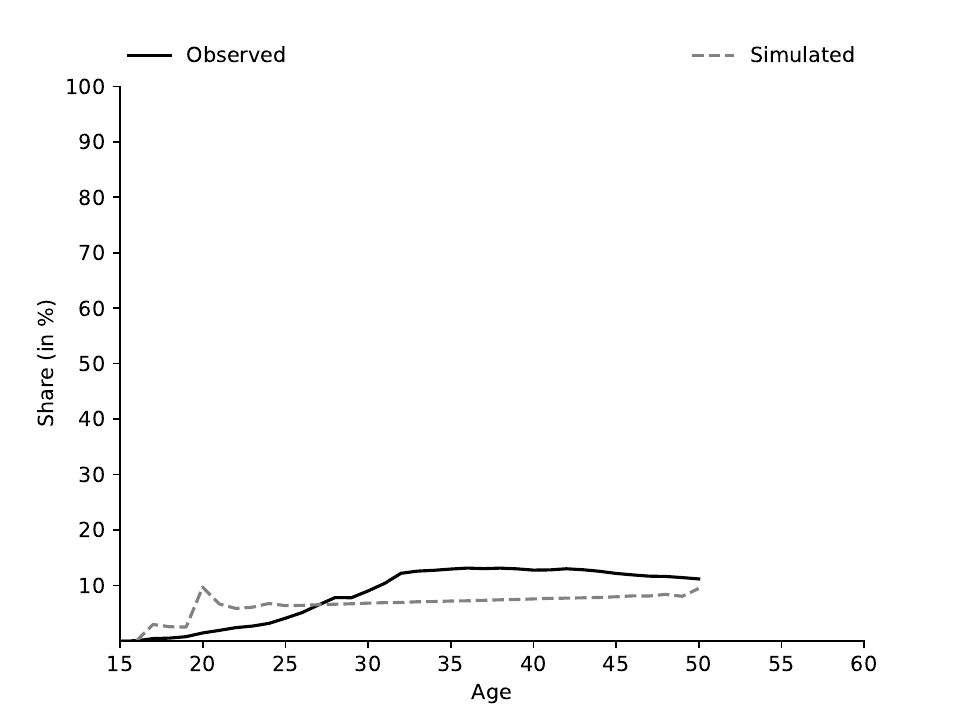}}}
\hspace{0.3cm}
\subfloat[Staying at Home]{\scalebox{0.4}{\includegraphics{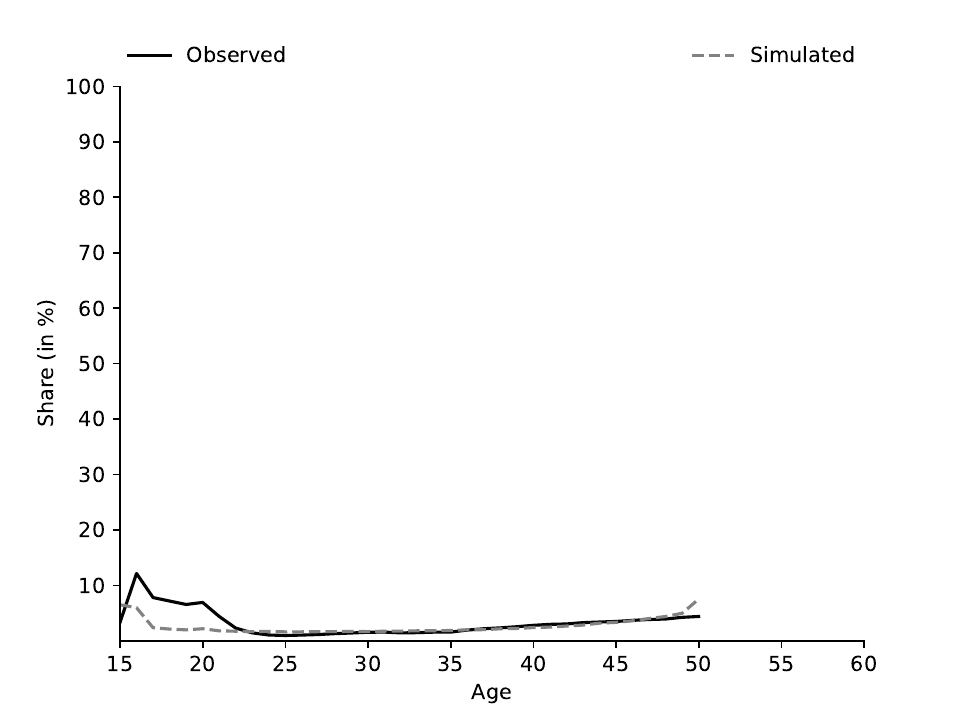}}}
\hspace{0.3cm}
\subfloat[Mean Earnings]{\scalebox{0.4}{\includegraphics{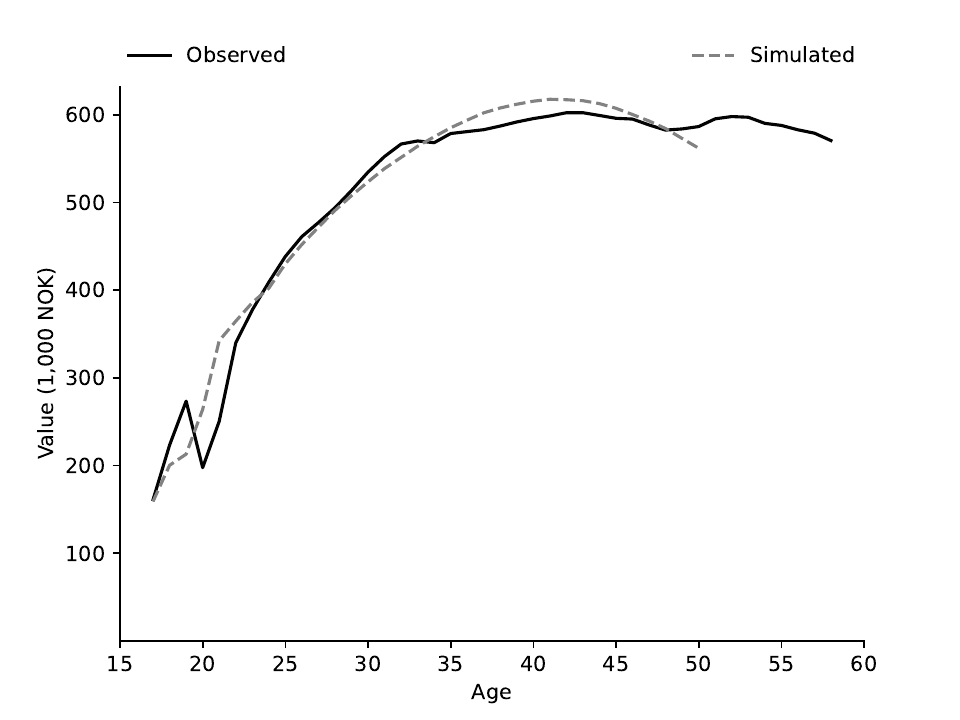}}}
\hspace{0.3cm}
\subfloat[Standard Deviation Earnings]{\scalebox{0.4}{\includegraphics{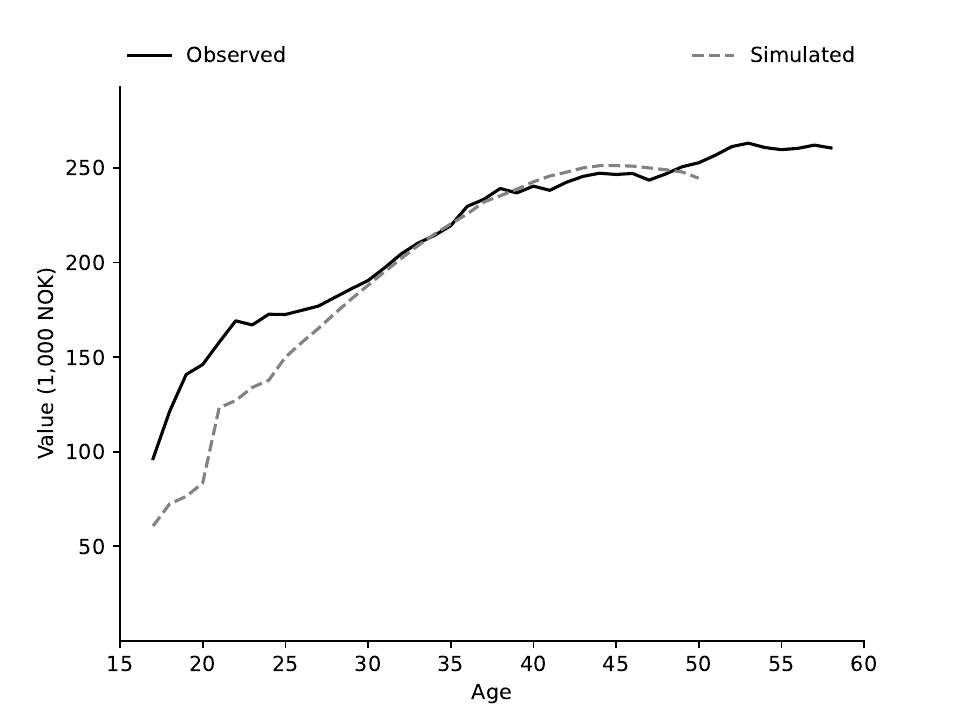}}}
\begin{center}
\begin{minipage}[t]{\columnwidth}
\emph{\scriptsize{}Note:}{\scriptsize{} The figure is based on averaging across 100,000 simulated life-cycle profiles using the estimated model.}{\scriptsize\par}
\end{minipage}
\end{center}
\end{figure}

\paragraph{In-Sample Model Fit}
We now assess our model's ability to reproduce the overall patterns of choices and earnings by comparing our simulated sample to the observed data. Figure \ref{Model fit for choices} shows the shares of individuals deciding to either attend academic (panel a) or vocational (panel b) school, private-sector work (panel c), public-sector work (panel d), self-employment (panel e) or stay at home (panel f). The model predictions (dotted gray lines) are in most cases closely aligned with the observed patterns in our data (black solid lines). Next, we show the model fit for the average (panel g) and standard deviation (panel h) of annual earnings by age. Our model does a decent job of reproducing these basic patterns over the life cycle as well.

\paragraph{Out-of-Sample Model Validation}
We now assess the out-of-sample performance of our model. For this purpose, we rely on variation in schooling choices coming from a compulsory schooling reform. As discussed in \cite{Black.2005}, since 1959, seven years of elementary education had been compulsory in Norway. However, each municipality--the lowest level of local administration--was allowed to enact nine years of compulsory school, i.e., \emph{two additional years beyond the national minimum} \emph{requirement}. In subsequent legislation in 1969, nine years of elementary education (\emph{grunnskole}) was made compulsory throughout Norway. Due to the lack of resources some municipalities nevertheless didn't enforce nine years of\emph{ }compulsory education before 1974. These features led to substantial geographic variation in compulsory education across Norway between 1960 and 1975. For more than a decade, Norwegian schools were divided into two separate systems, where the length of compulsory schooling depended on the birth year and the municipality of residence at age 14, i.e., the childhood municipality.\footnote{The staggered roll-out of the compulsory schooling reform in Norway has led to extensive literature relying on quasi-experimental designs to study the causal effects of schooling, see, e.g., \cite{Black.2005}, \cite{Monstad.2008}, \cite{Aakvik.2010}, \cite{Machin.2012},  \cite{Mogstad.2017}, and \cite{Aryal.2021}. None of these studies consider the ex-ante returns or the option values of schooling choices.}\\ 

\begin{figure}[h!]\centering
\caption{Out-of-Sample Validation Using Compulsory Schooling Reform.}\label{Model validation}
\subfloat[Overall]{\scalebox{0.45}{\includegraphics{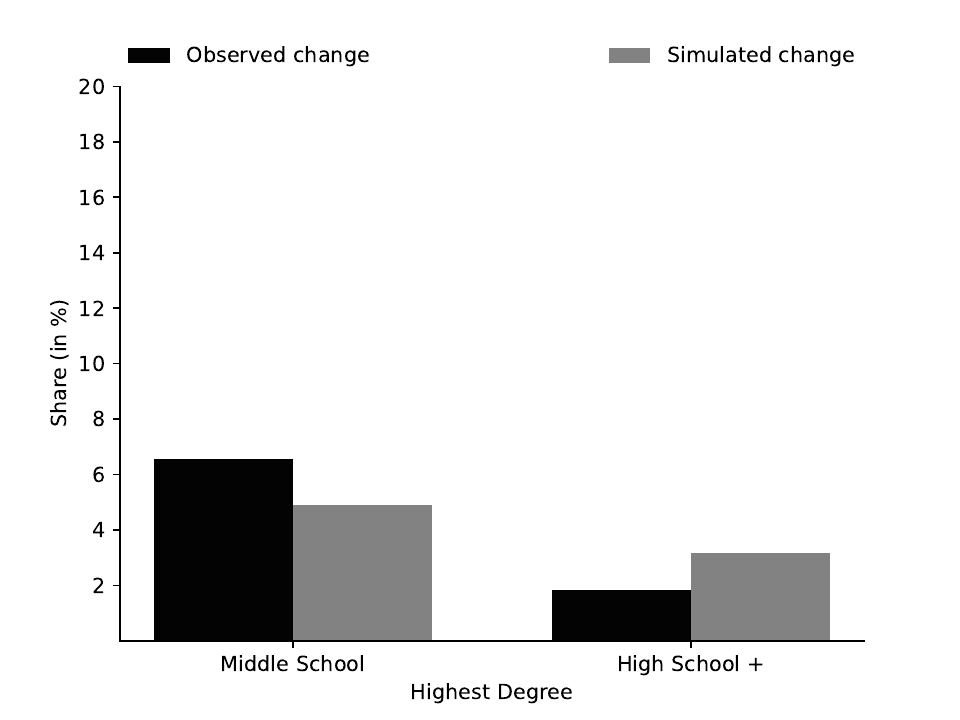}}}\hspace{0.3cm}
\subfloat[High Ability]{\scalebox{0.45}{\includegraphics{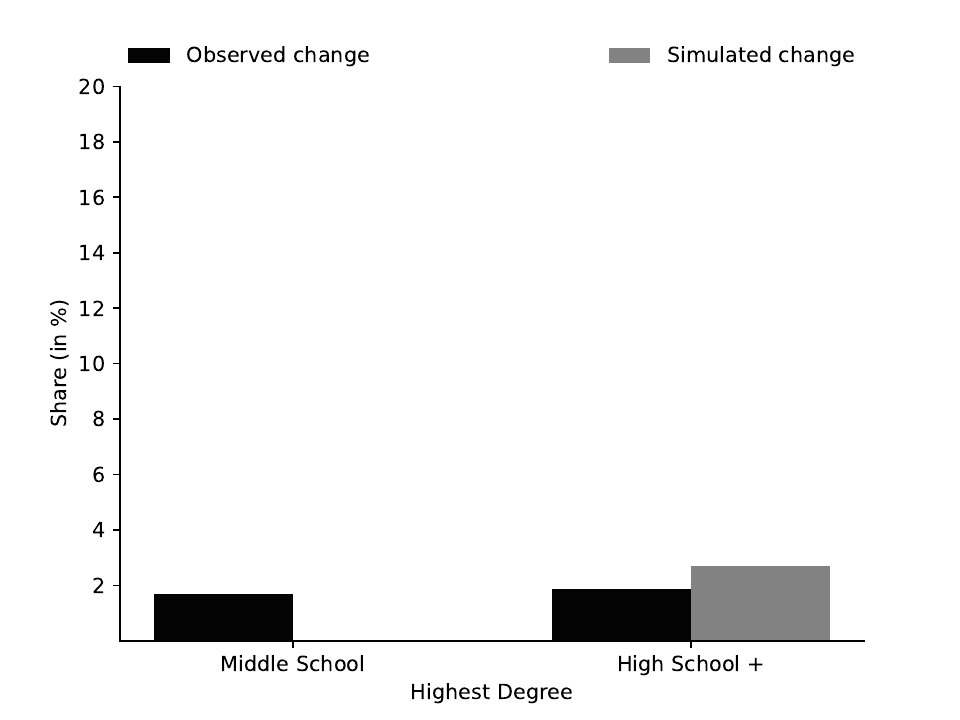}}}
\hspace{0.3cm}
\subfloat[Medium Ability]{\scalebox{0.45}{\includegraphics{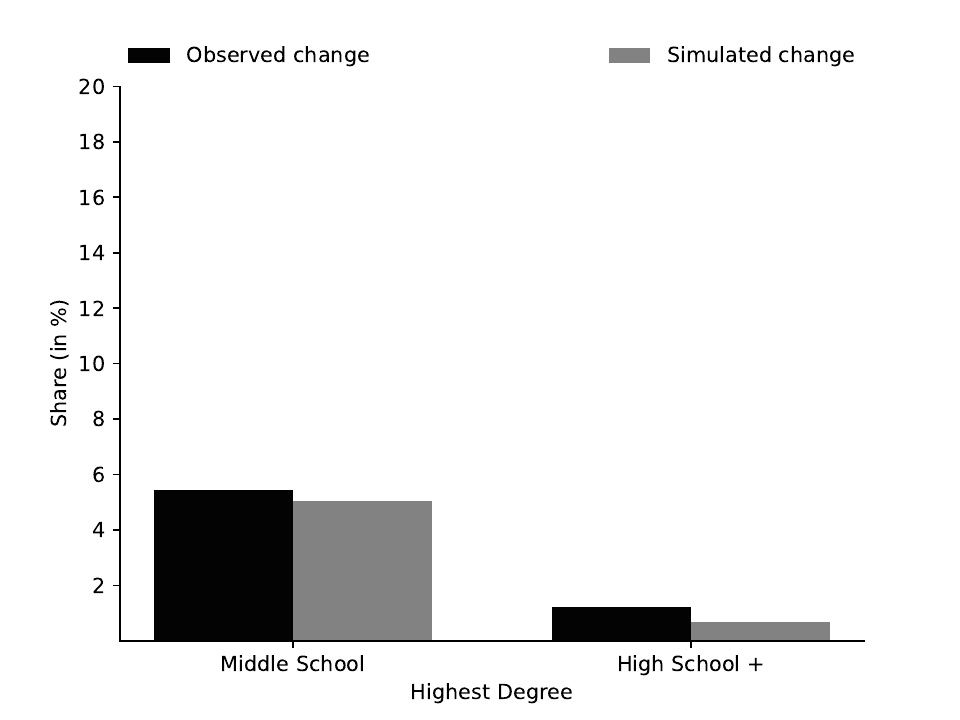}}}\hspace{0.3cm}
\subfloat[Low Ability]{\scalebox{0.45}{\includegraphics{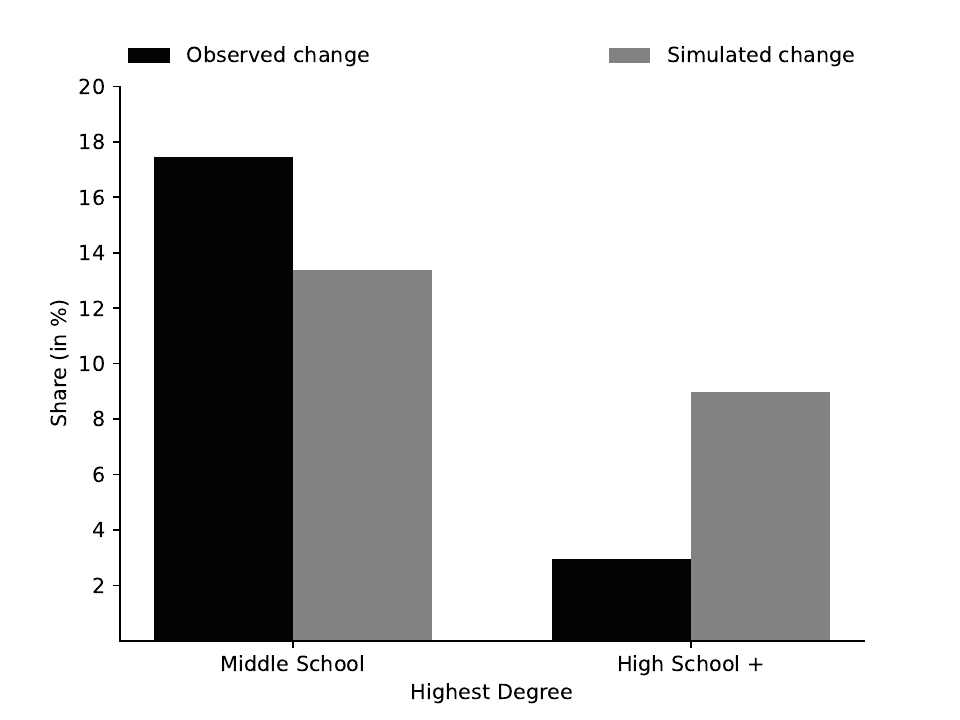}}}
\begin{center}
\begin{minipage}[t]{\columnwidth}
\emph{\scriptsize{}Note:}{\scriptsize{} The figure is based on two samples of 100,000 simulated schooling careers based on the estimated model. We first simulate the model with seven years of compulsory schooling, as in our baseline simulations. Next, we rerun the simulation but impose nine years of compulsory schooling. Throughout, we keep the random realizations of the productivity and taste shocks $\bm{\epsilon_t}$ fixed, and we are thus able to compare the schooling decisions of the same individual under the two different regimes. `Middle School' represents all individuals between 9 and 12 years of academic schooling and 9 and 11 years of vocational schooling. `High School +' represents all individuals with 12 or more years of academic schooling and with 11 or more years of vocational schooling. The black bars \textit{Simulated change} show the percentage points differences in the fraction of individuals who receive a middle school degree and those that receive at least a high school degree, respectively. The gray bars \textit{Observed change} show the same difference in the observed data before and after the reform. We do not distinguish between vocational and academic tracks in this calculation.
}{\scriptsize\par}
\end{minipage}
\end{center}
\end{figure}

\noindent We exploit the Norwegian compulsory schooling reform to validate our model in the following manner. First, as described in Section \ref{subsec:Data-Sample}, our model is estimated solely using data on individuals born 1955--1960 who were subject to the pre-reform education system, and is geared to capture the schooling system that exited pre-reform. Second, for the purposes of model validation, we constraint the choice sets in a modified version of our model where individuals cannot leave schooling before nine years to reflect the implementation of a nine year compulsory schooling. We then compare the predicted changes in education choices in our model to the observed differences in education choices across pre- and post-reform cohorts born 1955--1960, respectively. Finally, in order to provide an insightful and strong validation of our model, we focus on ``inframarginal'' responses beyond the new minimum schooling requirement. We focus on such responses as the reform is mechanically expected to induce increases in educational attainment up to the new minimum schooling requirement among those who otherwise would have had less than nine years of schooling, both in the observed data and in our model, while it is less clear how educational choices beyond the new minimum schooling requirement are affected.  \\

\noindent Figure \ref{Model validation} provides the results of our validation exercise. We compare the predicted changes in the fractions of individuals with nine years of completed schooling or more but less than a high school degree, i.e., middle school, and the fractions of individuals with at least a high school degree, respectively, in model simulations pre- and post-reform to the corresponding changes observed in actual data. The middle school margin is directly affected by the reform as the compulsory schooling reform forces individuals to attend at least nine years of schooling. The high school margin on the other hand is not directly affected by the reform. The validation exercise highlights the main motivation of our modelling approach as there are indeed changes in the distribution of education attainment across margins that are not directly affected by the policy reform, and the model does predict that there will be such ``inframarginal'' responses. In panel (a), our model predicts an increase in the share completing middle school, which is the margin that is directly affected by the reform, of around 5\%. Additionally we predict that the share completing a high school degree or more by around 3\%. In the validation sample, the share of individuals with a middle school degree increases by 6.5\% and the share of individuals with at least a high school degree increases by around 2\%. Thus, our model predictions slightly overestimate the ``inframarginal'' responses, but is capable of reproducing the general patterns found in the validation sample. Considering the different ability groups separately, we show that we are able to predict changes for medium and high ability groups quite well, while our model overestimates the ``inframarginal'' responses for low ability individuals.
Notably, the low ability group also has the fewest individuals who proceed until high school in the observed data, which renders the extrapolation that the reform induces in our baseline model more challenging for this group. In a nutshell, the predictions from our model line up with the observed changes. We return to these findings in Section \ref{subsec:Results-Policy}, where we provide additional evidence on economic mechanisms from our model that shed more light on the nature of these responses.


\section{Empirical Results}\label{sec:Results}
We now present our empirical findings based on the estimated model. We first document heterogeneity in ex-ante returns by year of schooling, choices of academic and vocational track, and ability, before we consider the importance of option values, wage risk and sectoral choices, and finally the impacts of alternative education reforms. In these calculations, we simulate life-cycle histories for 5,000 individuals for each ability group using the estimated model.

\subsection{Evidence on Ex-ante Returns}\label{subsec:Results-Exante}
To construct measures of ex-ante returns to schooling, we will compare the discounted lifetime value of attending schooling in a particular track in a given period in our model to the corresponding value associated with the best alternative choice.
Notably, since our model allows for re-enrollment in a flexible manner, the best alternative choice can include the possibility of attending more education at a later stage in the life-cycle. Individuals can thus reach a given level of final schooling at the end of their life-cycle through many different paths due to the opportunities of track switching and re-enrollment. To ease interpretation and tractability of our findings, we will therefore focus on individuals who in our model have had uninterrupted schooling careers in given track up to the period where we explore the ex-ante returns associated with different schooling choices. To avoid our results to be driven by a small number of individuals, we will drop particular transitions in our expositions for groups who have less than 0.5\% chance of reaching such transitions (e.g., low ability individuals attending college).

\begin{figure}[h!]\centering
\caption{Average Ex-ante Returns to Academic and Vocational Schooling.}\label{Ex-ante return}
\subfloat[Academic Schooling]{\scalebox{0.45}{\includegraphics{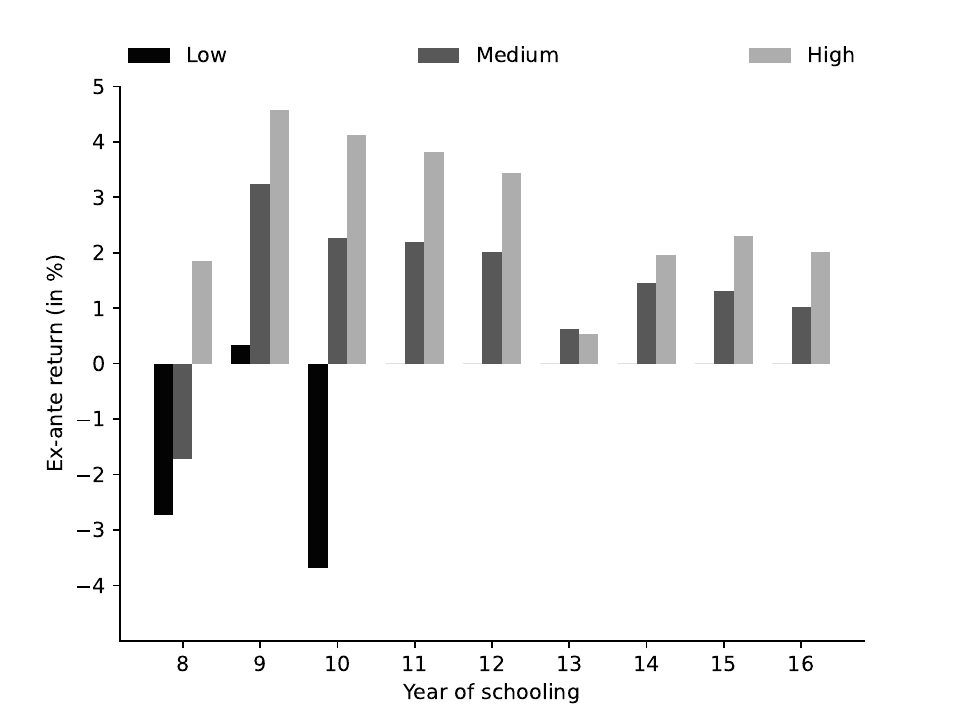}}}
\subfloat[Vocational Schooling]{\scalebox{0.45}{\includegraphics{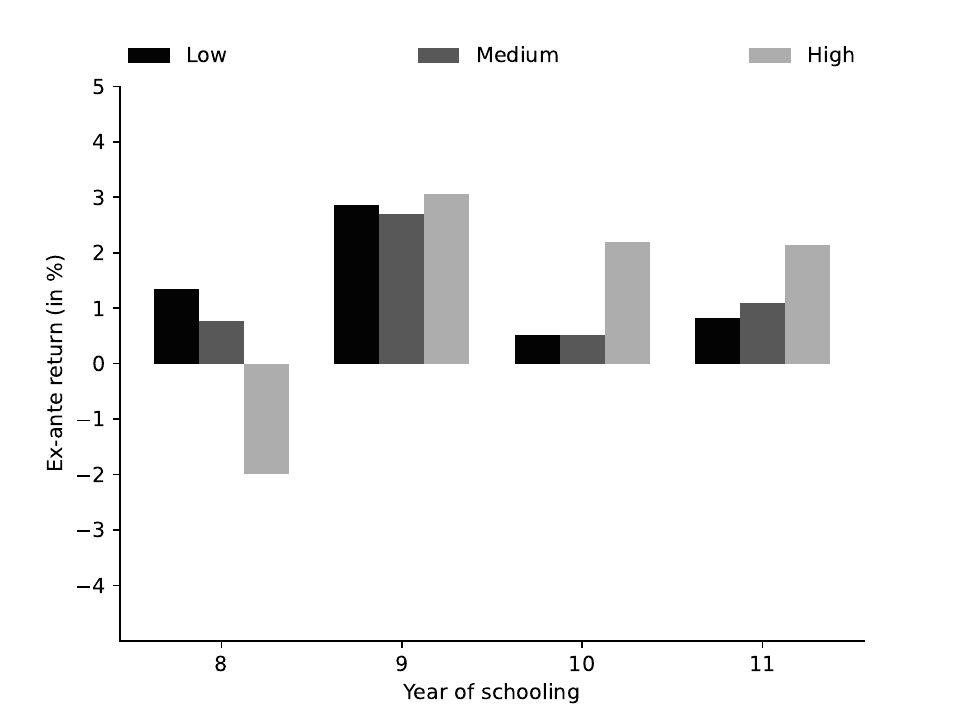}}}\\
\begin{center}
\begin{minipage}[t]{\columnwidth}
\emph{\scriptsize{}Note:}{\scriptsize{} The figure is based on samples of 5,000 simulated schooling careers for each ability group based on the estimated model. This figure contains average ex-ante returns in utility terms for each schooling year-track-ability cell. The left figure shows how ex-ante returns to academic schooling develop over time for each group while the right figure shows the same for the vocational track. Each bar shows the average ex-ante return to a particular year of schooling for the individuals in the respective ability group that reaches the relevant transition in our model. For instance, the bar for the high-ability group in the academic panel in year 11 shows the average ex-ante return of the 11th year overall high-ability individuals that have had an uninterrupted academic schooling career until the 10th year. We compute the ex-ante return as defined in Equation (\ref{Calculation ex-ante return}). Whenever there are only a few people of a particular ability group that reach a particular transition we do omit this group from the calculation.}{\scriptsize\par}
\end{minipage}
\end{center}
\end{figure}

\paragraph{Ex-ante Returns}
\noindent Figure \ref{Ex-ante return} shows our main evidence on the ex-ante returns to continue schooling in academic or vocational track for an extra year by individual ability type. The returns are shown for each year of schooling along the horizontal axis and are computed for individuals that have reached this particular stage and are faced with the decision to continue schooling, i.e., the bars at 10 years illustrate the average ex-ante returns of attending the 10th grade by track and ability for individuals who have completed 9 years of schooling in this track in an uninterrupted spell. As we move towards right in each panel, the illustrate returns pertain to only individuals who actually reach those stages in our model. At each stage, these returns capture both the immediate rewards and the discounted future rewards. In this sense, the model allows us to estimate the average treatment effect for those facing a particular treatment choice, as compared to their best alternative choice \citep{Heckman.2016, Heckman.2007c}.

\noindent We find substantial heterogeneity in the average ex-ante returns by ability, track choice and year of schooling. In the academic track, we estimate negative returns among low-ability individuals of $-2.75\%$ and $-3.5\%$, respectively, at the 8th and 10th grades, and positive returns among high-ability individuals at all years. By contrast, we estimate a negative return for high-ability individuals at the 8th grade in vocational track. Underlying this heterogeneity is a strong pattern of sorting into academic and vocational tracks by ability, which in our model is both caused by differences in wage rewards and preference heterogeneity. Indeed, the structure of the 8th grade returns reflects the separation of the ability groups in the different schooling tracks at an early stage of the educational pathways, consistent with the descriptive evidence in Figure \ref{fig:Desc-Stats}, panel (e). As the average return to vocational track at the 8th year is negative for high-ability individuals, the majority of them never enter vocational school. The opposite is true for low-ability individuals, for whom the returns to an academic track are negative initially, pushing them into vocational schooling instead. Indeed, most low-ability individuals never attend an academic school. As we move towards right in panel (a), we notice that the returns to academic schooling remain consistently positive only for high-ability individuals. In each track, we find the highest returns for transitions that entail a middle school degree at the 9th year, with gradually decreasing returns as we progress further.

\paragraph{Earnings Returns}
The above evidence on ex-ante average \textit{true} returns in terms of \textit{utility} may contrast ex-ante average \textit{lifetime earnings} returns associated with different schooling choices, as discussed in Section \ref{subsec:Model-objects}. To illustrate this, we also provide evidence on ex-ante earnings returns in the Appendix Figure \ref{Ex-ante return-wage} for each schooling year-track-ability cell, using our estimated model. Consistent with the notion of strong ability-related sorting in our data, we again find that the ex-ante earnings returns to academic (vocational) schooling are negative for low-ability (high-ability) individuals. However, we find different patterns across the schooling careers. Unlike the true returns, earnings returns in the academic track peak at the 13th year of schooling, reflecting that once an individual enters college their probability of receiving a college degree increases substantially. Notably, however, earnings returns may not align with individual choices, as opposed to the true (utility) returns, which also capture expected non-pecuniary rewards.

\begin{figure}[h!]\centering
\caption{The Final Years of Schooling for those Exiting School at the 8th Grade.}\label{Final schooling for 8th grade dropouts}
\scalebox{0.5}{\includegraphics{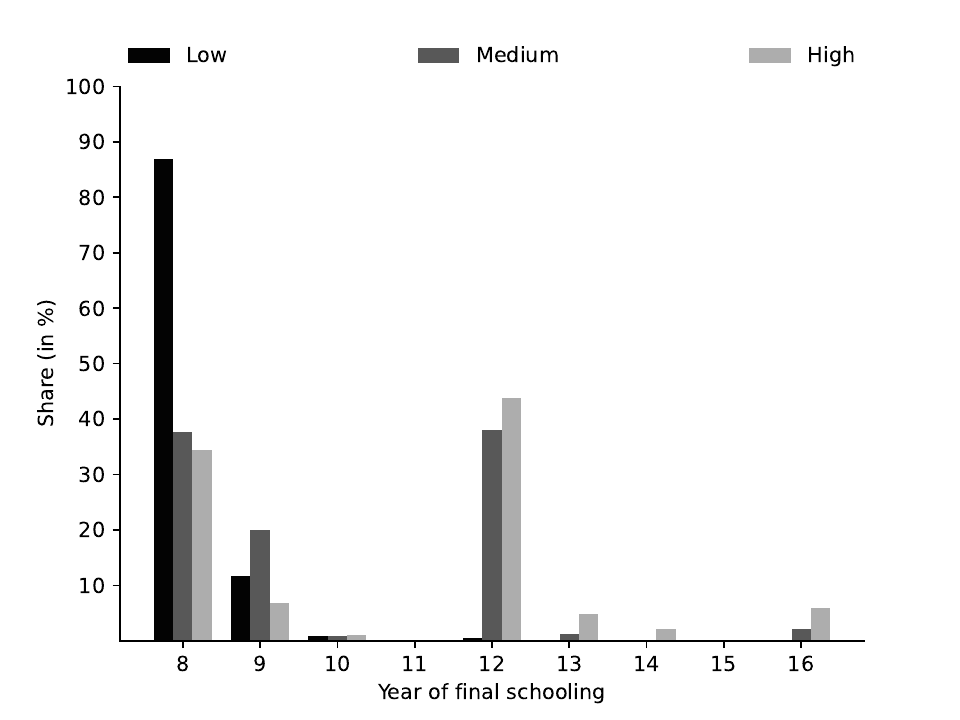}}
\begin{center}
\begin{minipage}[t]{\columnwidth}
\emph{\scriptsize{}Note:}{\scriptsize{} The figure is based on samples of 100,000 simulated schooling careers for each ability group based on the estimated model. We further restrict the sample to individuals who continue their schooling for only one additional year and then initially drop out of school at the 8th year of schooling, i.e., early drop-outs. We determine an individual's final schooling level as the sum of the years spent in academic and vocational school.}{\scriptsize\par}
\end{minipage}
\end{center}
\end{figure}

\paragraph{The Role of Re-enrollment}
A salient aspect of our model--crucial in interpreting the results above--is the individuals' ability to re-enroll in school after having exited and undergone a non-schooling spell. Indeed, the relatively low average ex-ante returns in Figure \ref{Ex-ante return} can be partly attributed to the fact that many individuals come back to school to pursue more education. In an attempt to illustrate this feature of our model, we show the final schooling level for individuals that initially leave school after the 8th grade in either track by ability in Figure \ref{Final schooling for 8th grade dropouts}. Among low ability individuals, dropping out at that stage determines the final schooling level for about 90\%. Only 10\% do still acquire their middle school degree at a later stage. For high-ability individuals, however, the large majority do continue their schooling at some point. Roughly 85\% do eventually end up with at least a middle school degree, and 30\% even continue to obtain at least a high school degree. We need to keep this feature of our model in mind when interpreting average ex-ante returns at each transition that are presented above.
\paragraph{The Distributions of Ex-ante Returns}
\noindent The average returns in Figure \ref{Ex-ante return} can mask considerable heterogeneity in returns by track choice across individuals with the same ability facing identical choices. For instance, \citet{Wiswall.2015a} and \citet{Attanasio.2017} document substantial heterogeneity in ex-ante returns using survey data on subjective expectations. In our model, heterogeneous returns across observationally similar individuals, i.e., those from the same ability group who have reached the same stage of the decision tree after an uninterrupted spell, are represented through the presence of heterogeneous latent types and different realizations of shocks to productivity and tastes associated with different choices.\\

\begin{figure}[h!]\centering
\caption{Distributions of Ex-Ante Returns to Academic and Vocational Schooling.}\label{Distribution of ex-ante returns}
\subfloat[Academic Schooling at 15th Year]{\scalebox{0.4}{\includegraphics{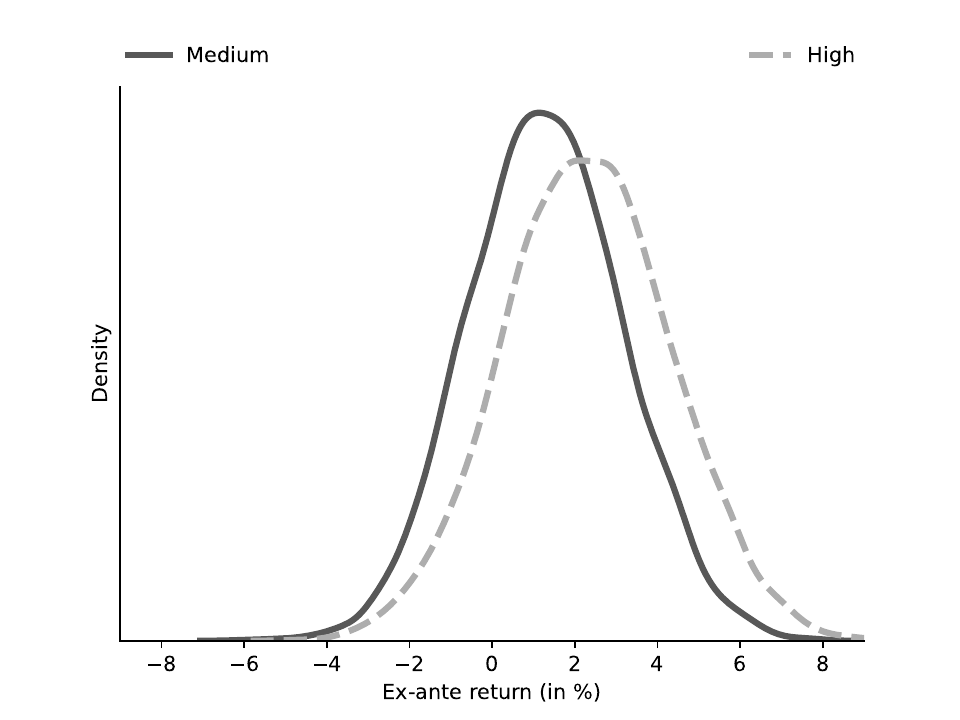}}}\hspace{0.3cm}
\subfloat[Academic Schooling at 11th Year]{\scalebox{0.4}{\includegraphics{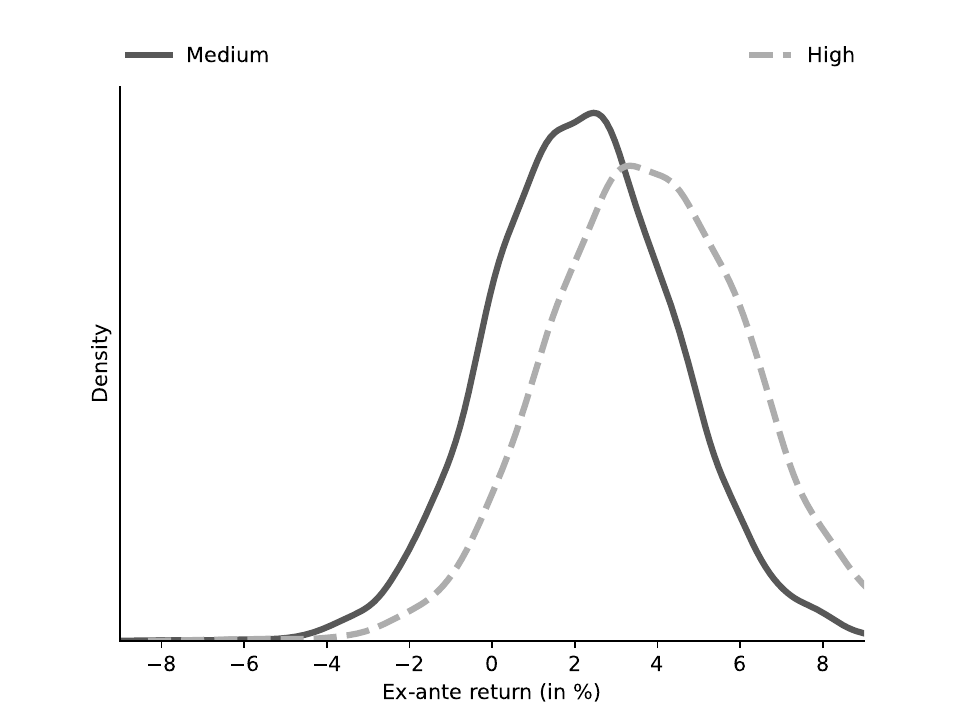}}}\\
\subfloat[Vocational Schooling at 11th Year]{\scalebox{0.4}{\includegraphics{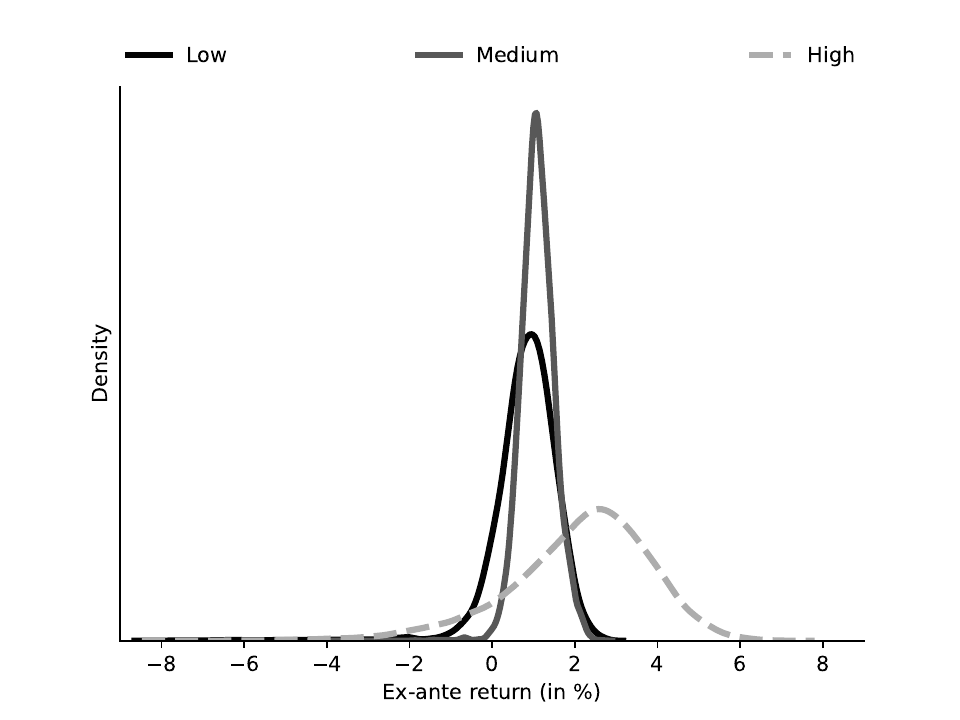}}}\hspace{0.3cm}
\subfloat[Vocational Schooling at 8th Year]{\scalebox{0.4}{\includegraphics{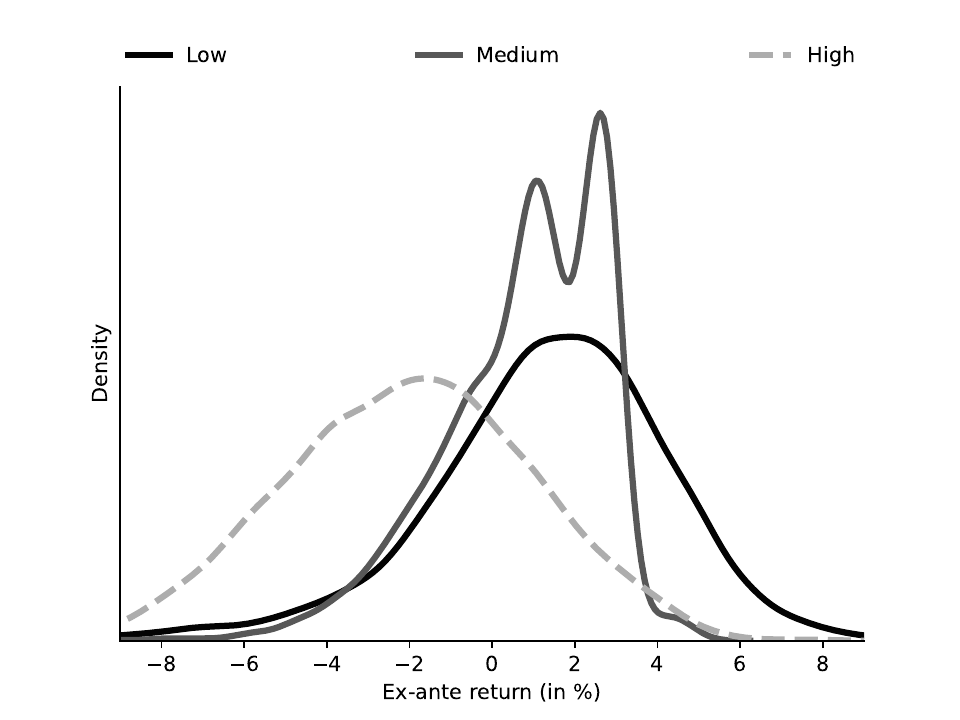}}}
\begin{center}
\begin{minipage}[t]{\columnwidth}
\emph{\scriptsize{}Note:}{\scriptsize{} The figure is based on samples of 5,000 simulated schooling careers for each ability group based on the estimated model. Each panel shows the distribution of ex-ante returns to a particular schooling choice by ability who those with an uninterrupted schooling career up to that point who have reached this transition. For instance, in panel (a), we show the distribution of ex-ante returns to the 15th year of academic schooling for all individuals with uninterrupted schooling careers up to that point. We compute the ex-ante return as defined in Equation (\ref{Calculation ex-ante return}). The heterogeneity in returns follow from the permanent differences across latent types $\bm{e}$ and the realizations of transitory shocks $\bm{\epsilon_t}$. Whenever there are only a few people of a particular ability group that reach a particular transition we do omit this group from the calculation.}{\scriptsize\par}
\end{minipage}
\end{center}
\end{figure}

\noindent To illustrate heterogeneity in ex-ante returns, we now focus in Figure \ref{Distribution of ex-ante returns} on individuals in each ability group who in our model faced the choices of 11th and 15th year of academic schooling and 8th and 11th year of vocational schooling, respectively.  In panel (b), we see that the average return of the 11th year of academic schooling is positive for both high- and medium-ability groups, but also that a considerable share within each group has negative returns. In our model, all individuals with negative returns to academic schooling would decide not to attend academic schooling as this choice is (in expectation) dominated by the best alternative choice. They might, however, pursue academic school later as this does not rule out the possibility of re-enrollment. In panel (b), we also see that the distributions overlap, so there are many individuals of medium-ability for whom returns to the 11th year of academic schooling are higher than for high-ability individuals. In panel (a), we see that the distributions of returns to the 15th year of academic schooling are shifted towards left as compared to panel (b), reflecting that fewer go on to attend the 15th year as compared to the 11th year. \\

\noindent Next, in Figure \ref{Distribution of ex-ante returns}, panels (c)-(d), we consider the distributions of returns to 11th and 8th year of vocational schooling, respectively. In these plots, we include all three ability groups as sufficiently many from each group are present at these transitions in our model. We find interesting ability-related patterns in panel (d), where the majority of low-ability individuals have positive returns to 8th year of vocational schooling, while the majority of high-ability individuals have negative returns. These results reflect the strong patterns of ability-related sorting at the 8th year schooling tracks. By contrast, the distributions of returns are much more similar across low- and medium-ability groups in panel (c), reflecting that conditional on having reached the 10th year of vocational schooling, the transition rates to 11th year of vocational school do not differ substantially across these groups. This finding may reflect that there are latent types in all ability groups with a high propensity to attend vocational schooling, and these types stay in vocational schooling through-out middle and high school, conditional on having enrolled.

\paragraph{Ex-ante and Ex-post Returns}
The preceding analysis have been focused on ex-ante returns, i.e., the relative rewards that agents in our model base their decisions on. We now contrast our estimates of ex-ante returns to another set of objects which we refer to as ex-post returns. The latter returns are based on our baseline model with taste and productivity shocks turned on, but where we use the actual realizations of shocks rather than the agents' expectations about these to calculate their returns to difference choices. Since our model assumes rational expectations, on average, ex-ante and ex-post returns must agree. However, there does exist a non-degenerate joint distribution of ex-ante and ex-post returns across agents. To construct measures of ex-post returns, we limit attention to individuals in our model who ended up selecting specific schooling choices and retain the realization of shocks they were exposed to as they traversed through their decision trees. By construction, the ex-ante returns for these individuals for the set of schooling choices they ended up making are strictly positive. To provide a comparison of the ex-ante and ex-post returns, we compare both the expected and the realized utility flows to the expected values of the next best alternatives the individuals faced.\\

\begin{figure}[h!]\centering
\caption{Joint Distributions of Ex-Ante and Ex-post Returns.}\label{Joint distribution of returns}
\subfloat[Academic Schooling at 15th Year]{\scalebox{0.4}{\includegraphics{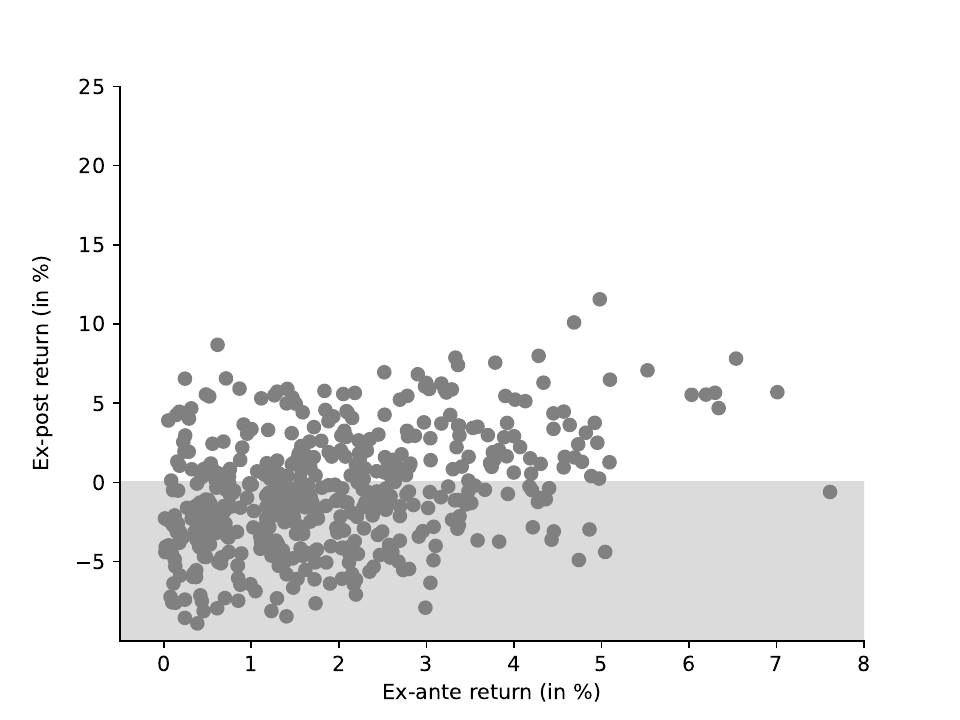}}}\hspace{0.3cm}
\subfloat[Academic Schooling at 11th Year]{\scalebox{0.4}{\includegraphics{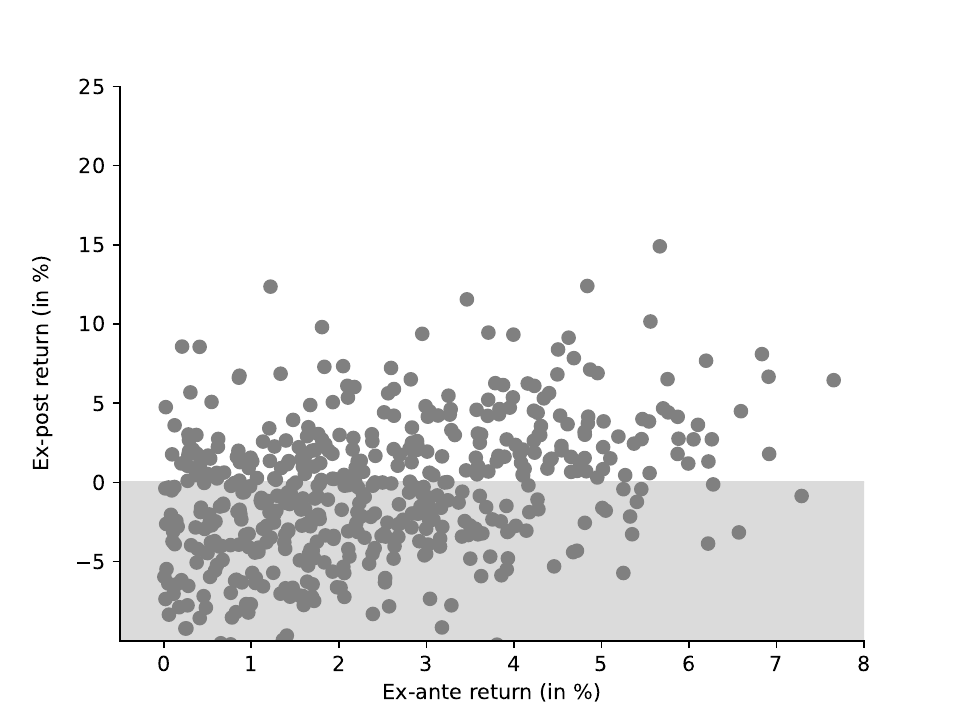}}}\\
\subfloat[Vocational Schooling at 11th Year]{\scalebox{0.4}{\includegraphics{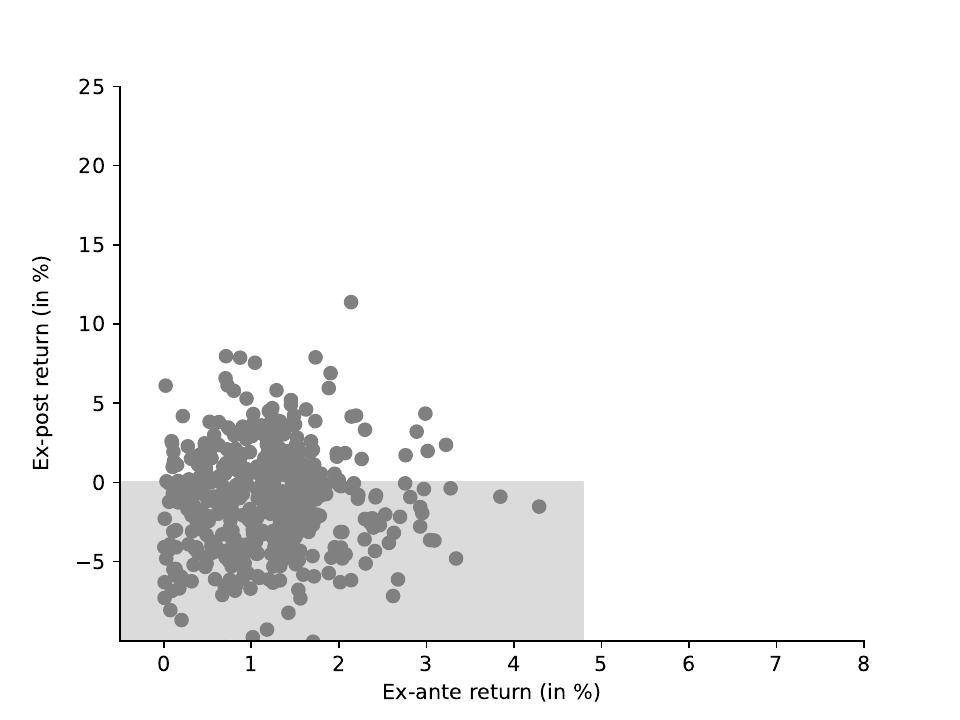}}}\hspace{0.3cm}
\subfloat[Vocational Schooling at 8th Year]{\scalebox{0.4}{\includegraphics{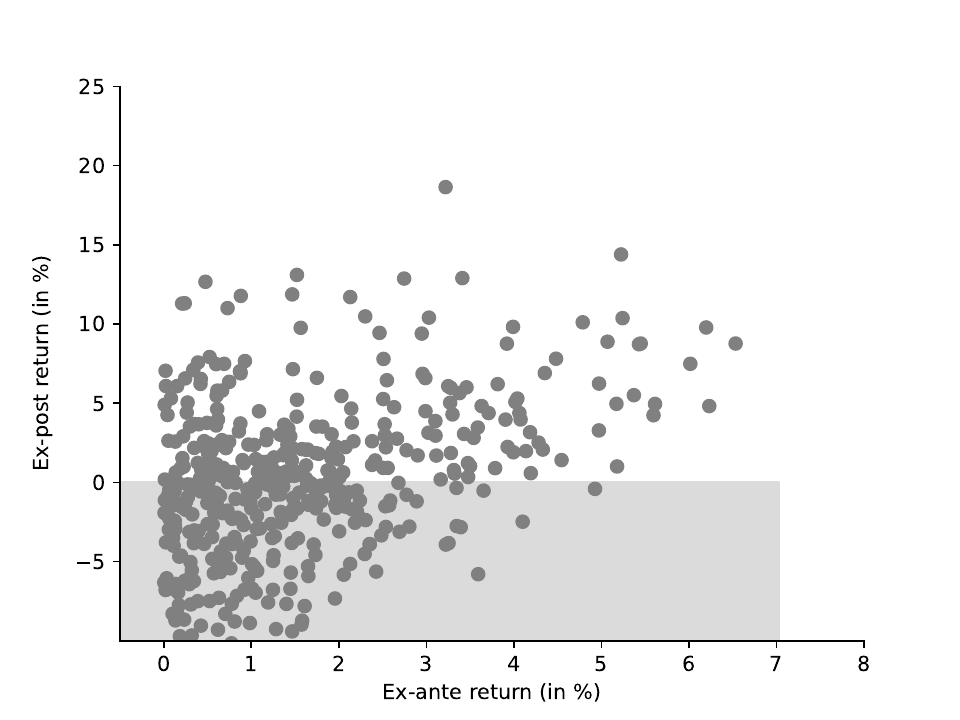}}}\begin{center}
\begin{minipage}[t]{\columnwidth}
\emph{\scriptsize{}Note:}{\scriptsize{} The figure is based on samples of 5,000 simulated schooling careers for each ability group based on the estimated model. We then restrict the sample to 500 random individuals with uninterrupted careers in the respective period for a particular track. Each panel shows the joint distributions of ex-ante and ex-post returns to a particular schooling choice in either academic or vocational track. Ex-post returns are the realized total discounted utilities over the remaining decision periods relative to the value function of the best alternative. The gray area shows all points where the realized return is smaller than the expected return from the second best option. Ex-ante return as defined in Equation (\ref{Calculation ex-ante return}). Whenever there are only a few people of a particular ability group that reaches a particular transition we do omit this group from the calculation.}{\scriptsize\par}
\end{minipage}
\end{center}
\end{figure}

\noindent Figure \ref{Joint distribution of returns} presents the joint distribution of ex-ante and ex-post returns for a random set of 500 individuals from our model at each transition. As expected, the ex-ante return are always positive by construction in each panel. However, the shaded areas indicate that the ex-post returns from pursuing an education were negative for some individuals, i.e., they faced regret due to the actual shock realizations. We also note that the ex-post returns to the 8th year of vocational schooling are relatively dispersed. As this decision is made early in the life-cycle, agents face very different life-time trajectories subject to the future shock realizations and choices. By contrast, the ex-post returns to the 15th year of academic schooling are more compressed. As most individuals would go on to attend the 16th year to attain a college degree and then enter the labor market, these ex-post returns are associated with more similar life-time trajectories. These findings also demonstrate how uncertainty is highest at the beginning of the life-cycle and the choices made early on can be more consequential in a dynamic setting.

\subsection{Evidence on Option Values}\label{subsec:Results-Option}
Part of the overall value to a schooling choice is the option to continue schooling further. We now provide quantitative evidence on such option values based on our estimated model. To construct measures of option values to schooling, we will compare the discounted lifetime value of attending schooling in a particular track in a given period in our model to the corresponding value of the same schooling track under a counterfactual policy where the individual is prohibited from making a schooling choice in any future period. As earlier, we will focus on individuals who in our model have had uninterrupted schooling careers in given track up to the period where we explore the option values associated with a schooling track choice.

\begin{figure}[h!]\centering
  \caption{The Option Value Contributions of Academic and Vocational Schooling.}\label{Option value contribution}
   \subfloat[Academic Schooling]{\scalebox{0.45}{\includegraphics{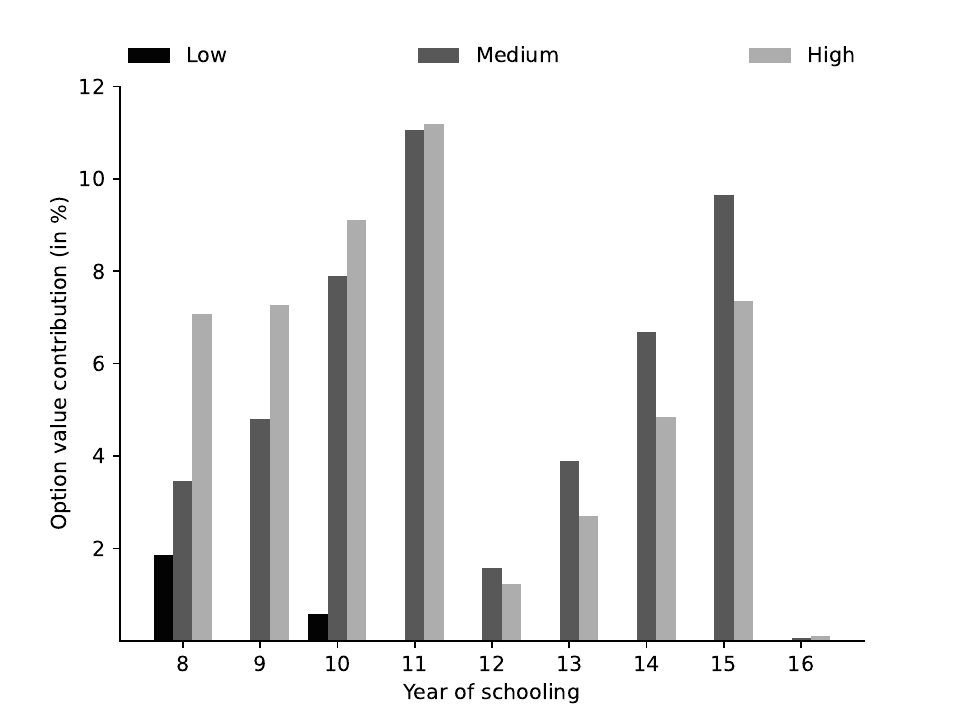}}}\hspace{0.3cm}
  \subfloat[Vocational Schooling]{\scalebox{0.45}{\includegraphics{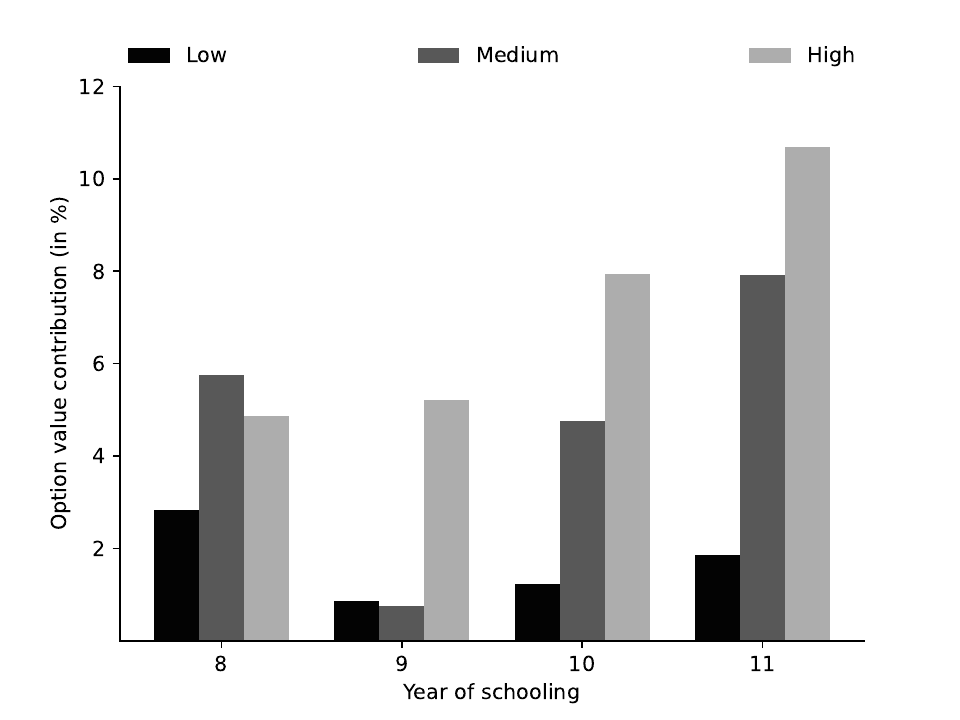}}}

  \begin{center}
\begin{minipage}[t]{\columnwidth}
\emph{\scriptsize{}Note:}{\scriptsize{} The figure is based on samples of 5,000 simulated schooling careers for each ability group based on the estimated model. We restrict the sample to individuals with uninterrupted schooling careers up the relevant transition using the estimated model. The left figure shows how option value contributions for academic schooling develop over time for each group while the right figure shows the same for the vocational track. The option value contribution is defined in Equation (\ref{Calculation option value contribution}). Whenever there are only a few people of a particular ability group that reaches a particular transition we do omit this group from the calculation.}{\scriptsize\par}
\end{minipage}  
  \end{center}
\end{figure}

\paragraph{Option Value Contributions}
Figure \ref{Option value contribution} shows the contribution of the option value to the overall value of a schooling track by the year of schooling and ability. The option value contributions in the academic track range from 11\% for the 11th year to almost zero beyond the 15th year of schooling. A recurring pattern is that the option value contribution is always the highest for the year of schooling right before the completion of an academic degree that entails considerable ``diploma'' effects. We also find sizeable heterogeneity by ability level. The option value contributions in the academic track tend to be the highest for medium- and high-ability individuals, as they are also likely to benefit the most from the additional schooling opportunities that open up from taking an extra year of schooling. By contrast, in the vocational track, the option value contributions for low-ability individuals are the highest at the 8th year of schooling and decline in later years. This pattern may reflect that most low-ability individuals attending this track go on to complete the 9th year in vocational school. Among those who progress further in vocational schooling, we again find a strong ability gradient. This likely reflects that among this group, also the medium- and high-ability individuals have the highest gain from attending the 10th and 11th year, and reach a vocational high school degree.\\

\begin{figure}[h!]\centering
  \caption{`Complier' Characterization -- Switching Off the Option Value.}\label{Option value importance}
  \subfloat[Academic at 11th Year (High)]{\scalebox{0.2}{\includegraphics{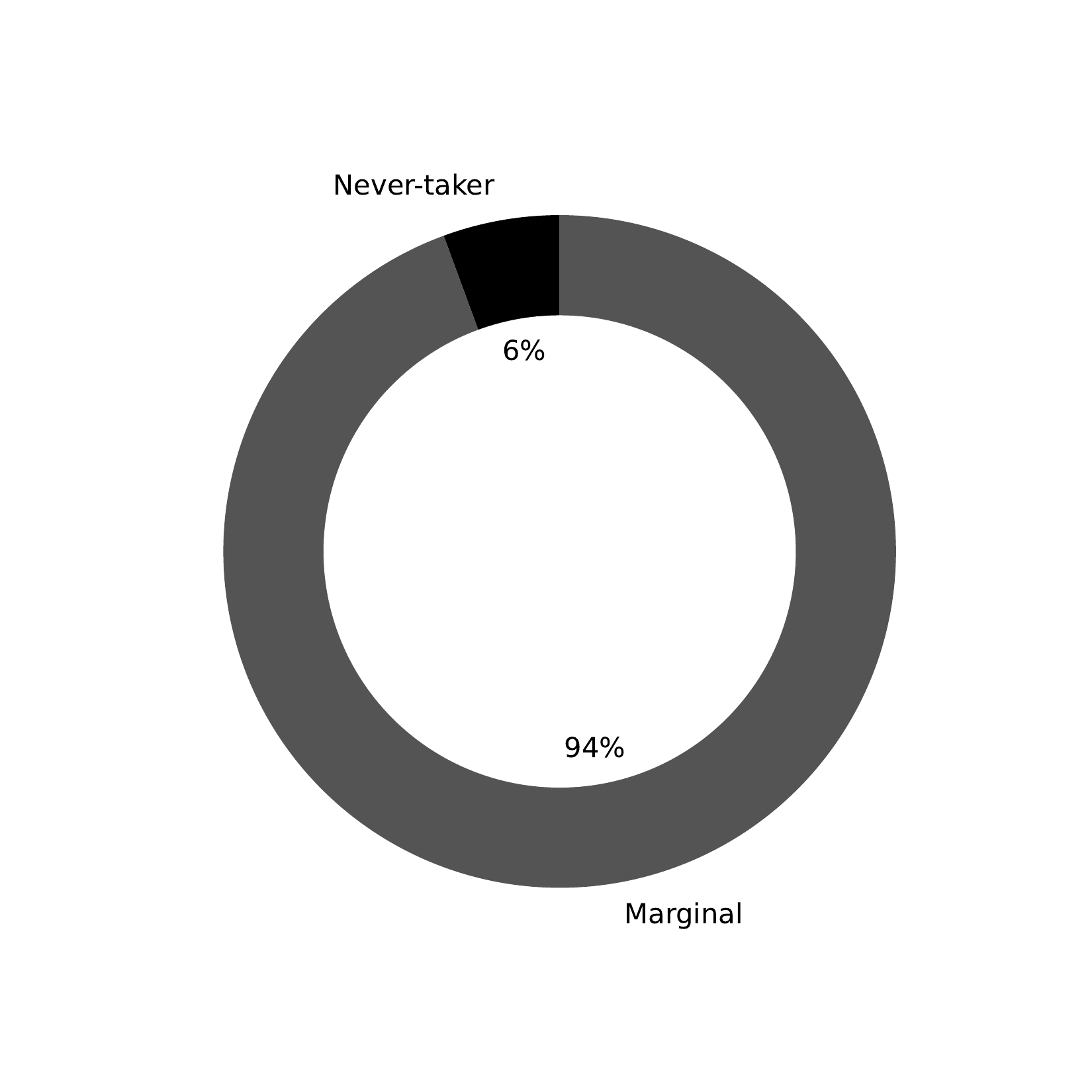}}}\hspace{0.3cm}
  \subfloat[Academic at 12th Year (High)]{\scalebox{0.2}{\includegraphics{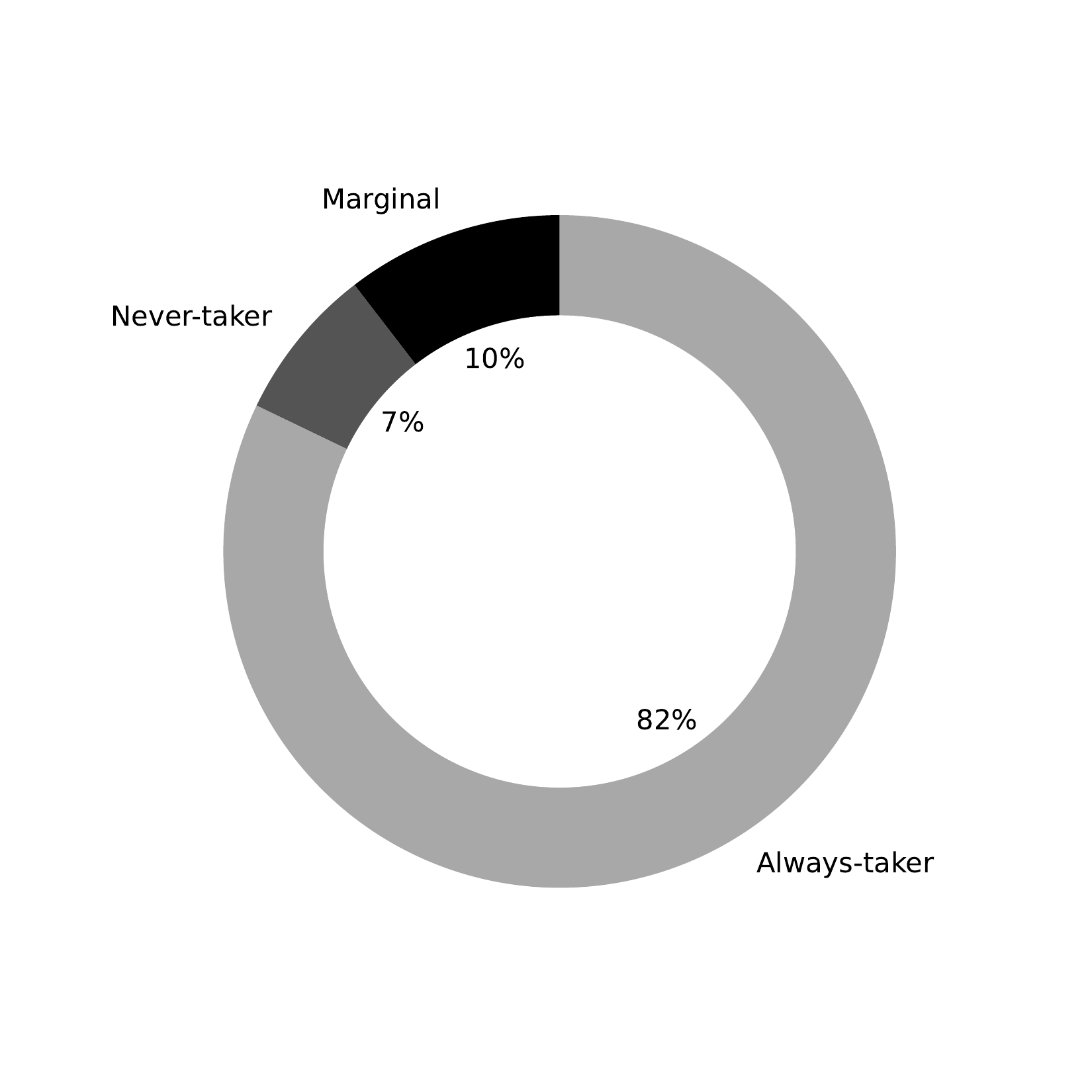}}}\\
  \subfloat[Vocational at 8th Year (All)]{\scalebox{0.2}{\includegraphics{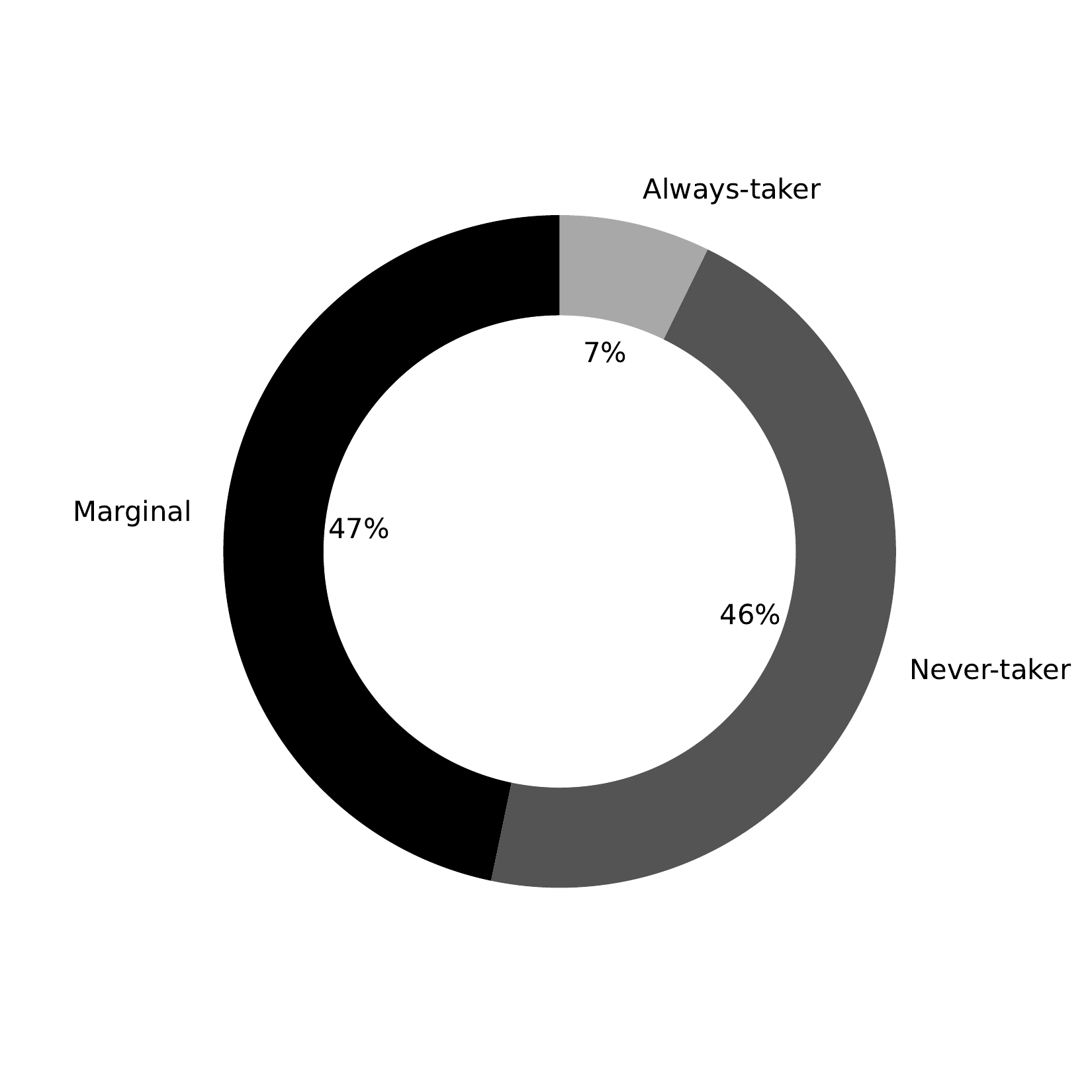}}}\hspace{0.3cm}
  \subfloat[Vocational at 9th Year (All)]{\scalebox{0.2}{\includegraphics{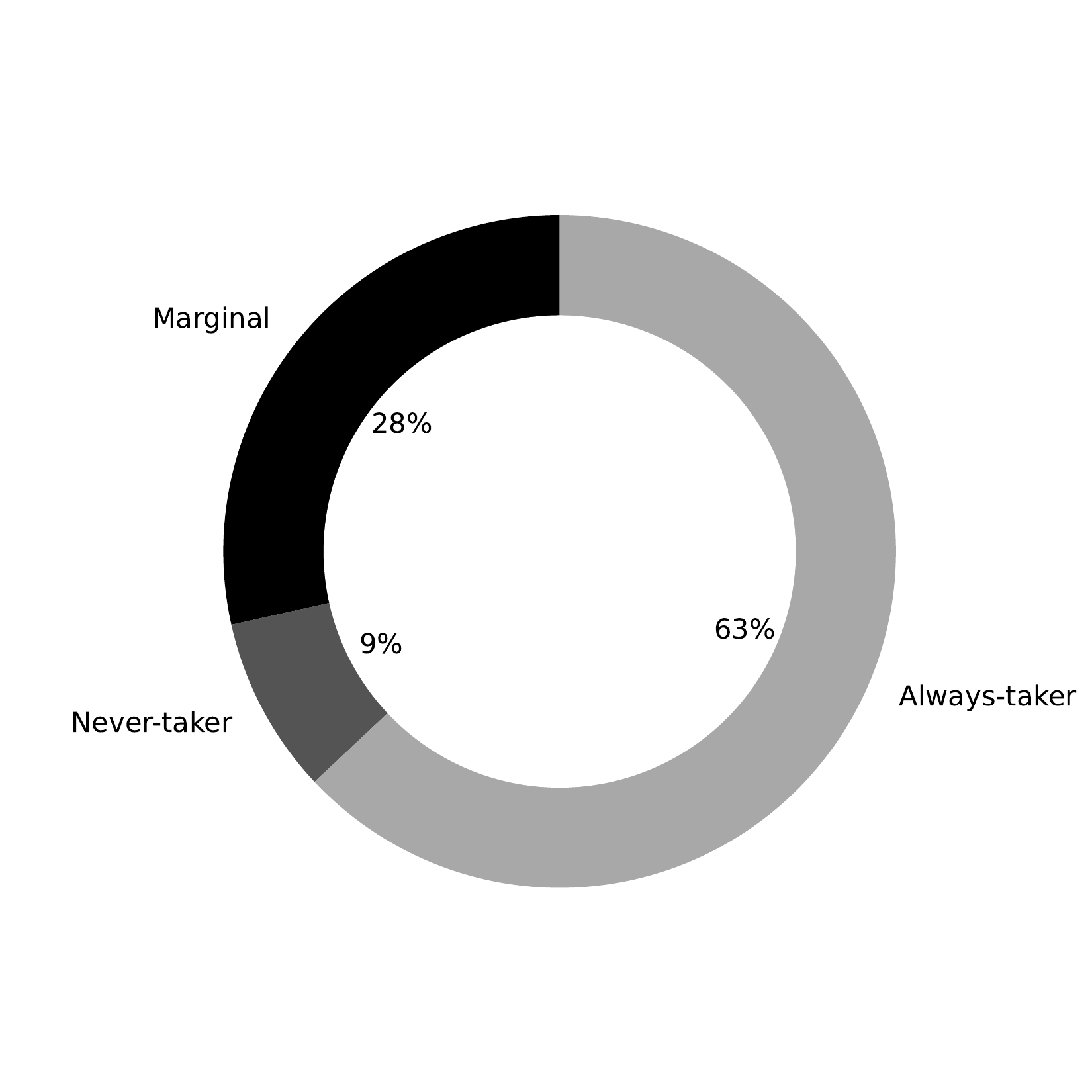}}}\
  \begin{center}
\begin{minipage}[t]{\columnwidth}
\emph{\scriptsize{}Note:}{\scriptsize{} The figure is based on samples of 5,000 simulated schooling careers for each ability group based on the estimated model. Panels (a)-(b) are restricted to high-ability individuals, while panels (c)-(d) average across all ability groups. Each panel provides a complier characterization based on model simulations where we turn off the option value of a particular schooling choice. This calculation compares the total value of schooling and the value of schooling net of the option value contribution to the next best alternative. The option value contribution is defined in Equation (\ref{Calculation option value contribution}). Always-takers (never-takers) always (never) choose to continue with another year of schooling, irrespective of the option value contribution, while marginal individuals take the additional year only because of the option value contribution. Whenever there are only a few people of a particular ability group that reaches a particular transition we do omit this group from the calculation.}{\scriptsize\par}
\end{minipage}    
 \end{center}
\end{figure}

\noindent While the previous illustration provides evidence on the option value contributions, measured as a fraction of the overall value of a choice, we now provide more direct evidence on how the option value channel can play a crucial role in shaping schooling careers. To get at this, we perform counterfactual experiments based on our model where we turn on and off the option value of a choice and characterize the schooling decisions made by the agents in our model under each scenario. Based on these comparisons and inspired by the IV/LATE complier characterizations done in the program evaluation literature (e.g., \cite{Angrist1996}), we perform a characterization of agents into three groups; always-takers, never-takers and marginal agents (or compliers). Agents that always (never) decide to take another year of schooling when faced with this choice, irrespective of the option value, are characterized as always-takers (never-takers). While, agents that decide to take another year of schooling when the option value is turned on but not when it is off are characterized as marginal (i.e., compliers). By construction, monotonicity is satisfied, as for given shock realizations, agents in our model are never more likely to take more schooling when the option value is turned off compared to when it is on.\\

\noindent Figure \ref{Option value importance} illustrates our evidence based on the complier characterizations described above. In panels (a)-(b), we focus on high-ability individuals who faced the decision to continue their schooling for the 11th and 12th year in the academic track. These figures provide interesting illustrations of how important option values can be close to the degree-rewarding schooling choices. Among those facing the decision to continue their academic schooling in the 11th year, we find that a large majority at 94\% among the high-ability individuals consists of marginal compliers, i.e., individuals who continue for another year of schooling only because of the option value stemming from being able to complete a high school degree right afterwards. By contrast, there are hardly any always-takers, who move ahead with their schooling even when no future schooling opportunities are available, and about 6\% are never-takers, who drop out regardless. The picture is somewhat different at the 12th year of academic schooling. The fraction of always-takers rises drastically to 82\% as completing the high school degree provides immediate considerable wage rewards. For 10\% the option value of the high school diploma is crucial to complete the 12th year, as receiving this diploma opens up the possibility of attending college. Still, 7\% drop-out and do not complete high school regardless of the option value.\\

\noindent Next, in panels (c)-(d) of Figure \ref{Option value importance}, we consider all individuals (irrespective of ability) who faced the decision to continue their schooling for the 8th and 9th year in the vocational track. Option values in the vocational track are highest at the 8th year of schooling as this gives the option to continue later on with the 9th year of vocational schooling, i.e., receive a two-year vocational diploma. At the 9th year of vocational schooling, a majority at 63\% of individuals facing this choice are characterized as always-takers, who attend this year due to the immediate wage gains associated with this choice, while the option value associated with the possibility to continue with a vocational high school degree drive this choice for 28\% of individuals.

\paragraph{Uncertainty and Option Values}
In our model, transitory shocks to productivity (i.e., wage risk) and tastes give rise to uncertainty in agents' decision-making. To shed light on these aspects through the lens of our model, we perform a series of comparative statics, where we re-compute option values contributions by ability and track shutting off various sources of shocks and compare the findings to our baseline model.\footnote{We summarize our findings on ex-ante returns from these exercises in Appendix Figure \ref{Ex-ante return-risk}. To facilitate comparison, panel A reports estimates of average ex-ante returns by ability and track from our baseline model (as in Figure \ref{Ex-ante return}), while panel B shows the corresponding estimates from a version of the model where we remove shocks to productivity (i.e., no wage risk), while in panel C we further also remove unobserved transitory shocks related to tastes for schooling, working or staying at home. While the ex-ante returns are broadly similar across panels A-B, we find substantially larger returns for low-ability (high-ability) individuals in vocational (academic) tracks in panel C. Thus, removing taste shocks would further strengthen the patterns of ability-related sorting into tracks in the early stage of educational careers in our model.}  We now consider how these sources of uncertainty contribute to the option values associated with different educational choices.\footnote{The existing literature on learning and educational choices in dynamic settings actually emphasizes uncertainty as the primary source of option values, i.e., individual learn about their own ability and preferences as they progress in their schooling career \citep{Arcidiacono.2016,Trachter.2015, Stinebrickner.2014, Stange.2012}. As the shocks in our model are distributed independently over time, individuals do not update their prior beliefs about their productivity or alternative-specific tastes, i.e., our model does not feature learning over time. We rather focus on the overall value attached to a schooling choice stemming from the possibility of pursuing further education, and not only the value associated with resolution of uncertainty and learning. The presence of transitory shocks may nonetheless affect individuals' schooling choices and alter the likelihood of attending further schooling, i.e., option values can depend on the presence of shocks.} \\

\noindent We present in Appendix Figure \ref{Option value uncertainty} two different scenarios where the presence of transitory shocks has opposite signed effects on the option value contribution in our model. In panel (a), we consider the option value of the 11th year in the academic track for individuals of medium ability under different sources of uncertainty. We first turn off the productivity shocks and then also turn off the taste shocks. In the baseline model, the option value contribution amounts to about $11\%$. When we turn off the productivity shocks alone, the option value contribution increases to $12\%$, and decreases to about $10\%$ in a scenario without any uncertainty. The latter pattern reflects that the continuation of schooling becomes less likely when we reduce the extent of uncertainty for some individuals. Specifically, some medium-ability individuals pursue the 11th year of academic schooling solely due to the realization of taste shocks. Once we remove these realizations, the option value of the 11th academic year declines for these individuals. \\

\noindent In panel (b) of Appendix Figure \ref{Option value uncertainty}, we consider another scenario showing how the presence of transitory shocks affects the option value contribution for the 8th year in the vocational track for low-ability individuals. In the baseline model, the option value contribution is at $3\%$. When we turn off the productivity shocks alone, the option value remains almost unchanged, but when we remove taste shocks this value increase to almost $6\%$. This pattern reflects that some among the low-ability individuals in our baseline model who drop-out at the 8th year of vocational schooling do so due to the realization of taste shocks. Once we remove these shocks from our model, their likelihood of continuing beyond the 8th year increases even further.

\subsection{Evidence on Sectoral Choices}\label{subsec:Results-Work}

An important feature of our model setup is that we allow individuals to work in one of three sectors--private, public and self-employment--and further allow sector-specific returns to schooling in academic and vocational tracks and to work experience. In the following, we use the estimated model to perform a series of comparative statics where we alter sectoral job opportunities, with the goal of informing how these options influence educational choices.

\begin{figure}[h!]\centering
\caption{Average Ex-ante Returns -- The Role of Sectoral Job Opportunities.}\label{Ex-ante return-sector}
\medskip
{\scalebox{0.90}{\textit{Panel A. Baseline Model}}} \\
\setcounter{subfigure}{0}
\renewcommand{\thesubfigure}{A-\arabic{subfigure}}
\subfloat[Academic Schooling]{\scalebox{0.40}{\includegraphics{material/results/fig-tr-transitions-iq-academic-bw}}}\hspace{0.3cm}
\subfloat[Vocational Schooling]{\scalebox{0.40}{\includegraphics{material/results/fig-tr-transitions-iq-vocational-bw}}}\hspace{0.3cm}
\medskip \\
{\scalebox{0.90}{\textit{Panel B. No Public Sector Choice}}} \\
\setcounter{subfigure}{0}
\renewcommand{\thesubfigure}{B-\arabic{subfigure}}
\subfloat[Academic Schooling]{\scalebox{0.40}{\includegraphics{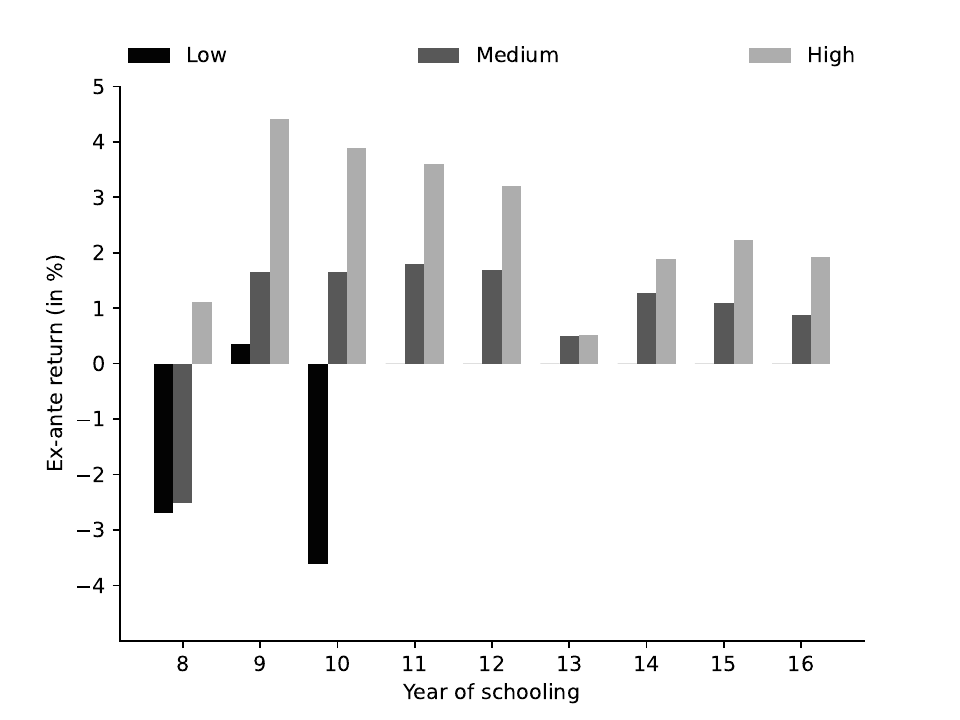}}}\hspace{0.3cm}
\subfloat[Vocational Schooling]{\scalebox{0.40}{\includegraphics{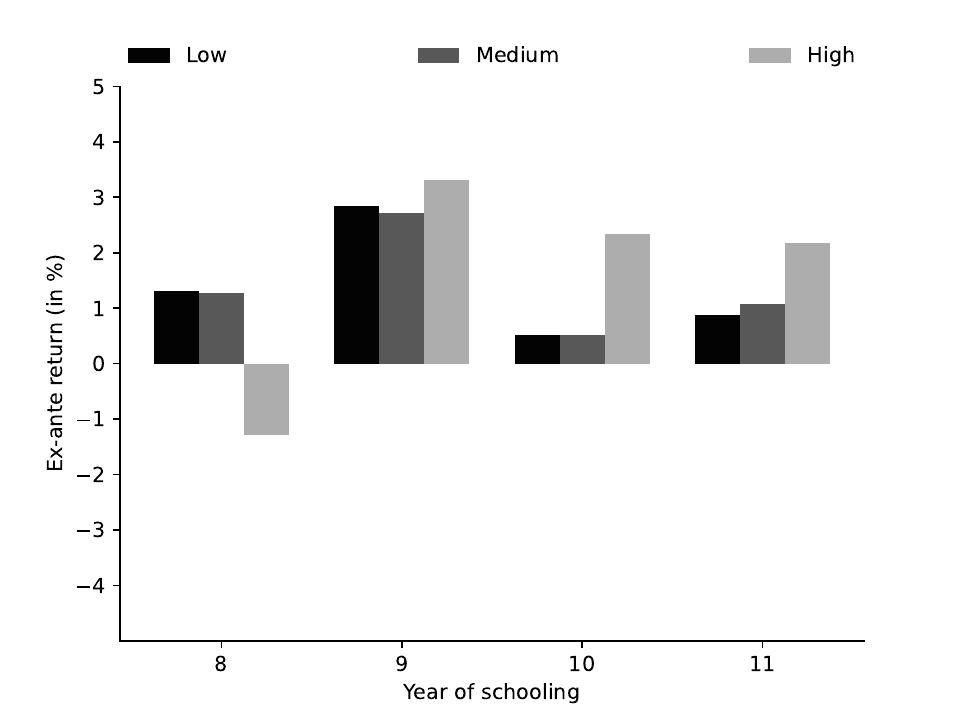}}}\hspace{0.3cm}
\medskip \\
{\scalebox{0.90}{\textit{Panel C. Only Private Sector Choice}}} \\
\renewcommand{\thesubfigure}{C-\arabic{subfigure}}
\setcounter{subfigure}{0}
\subfloat[Academic Schooling]{\scalebox{0.40}{\includegraphics{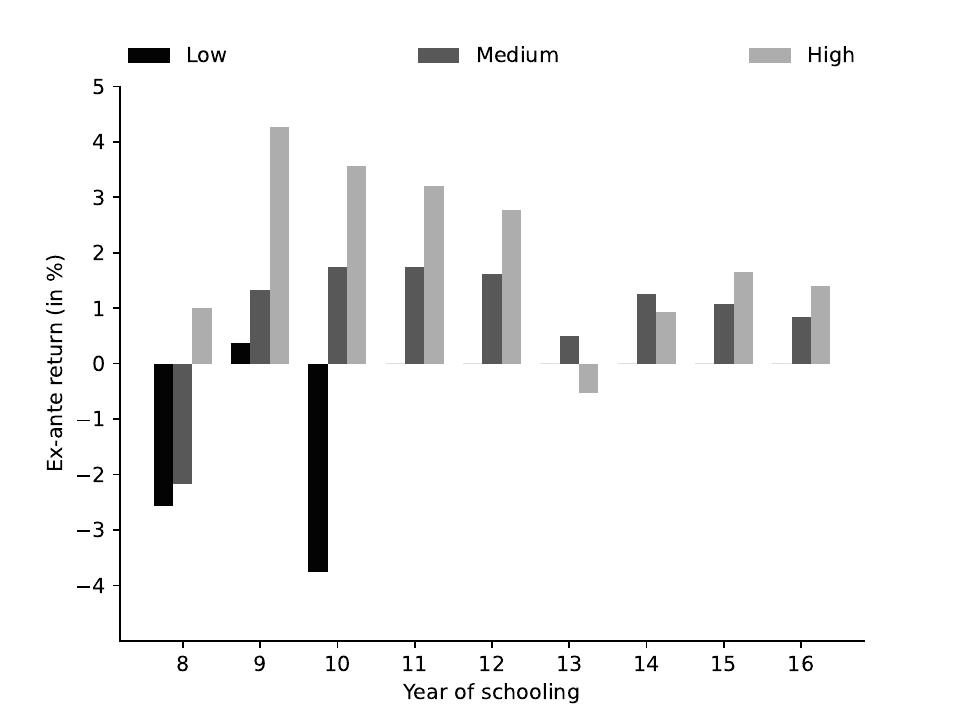}}}\hspace{0.3cm}
\subfloat[Vocational Schooling]{\scalebox{0.40}{\includegraphics{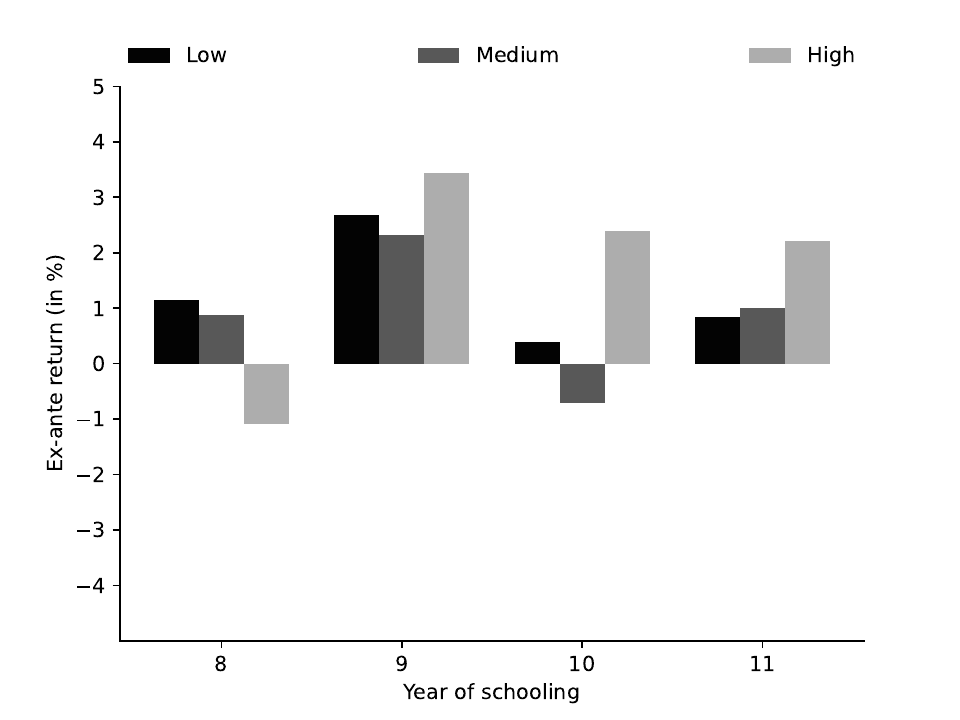}}}
\begin{center}
\begin{minipage}[t]{\columnwidth}
\emph{\scriptsize{}Note:}{\scriptsize{} The figure is based on samples of 5,000 simulated schooling careers for each ability group based on the estimated model under different specifications of transitory shocks. Each panel contains average ex-ante returns for each year-track-ability cell for alternative model specifications. Panel A shows results based on the baseline model, while panel B removes the public sector and panel C further also removes the option to work as self employed. In each panel, the left figure shows how ex-ante returns to academic schooling develop over time for each group while the right figure shows the same for the vocational track. Each bar shows the average ex-ante return to a particular year of schooling for the subset of the respective ability group that has reached thohe relevant transition in our model. For instance, the bar for the high-ability group in the academic panel in year 11 shows the average ex-ante return of the 11th year for those that have had an uninterrupted academic schooling career until the 10th year. We compute the ex-ante return as defined in Equation (\ref{Calculation ex-ante return}). Whenever there are only a few people of a particular ability group that reaches a particular transition we do omit this group from the calculation.}{\scriptsize\par}
\end{minipage}
\end{center}
\end{figure}

\paragraph{Sectoral Opportunities}
As noted above, we allow the returns to schooling tracks and work experience to differ across sectors. Consequentially, the ex-ante returns depend on the availability of job opportunities in different sectors. In Figure \ref{Ex-ante return-sector}, we compares ex-ante returns to each year of academic and vocational schooling by ability under different assumptions on sectoral job opportunities. For comparison, panel A shows the returns estimated in our baseline model. In panel B, we remove the option of public sector choice, and recompute the ex-ante returns associated with each educational choice. Notably, removing the public sector job opportunities decreases ex-ante returns to academic schooling considerably for high- and medium-ability individuals, who are most likely to benefit from these opportunities. \\


\noindent The most noticeable decline in the ex-ante return is observed for the 9th year of academic schooling. This change implies that removing the public sector opportunities alters the schooling track choice of some individuals. Indeed, in Appendix Figure \ref{Final Schooling Sectoral Choice} (\ref{Final Schooling Sectoral Choice Ability}), we show that removing public sector opportunities induces some (medium-ability) individuals towards a vocational high school degree instead. Further, in Figure \ref{Ex-ante return-sector}, panel C, we further remove the option to work as self-employment. The latter change, however, does not substantially alter returns for low- and medium-ability individuals, as these groups alternatively move into private sector jobs and retain similar premiums. More surprisingly, the return to academic schooling decreases for high-ability individuals once we remove self-employment choice. This phenomenon may reflect the presence of sizeable returns to academic college education also in the self-employment sector (e.g., high-paying consultancy work).


\paragraph{Sectoral Opportunities and the Option Value Contributions}
Sectoral opportunities shapes educational choices and thus also the size of option value contributions. In Figure \ref{Option value sector}, we compare option values across model specifications with different sectoral opportunities for medium- and low-ability individuals.
Panel (a) shows option value contributions for medium-ability individuals who consider enrolling in the 8th year of academic schooling. 
Removing the option to enter the public sector substantially reduces the option value of academic schooling.
The return to academic degrees is relatively large in the public sector which leads to a reduction in option values if this job opportunity is removed.
However, removing the option to work as self-employed does not change option values to academic schooling, as degree effects for self-employed are substantially smaller. Next, panel (b) shows how the option value of the 8th year of vocational schooling changes for low-ability individuals when sectoral opportunities are constrained. Notably, the option values for vocational schooling do not react to removing sectoral opportunities. Low-ability individuals who choose vocational schooling mostly work in the private sector, so the option values remain constant once the options to work in the public sector and/or be self-employed are removed. Thus, sectoral opportunities influence educational choices, partly by altering the composition of returns to different educational options.







\subsection{Educational Policy Evaluation}\label{subsec:Results-Policy}
We now use our model to analyze the impacts of compulsory schooling reforms. First, we provide further evidence on compliance to the Norwegian compulsory schooling reform that we earlier used to validate our model. We show who is affected by the policy along the distribution of schooling by ability and by early drop-out status. Second, we investigate the impacts of a high school enrollment policy, which requires everyone to attend ten years of schooling. Finally, we consider the impacts of introducing a tuition fees policy.

\begin{figure}[h!]\centering
\caption{Compliance to the Norwegian Compulsory Schooling Reform.}\label{Share affected}
\subfloat[By Ability]{\scalebox{0.45}{\includegraphics{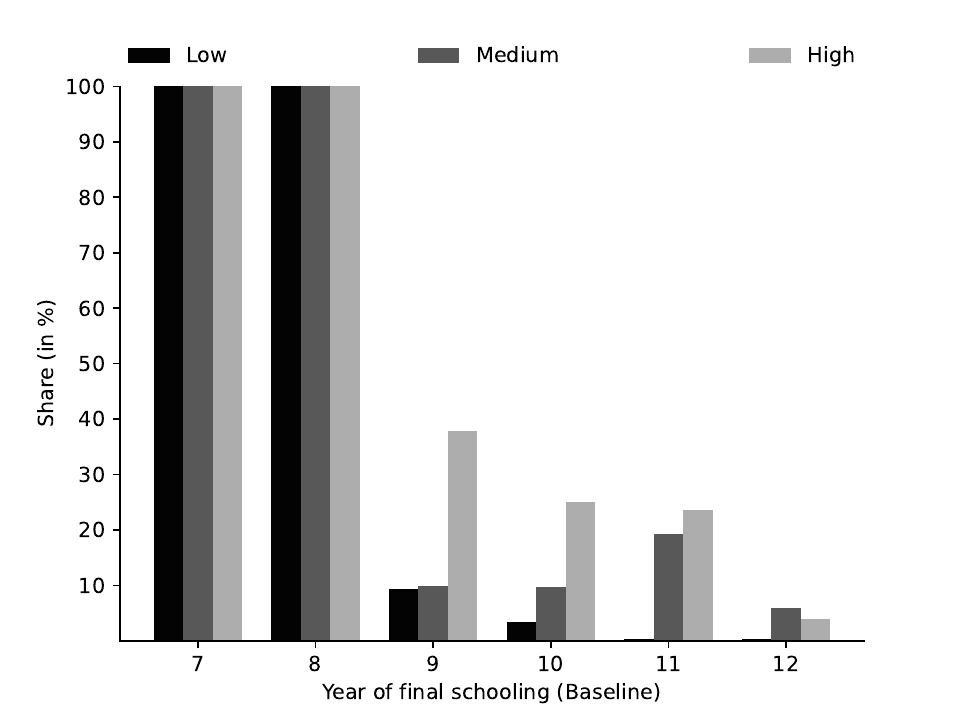}}}\hspace{0.3cm}
\subfloat[By Ability, Among Early Drop-outs]{\scalebox{0.45}{\includegraphics{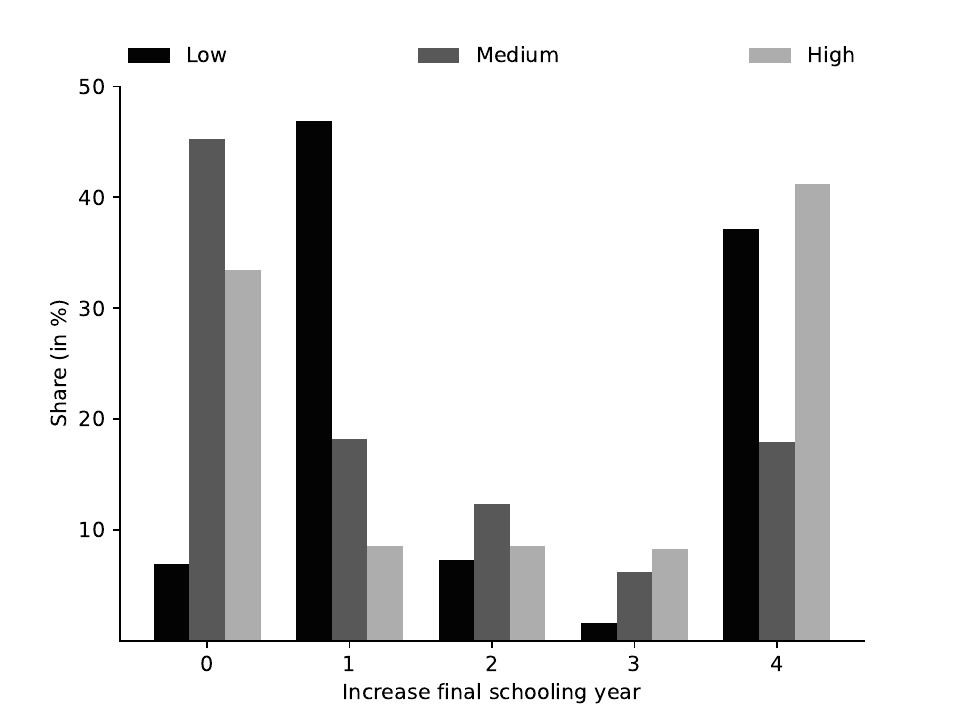}}}
\begin{center}
\begin{minipage}[t]{\columnwidth}
\emph{\scriptsize{}Note:}{\scriptsize{} The figure is based on two samples of 100,000 simulated schooling careers for each ability group under alternative scenarios. Using the point estimates, we first simulate the model with the original seven years of compulsory schooling. Next, we rerun the simulation but impose nine years of compulsory schooling. Throughout, we keep the random realizations of the productivity and taste shocks $\bm{\epsilon_t}$ fixed, and we are thus able to compare the schooling decisions of the same individual under the two different regimes. In panel (a), we plot the fractions of individuals who change their schooling decisions for varying levels of final schooling in the baseline scenario along the horizontal axis. In panel (b), we restrict our sample to individuals who initially dropped out after the 8th year of uninterrupted schooling in the baseline scenario and then illustrate the distribution of observed increases in their final schooling due to the policy reform.}{\scriptsize\par}
\end{minipage}   
\end{center}
\end{figure}

\paragraph{The Norwegian Compulsory Schooling Reform}
\noindent As described in Section \ref{subsec:Model-Validation}, the Norwegian compulsory schooling reform increased the minimum schooling requirement from seven to nine years, and was gradually introduced in different municipalities in different years. In our analysis thus far we used individuals born 1955-1960 who were not exposed to the reform and relied on the reform variation in an out-of-sample validation of our model. We now use our estimated model to shed light on the compliance to this reform by ability and early drop-out status. In panel (a) of Figure \ref{Share affected}, we show the fractions of individuals by their final year of schooling in the baseline scenario (i.e., pre-reform) along the horizontal axis that change their schooling choices due to the reform. By construction, since the post-reform compulsory schooling is nine years, all individuals that previously decided to stop after seven or eight years are affected. Notably, as  discussed in Section \ref{subsec:Model-Validation}, some of these individuals even increase their schooling beyond the new minimum requirement. Such ``inframarginal'' responses in our model can be explained by the presence of option values; by forcing individuals to attend nine years of schooling, we also bring them closer to transitions that make a high school diploma within reach.\footnote{Indeed, we relied on the extent of such ``inframarginal'' responses for the model validation in Section \ref{subsec:Model-Validation}.}

\noindent More interestingly, our model also predicts alterations in the educational trajectories among those who in the baseline scenario actually had attended nine or more years of schooling. While the presence of option values can trigger the ``inframarginal'' responses discussed above among those with less than nine years of schooling, an additional mechanism of re-enrollment possibilities is at play when we consider those having nine or more years of schooling pre-reform. Prior of the reform, 10\% of individuals had dropped out either after the 7th or 8th grade but then re-enrolled at a later time. Since the reform rules out any interruptions between 7th and 9th year of schooling, the educational trajectories individuals who earlier dropped out after the 7th or 8th grade and re-enrolled are also affected. Indeed, about 40\% of high-ability individuals who end up with only nine years of schooling in the baseline scenario increase their final schooling level after the reform. They do so because they no longer face the considerable re-enrollment costs they had to incur in the baseline scenario where they dropped out after the 7th or 8th year.\\

\noindent In panel (b) of Figure \ref{Share affected}, we show that the compulsory schooling reform affects individuals who initially dropped out after the 8th year of schooling in our model. Around 40\% of the early drop-outs with high- and medium-ability do not increase their final years of schooling post reform; these individuals all re-enrolled even in the baseline scenario and attained at least nine years of schooling in the end. However, among low-ability individuals, around 45\% increase their schooling by one year and thus only meet the new requirement, while around 35\% increase their schooling level by four years and thus attain a high school degree after the reform. Thus, option values and re-enrollment are important channels that explain these compliance patterns.

\paragraph{Compulsory High School Enrollment Policy}

\noindent Another policy we consider next is the introduction of compulsory high school enrollment, which requires all individuals to attend ten years of schooling, i.e., one more year than the Norwegian compulsory schooling reform. Appendix Figure \ref{Increase to 10 years of compulsory schooling} compares the distributions of final years of schooling in our model between the simulated reform with the nine year of compulsory schooling (`Reform 9') and the compulsory high school enrollment policy (`Reform 10'). Interestingly, we again find evidence of strong ``inframarginal'' responses; most individuals that are induced to change their schooling level by the compulsory high school enrollment policy indeed go on to complete a high school degree. Overall, the fraction of individuals who at least hold a high school degree increases from $82\%$ to more than $95\%$. Most of these increases are stem from low- and medium-ability individuals, who increase their high school shares from about $78\%$ to $97\%$ and $52\%$ to $82\%$, respectively. 

\paragraph{College Tuition Fees Policy}

\noindent The final policy we consider next is the introduction of a tuition fees corresponding to 50,000 Norwegian Kroner (2018) per year of college attendance, corresponding to around USD 6,000. As illustrated in Appendix Figure \ref{Tuition}, our model predicts that the introduction of this tuition fees would drive out virtually all medium-ability individuals from college, while this would apply to two-thirds of high-ability individuals who would attend college in the baseline. Overall college graduation share declines from $17.5$ to $3.5$ percent, with a corresponding increase in high school attendance. In particular, medium-ability individuals drawn away from college by this policy would instead pursue a vocational high school degree.

\section{Conclusion}\label{sec:Conclusion}
This paper has attempted to provide evidence on the ex-ante returns and option values to educational choices. To achieve this, we devised a dynamic model of education and work decisions in a life-cycle context that acknowledges uncertainty and sequential nature of schooling decisions. We estimated this model using Norwegian population panel data with long earnings histories, and validated this against variation in schooling choices induced by a compulsory schooling reform. Finally, we used the structure of our model to learn about the rich patterns of compliance observed in our data and the potential economic mechanisms driving these. \\

\noindent  Our analysis gave several interesting insights. The ex-ante returns to schooling vary across the different stages of educational careers, depend on the choice of track and the ability of individuals. Underlying these heterogeneities is a strong pattern of ability-related sorting into different educational tracks. We find that option values play a dominant role in shaping schooling decisions at several points in the educational career. We also document how the presence of option values and re-enrollment opportunities could explain the ``inframarginal'' impacts of compulsory schooling reforms across the distribution of schooling attainment. Further, we provide insights on how sectoral opportunities influence individuals' educational choices, reflecting differences in sector-specific returns to alternative educational tracks.\\

\noindent  While our paper provides several insights, one shortcoming is worth mentioning. Recent studies emphasize the role of ``experimentation'' in educational decisions, where individuals make such decisions in view of the returns generated through the subsequent resolution of uncertainty that they are initially faced with (see, e.g., \cite{Arcidiacono.2016} and references therein). We regard this as an important stream of research, which highlights another channel for why option values can matter in educational decisions. Our model however does not feature learning and updating of individuals' prior beliefs, but instead has focused on the analyzing educational decisions in a life-cycle context. We leave it for future work to develop a modelling framework for educational choices with learning in a life-cycle context with many periods.


\bibliographystyle{apalike}
\bibliography{main.bib}

\clearpage
\newpage{} 
\begin{center}
\setcounter{page}{1}
\end{center}

\section*{Appendix}\label{sec:App-A}
\setcounter{footnote}{0}
\setcounter{section}{0}
\setcounter{equation}{0}
\setcounter{table}{0}
\setcounter{figure}{0}
\renewcommand{\thesection}{A}
\renewcommand{\theequation}{A.\arabic{equation}}
\renewcommand{\thetable}{A.\arabic{table}}
\renewcommand{\thefigure}{A.\arabic{figure}}
\renewcommand{\thepage}{A-\arabic{page}}

\subsection{Specifications of Immediate Reward Functions}\label{subsec:App-A-utility functions}
We present here the parametrizations of immediate reward functions in our model. In the estimation, all parameters are allowed to vary freely across three observed ability types.
The specification of non-pecuniary utilities varies slightly across ability groups. Given the observed patterns in data, some parameters are well identified for one group but not for the other groups.

\paragraph{Work Choice: Wages and Nonpecuniary Utility}
\hfill \\  \\
The immediate reward of choosing the work alternative, i.e., $a \in \{PR, PU ,S\}$, consists a wage and a non-pecuniary component, where all parameters are allowed to vary flexibly across the private ($PR$), public ($PU$) and self-employment ($S$) sectors, as follows:
\begin{align*}
\underbrace{\zeta_a(\bm{k_t}, \bm{h_t}, t, a_{t -1}, e_{j,W})}_{\text{Non-Pecuniary Component}}  + \underbrace{w_a(\bm{k_t}, \bm{h_t}, t, a_{t -1}, e_{j,W}, \epsilon_{W,t})}_{\text{Wage Component}}               
\end{align*}
The wage reward is defined as follows:
\begin{align*}
x_a(\bm{k_t}, \bm{h_t}, t, a_{t-1}, e_{j, a}, \epsilon_{a,t}) & = \exp \big( \Gamma_a(\bm{k_t},  \bm{h_t}, t, a_{t-1}, e_{j,a}) \cdot \epsilon_{a,t} \big) \nonumber
\end{align*}
\\ 
\noindent \textbf{\textit{Wage Component}}
\begin{align*}
\Gamma_a(\bm{k_t}, \bm{h_t}, t, a_{t-1}, e_{j, W}) & = e_{j,a} \\ \nonumber
 & + \delta_{1,a} \cdot  h^A_{t} + \delta_{2,a} \cdot  h^V_{t} + \delta_{3,a} \cdot  k^{W}_{t} + \delta_{4,a} \cdot  (k^{W}_{t})^2+ \delta_{5,a} \cdot  k^{a}_{t} + \delta_{6,a} \cdot  (k^{a}_{t})^2 \\\nonumber
 & + \sum_{d\in \{9, 12, 16\}} \gamma^A_{d,a} \cdot \ind[h^A_t \geq d] + \sum_{d\in \{9, 12\}} \gamma^V_{d,a} \cdot \ind[h^V_t \geq d] \\ \nonumber
& + \eta_{1,a} \cdot \ind[a_{t-1} = W] + \nu_{1,a} \cdot  (t - 15) + \nu_{2,a} \cdot \ind[t < 17]
\end{align*}

\noindent \textbf{\textit{Nonpecuniary Component}} \\
$\beta_{9,a}$ is only included in the reward equations for self-employment and the public sector for low-ability individuals. 
$\beta_{6,a}$ is included in the self-employment and public sector rewards for medium- and high-ability individuals.
The choice is motivated by the fact that low-ability individuals tend to have high incentives to not work at all and the term counteracts this.
$\beta_{8,a}$, $\beta_{2,a}$ account for the importance of general work experience.
Since low-ability individuals work primarily in the private sector, these terms are not included for this group.

\begin{align*}
\zeta_{a}(\bm{k_t}, \bm{h_t}, t, a_{t-1})  & = e_{j,a}  + \beta_{1,a} \cdot \ind[t < 17] + \beta_{2,a} \cdot k^{W}_t+ \beta_{3,a} \cdot k^{a}_t + \beta_{4,a} \cdot h^A_t + \beta_{5,a} \cdot h^V_t \\ \nonumber 
& + \sum_{d\in \{9, 12, 16\}} \vartheta^A_{d,a} \cdot \ind[h^A_t \geq d]
 + \sum_{d\in \{9, 12\}} \vartheta^V_{d,a} \cdot \ind[h^V_t \geq d] \\ \nonumber
 & + \beta_{6,a} \cdot \ind[a_{t-1} = PR] + \beta_{7,a} \cdot \ind[k^{a}_t > 0]+ \beta_{8,a} \cdot \ind[k^{W}_t > 0] + \beta_{9,a} \cdot \ind[a_{t-1} = a] \\ \nonumber
\end{align*}

\paragraph{Academic Choice: Nonpecuniary Utility}
\begin{align*}
\zeta_{A}(\bm{k_t}, \bm{h_t}, t, a_{t-1})  & = e_{j,A}  + \beta_{1,A} \cdot (t-15) +  \beta_{4,A} \cdot \ind[h^V_{t} > 0]\\
& + \beta_{4,A} \cdot \ind[a_{t-1} = A]  + \beta_{5,A} \cdot \ind[a_{t-1} \in \{PR,PU,S,H\}] + \beta_{6,A} \cdot \ind[a_{t-1} = A] \cdot \ind[h^A_t \geq 12] \\
& + \sum_{d\in \{9, 12, 16\}} \vartheta^A_{d,A} \cdot \ind[h^A_t \geq d] + \vartheta^A_{9,A} \cdot \ind[h^V_t \geq 9] + \epsilon_{t,A}
\end{align*}
The coefficient $\beta_{4,A}$ is only included for high IQ individuals. The coefficients is included to avoid excessive dropping out and re-enrolling. The coefficient $\vartheta^A_{9,A}$ is mainly relevant in settings where individuals dropout early and potentially re-enroll later. Since this does not happen frequently for high IQ individuals the coefficient is only included for medium and low IQ individuals.

\paragraph{Vocational Choice: Nonpecuniary Utility}
\begin{align*}
\zeta_{V}(\bm{k_t}, \bm{h_t}, t, a_{t-1})  & = e_{j,a} + \beta_{1,a} \cdot (t-15)  + \beta_{2,a} \cdot \ind[t < 17 ] + \beta_{3,a} \cdot \ind[h^{A}_{t} > 0]\\ \nonumber
&  + \beta_{4,a} \cdot \ind[a_{t-1} = a] + \beta_{5,a} \cdot \ind[a_{t-1} \in \{PR,PU,S,H\}]\\ \nonumber
& +\vartheta^A_{9,a} \cdot \ind[h^V_t \geq 9]+ \vartheta^A_{9,a} \cdot \ind[h^A_t \geq 9] + \epsilon_{t,V}
\end{align*}
The coefficient $\beta_{5,V}$ is only included for high IQ individuals. The coefficient is included to avoid excessive dropping out and re-enrolling. The coefficient $\vartheta^A_{9,V}$ is mainly relevant in settings where individuals dropout early and potentially re-enroll later. Since this does not happen frequently for high IQ individuals the coefficient is only included for medium and low IQ individuals.

\paragraph{Home Choice: Nonpecuniary Utility}
\begin{align*}
\zeta_{H}(\bm{k_t}, \bm{h_t}, t, a_{t-1})  & = e_{j,H} + \beta_{1,H} \cdot \ind[t < 17] + \sum_{d\in \{12, 16\}} \vartheta^A_{d,H} \cdot \ind[h^A_t \geq d] + \vartheta^V_{12,H}  \cdot \ind[h^V_t \geq 12] + \epsilon_{t,H}
\end{align*}

\pagebreak
\subsection{Estimation Results}\label{subsec:App-A-estimation-results}
This section presents the estimated parameters and their corresponding standard errors. Whenever a parameter has no standard error, it means that it is not well-identified or irrelevant and that we have taken this parameter out of the corresponding specification. See Section \ref{subsec:App-A-utility functions} for a more detailed explanation.
In some cases, a particular unobserved latent type is not well identified for some choice options since the corresponding type is far from ever choosing this option. 
Another special case relates to the penalty for minors working, which are set so that people below 17 do not work as self-employed or in the public sector. 

\paragraph{\underline{Choice Alternative: Private Sector}} 
\hfill \\  \\
Table \ref{Parameter estimates for wages_private} presents the point estimates and the standard errors for the parameters in the wage component, while Table \ref{Parameter estimates for working_private} presents the point estimates and the standard errors for the parameters in the specification of non-pecuniary component.\\

\begin{table}[H]
\centering
  \caption{Choice Alternative: Private Sector -- Wage Component.}
  \label{Parameter estimates for wages_private}

\begin{tabularx}{\textwidth}{l c c c c}
\toprule
{}  &{} & \textbf{Low IQ} &\textbf{Medium IQ} &\textbf{High IQ}\\
\midrule

                Constant & $e_{1,PR}$ & 12.1 & 11.5 & 11.5\\
                &&(0.00469) & (0.01962)& (0.02496) \\
                Type 1 & $e_{2,PR}$ & 0.17628 & 0.07283 & 0.25511\\
                &&(0.02692) & (0.02309)& (0.05611) \\
                Type 2 & $e_{3,PR}$ & -0.05007 & 0.50311 & 0.37545\\
                &&(0.02414) & (0.01856)& (0.02000) \\
                Type 3 & $e_{4,PR}$ & 0.09268 & - & -\\
                &&(0.04047) & (-)& (-) \\
                Reward Years of Academic Schooling & $\delta_{1,PR}$ & 0.10764 & 0.14816 & 0.17273\\
                &&(0.00136) & (0.00078)& (0.00042) \\
                Reward Years of Vocational Schooling & $\delta_{2,PR}$ & 0.16961 & 0.12013 & 0.15994\\
                &&(0.00109) & (0.00250)& (0.00032) \\
                Reward Experience General Work & $\delta_{3,PR}$ & - & 0.10525 & 0.11110\\
                &&(-) & (0.00030)& (0.00301) \\
                Reward Experience General Work Square & $\delta_{4,PR}$ & - & -0.10907 & -0.09934\\
                &&(-) & (0.01408)& (0.00247) \\
                Reward Experience Work Private & $\delta_{5,PR}$ & 0.10496 & 0.08750 & 0.06362\\
                &&(0.00145) & (0.00029)& (0.00336) \\
                Reward Experience Work Private Square & $\delta_{6,PR}$ & -0.05487 & -0.04357 & -0.03549\\
                &&(0.00309) & (0.01399)& (0.00221) \\
                Reward Vocational High School Degree & $\gamma^{V}_{12,PR}$ & 0.05735 & 0.33053 & 0.35622\\
                &&(0.00603) & (0.00819)& (0.01879) \\
                Reward Vocational Middle School Degree & $\gamma^{V}_{9,PR}$ & 0.00600 & 0.24167 & 0.02969\\
                &&(0.00731) & (0.00914)& (0.01233) \\
                Reward Middle School Degree & $\gamma^{A}_{9,PR}$ & 0.10091 & -0.00008 & 0.05080\\
                &&(0.00554) & (0.04361)& (0.01109) \\
                Reward College Degree & $\gamma^{A}_{15,PR}$ & 0.02324 & 0.28610 & 0.14156\\
                &&(-) & (0.00993)& (0.00672) \\
                Reward High School Degree & $\gamma^{A}_{12,PR}$ & 0.05448 & 0.31063 & 0.27187\\
                &&(0.09879) & (0.04340)& (0.01993) \\
                Period & $\nu_{1,PR}$ & -0.07769 & -0.13118 & -0.11556\\
                &&(0.00071) & (0.00015)& (0.00045) \\
                Minor Age & $\nu_{2,PR}$ & - & -2.0 & -1.6\\
                &&(-) & (2.5)& (0.65087) \\
                Work Lagged & $\eta_{1,PR}$ & 0.34226 & 0.36983 & 0.25920\\
                &&(0.00565) & (0.02131)& (0.01372) \\
                \bottomrule
 \end{tabularx}

\end{table}

\begin{table}[H]
\centering
  \caption{Choice Alternative: Private Sector -- Non-Pecuniary Component.}
  \label{Parameter estimates for working_private}

\begin{tabularx}{\textwidth}{l c c c c}
\toprule
{}  &{} & \textbf{Low IQ} &\textbf{Medium IQ} &\textbf{High IQ}\\
\midrule

                Constant & $e_{1,PR}$ & 240864.3 & 302097.3 & 280533.2\\
                &&(8270.5) & (6917.2)& (15478.1) \\
                Type 1 & $e_{2,PR}$ & -8427.4 & -654.3 & -11050.5\\
                &&(10530.6) & (32282.4)& (12603.6) \\
                Type 2 & $e_{3,PR}$ & -8005.3 & 24773.1 & 130.8\\
                &&(6211.8) & (9107.2)& (11889.3) \\
                Type 3 & $e_{4,PR}$ & -1410.1 & - & -\\
                &&(8725.7) & (-)& (-) \\
                Reward Experience General Work & $\beta_{2,PR}$ & - & 3218.4 & 2471.5\\
                &&(-) & (7597.3)& (850.0) \\
                Reward Experience Work Private & $\beta_{3,PR}$ & 11547.3 & 3763.4 & 3195.7\\
                &&(198.9) & (7593.3)& (853.8) \\
                Reward Years of Academic Schooling & $\beta_{4,PR}$ & 1687.0 & 11947.7 & 14105.7\\
                &&(1704.2) & (1409.3)& (1065.5) \\
                Reward Years of Vocational Schooling & $\beta_{5,PR}$ & 11593.5 & 2753.6 & 13330.3\\
                &&(933.2) & (823.3)& (1322.7) \\
                Any Private Work Experience & $\beta_{7,PR}$ & 137907.6 & 32540.4 & 33311.2\\
                &&(8478.6) & (13397.0)& (55400.4) \\
                Any Work Experience & $\beta_{8,PR}$ & - & 57641.4 & 50964.2\\
                &&(-) & (12362.3)& (55540.5) \\
                Reward Vocational High School Degree & $\vartheta^{V}_{12,PR}$ & -2896.1 & 12191.0 & 13776.8\\
                &&(3909.6) & (3097.0)& (7457.6) \\
                Reward Vocational Middle School Degree & $\vartheta^{V}_{9,PR}$ & -3005.7 & 16408.3 & 13558.2\\
                &&(2861.5) & (4014.0)& (7418.2) \\
                Reward Middle School Degree & $\vartheta^{A}_{9,PR}$ & 4049.9 & -4154.2 & 15394.2\\
                &&(3725.5) & (7603.9)& (5103.6) \\
                Reward College Degree & $\vartheta^{A}_{16,PR}$ & - & 24647.0 & 15457.1\\
                &&(-) & (6302.7)& (4607.5) \\
                Reward High School Degree & $\vartheta^{A}_{12,PR}$ & 38683.3 & 12760.1 & 35410.9\\
                &&(10713.2) & (7812.6)& (5174.1) \\
                \bottomrule
 \end{tabularx}

\end{table}

\pagebreak
\paragraph{\underline{Choice Alternative: Public Sector}} 
\hfill \\  \\
Table \ref{Parameter estimates for wages_public} presents the point estimates and the standard errors for the parameters in the wage component, while Table \ref{Parameter estimates for working_public} presents the point estimates and the standard errors for the parameters in the specification of non-pecuniary component.\\

\begin{table}[H]
\centering
  \caption{Choice Alternative: Public Sector -- Wage Component.}
  \label{Parameter estimates for wages_public}

\begin{tabularx}{\textwidth}{l c c c c}
\toprule
{}  &{} & \textbf{Low IQ} &\textbf{Medium IQ} &\textbf{High IQ}\\
\midrule

                Constant & $e_{1,PU}$ & 11.8 & 11.1 & 10.9\\
                &&(0.02055) & (0.00478)& (0.07144) \\
                Type 1 & $e_{2,PR}$ & - & 0.33312 & 0.25268\\
                &&(-) & (0.02139)& (0.02767) \\
                Type 2 & $e_{3,PR}$ & 0.63879 & -0.01928 & 0.00786\\
                &&(0.03723) & (-)& (0.02514) \\
                Reward Years of Academic Schooling & $\delta_{1,PU}$ & 0.07376 & 0.14692 & 0.17320\\
                &&(0.00285) & (0.00053)& (0.00018) \\
                Reward Years of Vocational Schooling & $\delta_{2,PU}$ & 0.15731 & 0.10570 & 0.15139\\
                &&(0.00610) & (0.00793)& (0.00163) \\
                Reward Experience General Work & $\delta_{3,PU}$ & -0.00313 & 0.10402 & 0.11290\\
                &&(0.00315) & (0.00028)& (0.00018) \\
                Reward Experience General Work Square & $\delta_{4,PU}$ & -0.00199 & -0.12492 & -0.13500\\
                &&(0.09412) & (0.00414)& (0.00405) \\
                Reward Experience Public Work & $\delta_{5,PU}$ & 0.09588 & 0.04119 & 0.04733\\
                &&(0.00302) & (0.00031)& (0.00029) \\
                Reward Experience Public Work Square & $\delta_{6,PU}$ & -0.06687 & 0.00276 & -0.01695\\
                &&(0.10340) & (0.00513)& (0.00512) \\
                Reward Vocational High School Degree & $\gamma^{V}_{12,PU}$ & 0.02690 & 0.52160 & 0.10028\\
                &&(0.01088) & (1.8)& (1.5) \\
                Reward Vocational Middle School Degree & $\gamma^{V}_{9,PU}$ & 0.00773 & 0.03469 & 0.08647\\
                &&(0.01425) & (1.8)& (0.00874) \\
                Reward Middle School Degree & $\gamma^{A}_{9,PU}$ & 0.10618 & 0.02698 & 0.02118\\
                &&(0.01650) & (0.13418)& (0.09915) \\
                Reward College Degree & $\gamma^{A}_{15,PU}$ & - & 0.17475 & 0.11769\\
                &&(-) & (0.01211)& (0.00889) \\
                Reward High School Degree & $\gamma^{A}_{12,PU}$ & 0.04767 & 0.47086 & 0.50295\\
                &&(0.58908) & (0.12628)& (0.08350) \\
                Period & $\nu_{1,PU}$ & -0.06696 & -0.09688 & -0.10288\\
                &&(0.00033) & (0.00051)& (0.00015) \\
                Work Lagged & $\eta_{1,PU}$ & 0.32174 & 0.60734 & 0.65906\\
                &&(0.01761) & (0.01458)& (0.06599) \\
                \bottomrule
 \end{tabularx}

\begin{center}
\begin{minipage}[t]{\columnwidth}
\emph{\scriptsize{}Note:}{\scriptsize{} Type 3 is missing in this table as it is not well identified in any of the three specifications. Type 3 is only included for low IQ individuals and the third type always works in the private sector in this specification.}{\scriptsize\par}
\end{minipage}
\end{center}
\end{table}

\begin{table}[H]
\centering
  \caption{Choice Alternative: Public Sector -- Non-Pecuniary Component.}
  \label{Parameter estimates for working_public}

\begin{tabularx}{\textwidth}{l c c c c}
\toprule
{}  &{} & \textbf{Low IQ} &\textbf{Medium IQ} &\textbf{High IQ}\\
\midrule

                Constant & $e_{1,PU}$ & 264882.5 & 238795.9 & 308910.4\\
                &&(5465.2) & (14060.6)& (12454.8) \\
                Type 1 & $e_{2,PR}$ & - & 4911.4 & 36709.8\\
                &&(-) & (14060.6)& (9129.3) \\
                Type 2 & $e_{3,PR}$ & -8241.0 & - & -\\
                &&(5465.2) & (-)& (-) \\
                Reward Experience Public Work & $\beta_{2,PU}$ & 5990.7 & 3538.9 & 3723.8\\
                &&(280.8) & (4247.8)& (783.8) \\
                Reward Experience General Work & $\beta_{3,PU}$ & - & 3993.8 & 3968.8\\
                &&(-) & (3978.4)& (811.6) \\
                Reward Years of Academic Schooling & $\beta_{4,PU}$ & 2535.7 & 2427.6 & 9569.9\\
                &&(2850.4) & (1618.7)& (999.9) \\
                Reward Years of Vocational Schooling & $\beta_{5,PU}$ & 10258.5 & 12569.4 & 13465.2\\
                &&(754.8) & (1886.2)& (3871.4) \\
                Work Lagged & $\beta_{6,PU}$ & - & - & 94855.7\\
                &&(-) & (-)& (8329.8) \\
                Any Public Work Experience & $\beta_{7,PU}$ & 125227.1 & 88779.2 & 5227.3\\
                &&(10162.9) & (12429.7)& (3458.3) \\
                Any Work Experience & $\beta_{8,PU}$ & - & 23630.2 & 114214.1\\
                &&(-) & (12429.7)& (11984.3) \\
                Public Work Lagged & $\beta_{9,PU}$ & 1644.7 & - & -\\
                &&(4891.2) & (-)& (-) \\
                Reward Vocational High School Degree & $\vartheta^{V}_{12,PU}$ & -2905.2 & 21144.7 & -\\
                &&(2525.8) & (6018.8)& (-) \\
                Reward Vocational Middle School Degree & $\vartheta^{V}_{9,PU}$ & -4699.7 & 10122.4 & 13544.8\\
                &&(3369.8) & (6018.8)& (6406.3) \\
                Reward Middle School Degree & $\vartheta^{A}_{9,PU}$ & 9307.8 & 842.7 & -4233.7\\
                &&(5019.9) & (20783.7)& (15409.8) \\
                Reward College Degree & $\vartheta^{A}_{16,PU}$ & - & 20541.4 & 74809.3\\
                &&(-) & (6433.7)& (3953.1) \\
                Reward High School Degree & $\vartheta^{A}_{12,PU}$ & 38956.0 & 425.0 & 17722.6\\
                &&(4327.4) & (14547.8)& (14975.3) \\
                \bottomrule
 \end{tabularx}

\begin{center}
\begin{minipage}[t]{\columnwidth}
\emph{\scriptsize{}Note:}{\scriptsize{} Type 3 is missing in this table as it is not well identified in any of the three specifications. Type 3 is only included for low IQ individuals and the third type always works in the private sector in this specification.}{\scriptsize\par}
\end{minipage}
\end{center}
\end{table}

\paragraph{\underline{Choice Alternative: Self Employment}} 
\hfill \\  \\
Table \ref{Parameter estimates for wages_self} presents the point estimates and the standard errors for the parameters in the wage component, while Table \ref{Parameter estimates for working_self} presents the point estimates and the standard errors for the parameters in the specification of non-pecuniary component.\\

\begin{table}[H]
\centering
  \caption{Choice Alternative: Self Employment -- Wage Component.}
  \label{Parameter estimates for wages_self}

\begin{tabularx}{\textwidth}{l c c c c}
\toprule
{}  &{} & \textbf{Low IQ} &\textbf{Medium IQ} &\textbf{High IQ}\\
\midrule

                Constant & $e_{1,S}$ & 11.9 & 11.5 & 10.9\\
                &&(0.00259) & (0.02111)& (0.09371) \\
                Type 1 & $e_{2,PR}$ & 0.11961 & 0.14859 & 0.17965\\
                &&(0.06508) & (0.00475)& (0.02237) \\
                Type 2 & $e_{3,PR}$ & -0.00542 & 0.38101 & 0.06762\\
                &&(0.04897) & (0.02913)& (0.01721) \\
                Type 3 & $e_{4,PR}$ & 0.00229 & - & -\\
                &&(0.08717) & (-)& (-) \\
                Reward Years of Academic Schooling & $\delta_{1,S}$ & 0.11141 & 0.12539 & 0.16501\\
                &&(0.00033) & (0.00048)& (0.00013) \\
                Reward Years of Vocational Schooling & $\delta_{2,S}$ & 0.16222 & 0.11082 & 0.11707\\
                &&(0.00268) & (0.00026)& (0.00052) \\
                Reward Experience General Work & $\delta_{3,S}$ & -0.00098 & 0.08185 & 0.10290\\
                &&(0.00450) & (0.00089)& (0.00026) \\
                Reward Experience General Work Square & $\delta_{4,S}$ & -0.00024 & -0.06265 & -0.12382\\
                &&(0.16097) & (0.00022)& (0.00264) \\
                Reward Experience Self Employed & $\delta_{5,S}$ & 0.10448 & 0.04706 & 0.03522\\
                &&(0.00459) & (0.00063)& (0.00082) \\
                Reward Self Employed Square & $\delta_{6,S}$ & -0.05367 & -0.02710 & -0.06122\\
                &&(0.16061) & (0.00091)& (0.00278) \\
                Reward Vocational High School Degree & $\gamma^{V}_{12,S}$ & 0.10238 & 0.14908 & 0.38303\\
                &&(0.01694) & (0.01239)& (0.08259) \\
                Reward Vocational Middle School Degree & $\gamma^{V}_{9,S}$ & 0.01859 & 0.04600 & 0.03953\\
                &&(0.01575) & (0.01266)& (0.08822) \\
                Reward Middle School Degree & $\gamma^{A}_{9,S}$ & 0.09662 & 0.07674 & 0.09875\\
                &&(0.01785) & (0.00982)& (0.03282) \\
                Reward College Degree & $\gamma^{A}_{15,S}$ & 0.01803 & 0.13065 & 0.16817\\
                &&(-) & (0.10580)& (0.02567) \\
                Reward High School Degree & $\gamma^{A}_{12,S}$ & 0.25517 & 0.19684 & 0.19378\\
                &&(0.02935) & (0.01131)& (0.05081) \\
                Period & $\nu_{1,S}$ & -0.07432 & -0.08945 & -0.06156\\
                &&(0.00014) & (0.00038)& (0.00021) \\
                Work Lagged & $\eta_{1,S}$ & 0.34353 & 0.54149 & 0.56502\\
                &&(0.00312) & (0.02402)& (0.08108) \\
                \bottomrule
 \end{tabularx}

\begin{center}
\begin{minipage}[t]{\columnwidth}
\emph{\scriptsize{}Note:}{\scriptsize{} Type 3 is missing in this table as it is not well identified in any of the three specifications. Type 3 is only included for low IQ individuals and the third type always works in the private sector in this specification.}{\scriptsize\par}
\end{minipage}
\end{center}
\end{table}

\begin{table}[H]
\centering
  \caption{Choice Alternative: Self Employment -- Non-Pecuniary Component.}
  \label{Parameter estimates for working_self}

\begin{tabularx}{\textwidth}{l c c c c}
\toprule
{}  &{} & \textbf{Low IQ} &\textbf{Medium IQ} &\textbf{High IQ}\\
\midrule

                Constant & $e_{1,S}$ & 295981.5 & 180293.9 & 230336.8\\
                &&(9347.0) & (9505.6)& (31204.5) \\
                Type 1 & $e_{2,PR}$ & - & 2221.2 & 3319.5\\
                &&(-) & (12988.7)& (12468.2) \\
                Type 2 & $e_{3,PR}$ & -3809.4 & 73325.3 & -880.2\\
                &&(9340.8) & (9990.1)& (21064.2) \\
                Reward Experience Self Employed & $\beta_{2,S}$ & 12380.7 & 3998.5 & 2366.1\\
                &&(277.3) & (4407.8)& (1256.7) \\
                Reward Experience General Work & $\beta_{3,S}$ & - & 4753.4 & 3060.0\\
                &&(-) & (4416.7)& (641.9) \\
                Reward Years of Academic Schooling & $\beta_{4,S}$ & 1667.3 & 4520.6 & 1835.7\\
                &&(2882.2) & (1645.4)& (1110.8) \\
                Reward Years of Vocational Schooling & $\beta_{5,S}$ & 12887.7 & 13212.8 & 12496.1\\
                &&(778.1) & (831.7)& (1888.6) \\
                Work Lagged & $\beta_{6,S}$ & - & 50218.3 & 200240.3\\
                &&(-) & (38470.0)& (3527.5) \\
                Any Self Employed Experience & $\beta_{7,S}$ & 84565.9 & 40700.5 & 46430.6\\
                &&(9372.8) & (23442.9)& (9830.4) \\
                Any Work Experience & $\beta_{8,S}$ & - & 105147.0 & 95528.2\\
                &&(-) & (23733.8)& (28781.1) \\
                Self Employed Lagged & $\beta_{9,S}$ & -119.1 & - & -\\
                &&(6091.9) & (-)& (-) \\
                Reward Vocational High School Degree & $\vartheta^{V}_{12,S}$ & 308.8 & 48114.5 & 18121.9\\
                &&(3169.3) & (3586.9)& (20039.2) \\
                Reward Vocational Middle School Degree & $\vartheta^{V}_{9,S}$ & 2565.6 & 29776.4 & 12126.4\\
                &&(2902.6) & (4919.8)& (19452.7) \\
                Reward Middle School Degree & $\vartheta^{A}_{9,S}$ & 2731.2 & -5594.5 & -6087.2\\
                &&(5757.4) & (6172.8)& (8189.4) \\
                Reward College Degree & $\vartheta^{A}_{16,S}$ & - & 25394.5 & 44290.8\\
                &&(-) & (23798.4)& (6371.5) \\
                Reward High School Degree & $\vartheta^{A}_{12,S}$ & -28042.9 & 14707.1 & 40264.8\\
                &&(10708.8) & (6758.2)& (7201.0) \\
                \bottomrule
 \end{tabularx}

\begin{center}
\begin{minipage}[t]{\columnwidth}
\emph{\scriptsize{}Note:}{\scriptsize{} Type 3 is missing in this table as it is not well identified in any of the three specifications. Type 3 is only included for low IQ individuals and the third type always works in the private sector in this specification.}{\scriptsize\par}
\end{minipage}
\end{center}
\end{table}

\pagebreak
\paragraph{\underline{Choice Alternative: Academic Schooling}}
\hfill \\  \\ 
Table \ref{Parameter estimates for academic schooling} presents the point estimates and the standard errors for the parameters determining the immediate utility from academic schooling.\\

\begin{table}[H]
\centering
  \caption{Choice Alternative: Academic Schooling.}
  \label{Parameter estimates for academic schooling}

\begin{tabularx}{\textwidth}{l c c c c}
\toprule
{}  &{} & \textbf{Low IQ} &\textbf{Medium IQ} &\textbf{High IQ}\\
\midrule

                Constant & $e_{1,A}$ & -98271.5 & -40476.2 & 57904.7\\
                &&(9596.7) & (14339.9)& (9465.8) \\
                Type 1 & $e_{2,PR}$ & 2024.4 & 16843.7 & -4009.4\\
                &&(10347.7) & (16143.9)& (10135.0) \\
                Type 2 & $e_{3,PR}$ & 19937.8 & 41294.3 & 60647.6\\
                &&(9580.2) & (15131.9)& (11910.4) \\
                Type 3 & $e_{4,PR}$ & -59833.1 & - & -\\
                &&(9486.5) & (-)& (-) \\
                Period & $\beta_{1,A}$ & -14372.3 & -39389.8 & -45497.9\\
                &&(3861.3) & (2568.7)& (1470.2) \\
                Any Vocational Experience & $\beta_{3,A}$ & -97028.4 & -120161.0 & -114535.1\\
                &&(10997.0) & (33595.5)& (9797.8) \\
                Academic Lagged & $\beta_{4,A}$ & 53576.5 & 83838.9 & 74942.5\\
                &&(9334.5) & (10075.2)& (7399.6) \\
                Lagged Choice not Education & $\beta_{5,A}$ & - & - & -269762.2\\
                &&(-) & (-)& (10205.2) \\
                Post Degree Return & $\beta_{6,A}$ & - & 40926.1 & 71997.1\\
                &&(-) & (10992.6)& (14822.9) \\
                Reward High School Degree & $\vartheta^{A}_{12,A}$ & - & 146293.7 & 46283.7\\
                &&(-) & (12784.7)& (7794.7) \\
                Reward College Degree & $\vartheta^{A}_{16,A}$ & - & 145380.5 & 147851.6\\
                &&(-) & (34924.4)& (18013.2) \\
                Reward Middle School Degree & $\vartheta^{A}_{9,A}$ & 79305.3 & 83377.3 & 92611.7\\
                &&(10867.9) & (10443.0)& (9277.5) \\
                Reward Vocational Middle School Degree & $\vartheta^{A}_{9,V}$ & - & -247748.4 & -\\
                &&(-) & (18167.5)& (-) \\
                \bottomrule
 \end{tabularx}

\end{table}

\pagebreak
\paragraph{\underline{Choice Alternative: Vocational Schooling}}
\hfill \\  \\  
Table \ref{Parameter estimates for vocational schooling} presents the point estimates and the standard errors for the parameters determining the immediate utility from vocational schooling.

\begin{table}[H]
\centering
  \caption{Choice Alternative: Vocational Schooling.}
  \label{Parameter estimates for vocational schooling}

\begin{tabularx}{\textwidth}{l c c c c}
\toprule
{}  &{} & \textbf{Low IQ} &\textbf{Medium IQ} &\textbf{High IQ}\\
\midrule

                Constant & $e_{1,V}$ & -80064.2 & -5157.2 & 76784.2\\
                &&(10198.2) & (13003.6)& (11992.3) \\
                Type 1 & $e_{2,PR}$ & 12692.6 & 3365.8 & -10485.6\\
                &&(9187.6) & (16773.2)& (17157.2) \\
                Type 2 & $e_{3,PR}$ & -5714.3 & 43587.5 & 75185.9\\
                &&(8374.0) & (13389.6)& (15083.3) \\
                Type 3 & $e_{4,PR}$ & -59483.1 & - & -\\
                &&(10103.0) & (-)& (-) \\
                Period & $\beta_{1,V}$ & -28210.9 & -15538.9 & -15306.9\\
                &&(3265.9) & (3563.6)& (4470.3) \\
                Minor Age & $\beta_{2,V}$ & - & 68987.5 & 91075.1\\
                &&(-) & (13400.4)& (10623.9) \\
                Any Academic Experience & $\beta_{3,V}$ & -101719.8 & -100112.4 & -115946.1\\
                &&(11000.8) & (21400.8)& (9184.2) \\
                Vocational Lagged & $\beta_{4,V}$ & 1606.0 & 8723.5 & 47514.7\\
                &&(6611.7) & (9392.2)& (9853.1) \\
                Lagged Choice not Education & $\beta_{5,V}$ & - & - & -260018.9\\
                &&(-) & (-)& (24414.3) \\
                Reward Vocational Middle School Degree & $\vartheta^{V}_{9,V}$ & 168603.7 & 75011.9 & 71910.2\\
                &&(10419.5) & (11311.6)& (12826.3) \\
                \bottomrule
 \end{tabularx}

\end{table}

\pagebreak
\paragraph{\underline{Choice Alternative: Staying at Home}}
\hfill \\  \\
Table \ref{Parameter estimates for staying at home} presents the point estimates and the standard errors for the parameters determining the immediate utility of staying at home.

\begin{table}[H]
\centering
  \caption{Choice Alternative: Staying at Home.}
  \label{Parameter estimates for staying at home}

\begin{tabularx}{\textwidth}{l c c c c}
\toprule
{}  &{} & \textbf{Low IQ} &\textbf{Medium IQ} &\textbf{High IQ}\\
\midrule

                Constant & $e_{1,H}$ & -61993.5 & 39523.1 & 134951.0\\
                &&(7897.9) & (12138.7)& (8204.2) \\
                Type 1 & $e_{2,PR}$ & -109584.5 & 14010.0 & 80035.0\\
                &&(11445.5) & (19666.1)& (11524.8) \\
                Type 2 & $e_{3,PR}$ & -136231.4 & 13232.0 & -\\
                &&(7848.4) & (17915.7)& (-) \\
                Type 3 & $e_{4,PR}$ & 35911.6 & - & -\\
                &&(8541.0) & (-)& (-) \\
                Minor Age & $\beta_{1,H}$ & 205531.2 & 418226.6 & 111000.0\\
                &&(10388.8) & (15531.8)& (15939.4) \\
                Period & $\beta_{2,H}$ & -2789.1 & - & -\\
                &&(550.9) & (-)& (-) \\
                Reward High School Degree & $\vartheta^{A}_{12,H}$ & 153457.4 & 79111.3 & 67771.2\\
                &&(6781.8) & (29987.2)& (10681.0) \\
                Reward College Degree & $\vartheta^{A}_{16,H}$ & - & 63060.1 & 67147.8\\
                &&(-) & (36040.7)& (27959.3) \\
                Reward Vocational High School Degree & $\vartheta^{V}_{12,H}$ & 4752.2 & 48480.3 & 47196.8\\
                &&(8417.2) & (13685.5)& (10132.1) \\
                \bottomrule
 \end{tabularx}

\end{table}

\newpage
\paragraph{\underline{Time Preferences and the Distribution of Shocks}}

\hfill \\  \\
Table \ref{Parameter estimates for discount rate and distribution of shocks} presents the point estimates and the standard errors for the parameters determining the discount rate and distribution of the shocks.

\begin{table}[H]
\centering
  \caption{Time Preferences and Distribution of Shocks.}
  \label{Parameter estimates for discount rate and distribution of shocks}

\begin{tabularx}{\textwidth}{l c c c c}
\toprule
{}  &{} & \textbf{Low IQ} &\textbf{Medium IQ} &\textbf{High IQ}\\
\midrule

                Discount Rate & Discount Rate & 0.96391 & 0.96120 & 0.95399\\
                &&(0.00060) & (0.00082)& (0.00045) \\
                Shock SD Work & Shock SD Work & 0.19820 & 0.22810 & 0.25433\\
                &&(0.00397) & (0.01502)& (0.00728) \\
                Shock SD Public & Shock SD Public & 0.14752 & 0.20060 & 0.22360\\
                &&(0.01042) & (0.00901)& (0.00847) \\
                Shock SD Self & Shock SD Self & 0.22307 & 0.32006 & 0.31039\\
                &&(0.02342) & (0.01320)& (0.01087) \\
                Shock SD Academic & Shock SD Academic & 351343.2 & 357454.0 & 372697.7\\
                &&(5547.8) & (1402.4)& (1009.1) \\
                Shock SD Vocational & Shock SD Vocational & 99114.8 & 42503.6 & 173032.0\\
                &&(3365.6) & (1751.0)& (1293.4) \\
                Shock SD Home & Shock SD Home & 1041070.1 & 652166.2 & 736163.6\\
                &&(3525.2) & (866.5)& (839.7) \\
                \bottomrule
 \end{tabularx}

\end{table}

\pagebreak
\subsection{Additional Figures and Tables}\label{subsec:App-C-figures}

\begin{figure}[h!]
\begin{centering}
\caption{Illustration of the Norwegian Education System in 1960s.\label{fig:Education-System}}
{\includegraphics[width=1\columnwidth]{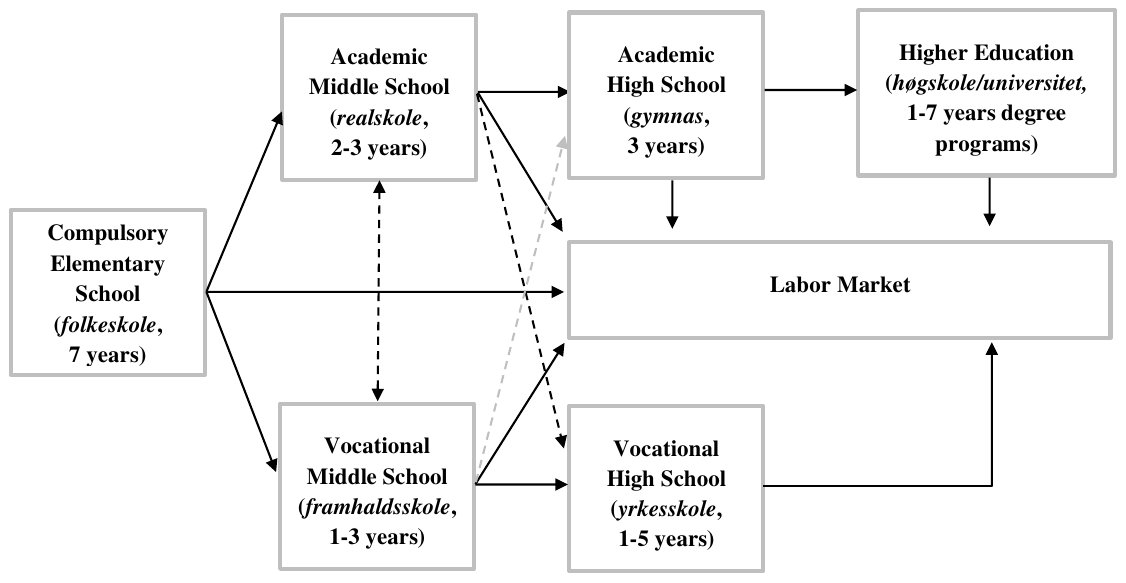}}
\par\end{centering}
\emph{\scriptsize{}Note:}{\scriptsize{} This figure illustrates the Norwegian education system following the 1959 legislation (}\emph{\scriptsize{}Lov om folkeskolen 1959}{\scriptsize{}), when only seven years of elementary schooling was compulsory. Solid black arrows indicate the typical paths that pupils could take as they traverse through the school system, while dotted black (gray) arrows indicate switching between tracks (considered particularly difficult or associated with additional requirements). Following the subsequent legislation in 1969 (\emph{Lov om grunnskolen 1969}), the compulsory schooling was extended to nine years, which was rolled-out in a staggered manner between 1960 and 1975 (see details in Section \ref{subsec:Model-Validation}). Our baseline analysis uses individuals born between 1955-1960,  who faced the pre-reform education system. In 1974, a new type of comprehensive high school (\textit{videreg{\aa}ende} skole) was introduced, which made track switching easier. The latter system remained in place up to the mid-1990s, when two reforms (Reform 94 and Reform 97) were enacted, which altered the structure of high school education and lowered the school starting age to six, respectively. See further details in \cite{Bertrand.2020}}.{\scriptsize\par}
\end{figure}

\begin{figure}[h!]\centering
\caption{Illustration of the Decision Tree and Transitions at Ages 14--17.}\label{Decision tree and data app}

\tikzset{
	treenode/.style = {shape=rectangle, rounded corners, draw, align=center, bottom color=blue!20},
	root/.style     = {treenode, font=\small, draw=none},
	env/.style      = {treenode, font=\small, draw=none},
	dummy/.style    = {circle,draw}
}

\begin{tikzpicture}
[
	x=30pt,
	y=26pt,
	yscale=-1,
	xscale=1,
	baseline=-120pt,
	grow                    = right,
	edge from parent/.style = {draw, -latex},
	every node/.style       = {font=\footnotesize, minimum width={2cm}},
	sloped
]

\node [root, top color = white, bottom color=white, draw] (0) at (-10,0) {\textbf{Compulsory Schooling}};

\node [env, top color = bwtabred, bottom color = bwtabred, color = bwtaborange, scale=0.8] (2) at (-5,-3.1) {\textbf{Academic}};
\draw[->, thick] (0) edge node[right of = 0, yshift=0.35cm,  rotate = 45.0, node distance = 0cm]{\textbf{37 \%}} (2);

\node [env, top color = bwtabred, bottom color = bwtabred, color = bwtaborange, scale = 0.55] (22) at (0,-3.6) {Academic};
\draw[->] (2) edge node[right of = 4, yshift=0.3cm, rotate=10, node distance=1cm]{\tiny 96\%}(22);
\node [env, top color = bwtabpurple, bottom color = bwtabpurple, scale = 0.55] (23) at (0,-3.1) {Home};
\draw[->] (2) edge node[right of = 4, yshift=0.13cm, rotate=-2, node distance=1cm]{\tiny 4\%}(23);
\node [env, top color = bwtaborange, bottom color = bwtaborange, scale = 0.55] (24)  at (0,-2.6) {Vocational};
\draw[->] (2) edge node[right of = 4, yshift=-0.05cm, rotate=-10, node distance=1cm]{\tiny 0\%}(24);

\node [env, top color = bwtabblue, bottom color = bwtabblue, color = bwtaborange, scale = 0.55] (201) at (4,-4.6) {Work};
\draw [->] (22) edge node[right of = 4, yshift=0.6cm, rotate=16, node distance=1cm] {\tiny 7\%} (201);
\node [env, top color = bwtabred, bottom color = bwtabred, color = bwtaborange, scale = 0.55] (202) at (4,-4.1) {Academic};
\draw [->] (22) edge node[right of = 4, yshift=0.35cm, rotate=5, node distance=1cm] {\tiny 77\%} (202);
\node [env, top color = bwtabpurple, bottom color = bwtabpurple, scale = 0.55] (203) at (4,-3.6) {Home};
\draw [->] (22) edge node[right of = 4, yshift=0.13cm, rotate=-7, node distance=1cm] {\tiny 9\%} (203);
\node [env, top color = bwtabpurple, bottom color = bwtabpurple, scale = 0.55] (204) at (4,-3.1) {Vocational};
\draw [->] (22) edge node[right of = 4, yshift=-0.05cm, rotate=-15, node distance=1cm] {\tiny 7\%} (204);

\node [env, top color = bwtabpurple, bottom color = bwtabpurple, scale = 0.8] (3) at (-5,0) {\textbf{Home}};
\draw[->, thick] (0) edge node[right of = 0, rotate=0, yshift = 0.25cm, node distance = 0.2cm]{\textbf{ 5 \%}} (3);

\node [env, top color = bwtabred, bottom color = bwtabred, color = bwtaborange, scale = 0.55] (32) at (0,-0.5) {Academic};
\draw[->] (3) edge node[right of = 4, yshift=0.3cm, rotate=10, node distance=1cm]{\tiny 13\%}(32);
\node [env, top color = bwtabpurple, bottom color = bwtabpurple, scale = 0.55] (33) at (0,0) {Home};
\draw[->] (3) edge node[right of = 4, yshift=0.13cm, rotate=-2, node distance=1cm]{\tiny 86\%}(33);
\node [env, top color = bwtaborange, bottom color = bwtaborange, scale = 0.55] (34)  at (0,0.5) {Vocational};
\draw[->] (3) edge node[right of = 4, yshift=-0.05cm, rotate=-10, node distance=1cm]{\tiny 1\%}(34);

\node [env, top color = bwtaborange, bottom color = bwtaborange, scale = 0.8] (4)  at (-5,3.1) {\textbf{Vocational}};
\draw[->, thick] (0) edge node[right of = 0, yshift=0.1cm, rotate = -45, node distance = 0.25cm]{\textbf{58 \%}} (4);

\node [env, top color = bwtabred, bottom color=bwtabred, color = bwtaborange, scale = 0.55] (42) at (0,2.6) {Academic};
\draw[->] (4) edge node[right of = 4, yshift=0.3cm, rotate=10, node distance=1cm]{\tiny 5\%}(42);
\node [env, top color = bwtabpurple, bottom color=bwtabpurple, scale = 0.55] (43) at (0,3.1) {Home};
\draw[->] (4) edge node[right of = 4, yshift=0.13cm, rotate=-2, node distance=1cm]{\tiny 6\%}(43);
\node [env, top color = bwtaborange, bottom color=bwtaborange, scale = 0.55] (44)  at (0,3.6)
{Vocational};
\draw[->] (4) edge node[right of = 4, yshift=-0.05cm, rotate=-10, node distance=1cm]{\tiny 89\%}(44);

\node [env, top color = bwtabblue, bottom color = bwtabblue, color = bwtaborange, scale = 0.55] (221) at (4,2.1) {Work};
\draw [->] (44) edge node[right of = 4, yshift=0.6cm, rotate=16, node distance=1cm] {\tiny 8\%} (221);
\node [env, top color = bwtabred, bottom color = bwtabred, color = bwtaborange, scale = 0.55] (222) at (4,2.6) {Academic};
\draw [->] (44) edge node[right of = 4, yshift=0.35cm, rotate=6, node distance=1cm] {\tiny 10\%} (222);
\node [env, top color = bwtabpurple, bottom color = bwtabpurple, scale = 0.55] (223) at (4,3.1) {Home};
\draw [->] (44) edge node[right of = 4, yshift=0.14cm, rotate=0, node distance=1cm] {\tiny 3\%} (223);
\node [env, top color = bwtabpurple, bottom color = bwtabpurple, scale = 0.55] (224) at (4,3.6) {Vocational};
\draw [->] (44) edge node[right of = 4, yshift=-0.06cm, rotate=-15, node distance=1cm] {\tiny 79\%} (224);

\end{tikzpicture}
\vspace{-1cm}
\begin{center}
\begin{minipage}[t]{\columnwidth}
\emph{\scriptsize{}Note:}{\scriptsize{} This figure shows the decision tree between ages 14 to 17, along with the fraction transiting from one state to another at each age. All individuals start with compulsory schooling at age 14, and transit into either academic schooling, vocational schooling or home at ages 15-16. From age 16 to 17, individuals may also transit into work. To ease the illustration of possible decision paths at age 17, the figure plots only the transitions from the academic--academic and vocational--vocational branches at age 15--16.}{\scriptsize\par}
\end{minipage}
\end{center}
\end{figure}

\begin{figure}[h!]\centering
\caption{Average Ex-ante Earnings Returns to Academic and Vocational Schooling.}\label{Ex-ante return-wage}
\subfloat[Academic Schooling]{\scalebox{0.50}{\includegraphics{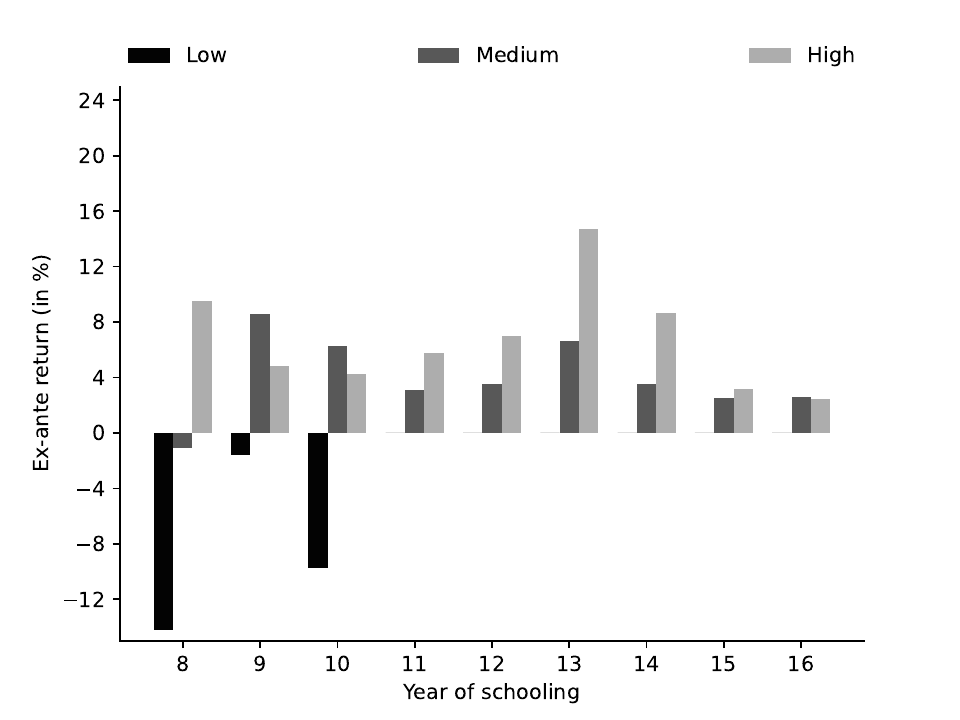}}}
\subfloat[Vocational Schooling]{\scalebox{0.50}{\includegraphics{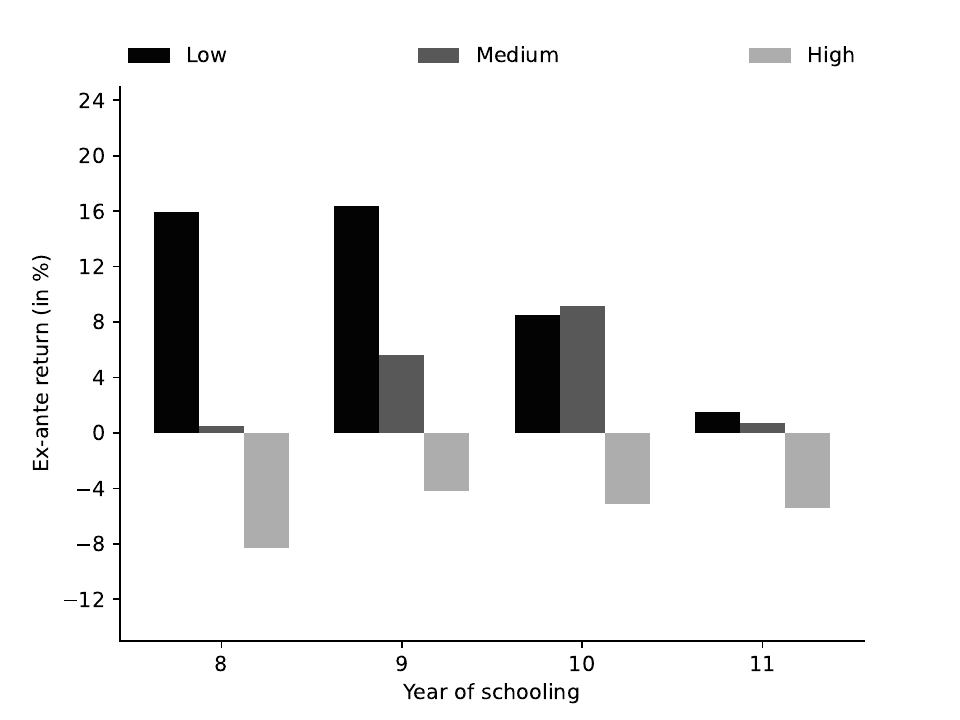}}}\\
\begin{center}
\begin{minipage}[t]{\columnwidth}
\emph{\scriptsize{}Note:}{\scriptsize{} The figure is based on samples of 5,000 simulated schooling careers for each ability group based on the estimated model. This figure contains average ex-ante lifetime earnings returns for each schooling year-track-ability cell. The left figure shows how ex-ante returns to academic schooling develop over time for each group while the right figure shows the same for the vocational track. Each bar shows the average ex-ante return to a particular year of schooling for the individuals in the respective ability group that reaches the relevant transition in our model. For instance, the bar for the high-ability group in the academic panel in year 11 shows the average ex-ante return of the 11th year overall high-ability individuals that have had an uninterrupted academic schooling career until the 10th year. We compute the ex-ante return as defined in Equation (\ref{lt_wage_return}). Whenever there are only a few people of a particular ability group that reach a particular transition we do omit this group from the calculation.}{\scriptsize\par}
\end{minipage}
\end{center}
\end{figure}

\begin{figure}[h!]\centering
\caption{Average Ex-ante Returns -- The Role of Shocks to Productivity and Tastes.}\label{Ex-ante return-risk}
\medskip
{\scalebox{0.90}{\textit{Panel A. Baseline Model}}} \\
\setcounter{subfigure}{0}
\renewcommand{\thesubfigure}{A-\arabic{subfigure}}
\subfloat[Academic Schooling]{\scalebox{0.4}{\includegraphics{material/results/fig-tr-transitions-iq-academic-bw}}}\hspace{0.3cm}
\subfloat[Vocational Schooling]{\scalebox{0.4}{\includegraphics{material/results/fig-tr-transitions-iq-vocational-bw}}}\hspace{0.3cm}
\medskip \\
{\scalebox{0.90}{\textit{Panel B. No Shocks to Productivity}}} \\
\setcounter{subfigure}{0}
\renewcommand{\thesubfigure}{B-\arabic{subfigure}}
\subfloat[Academic Schooling]{\scalebox{0.4}{\includegraphics{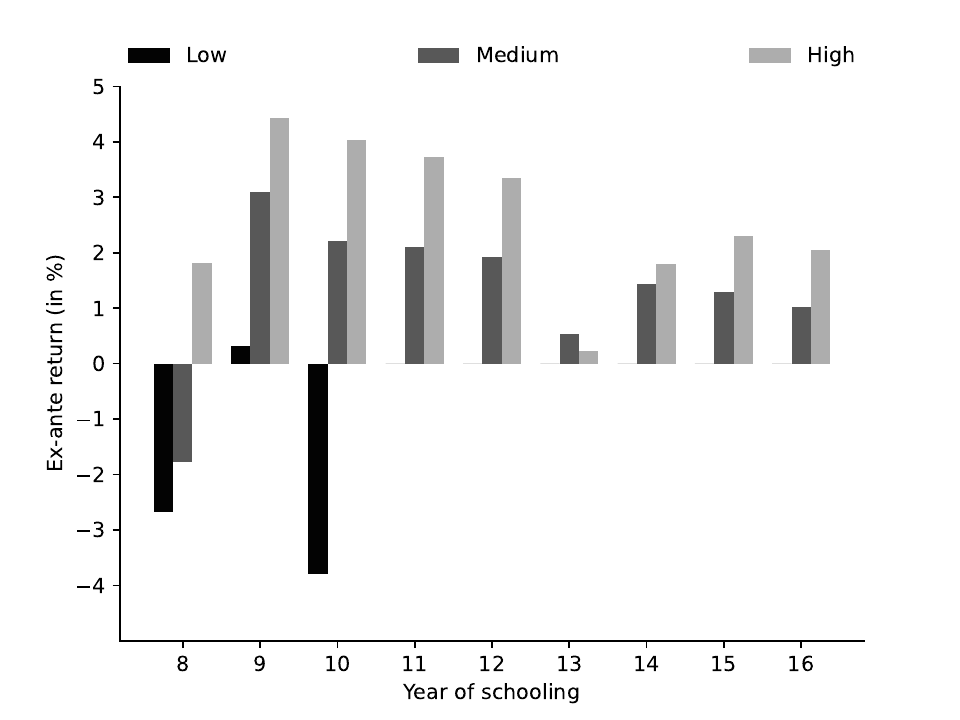}}}\hspace{0.3cm}
\subfloat[Vocational Schooling]{\scalebox{0.4}{\includegraphics{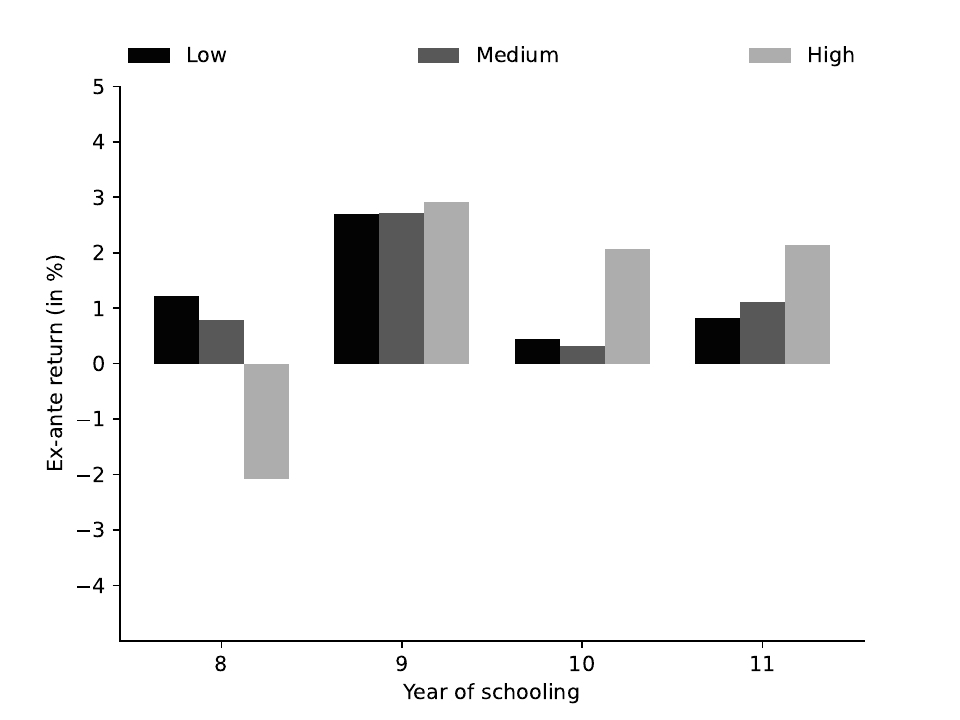}}}\hspace{0.3cm}
\medskip \\
{\scalebox{0.90}{\textit{Panel C. No Shocks to Productivity and Tastes}}} \\
\renewcommand{\thesubfigure}{C-\arabic{subfigure}}
\setcounter{subfigure}{0}
\subfloat[Academic Schooling]{\scalebox{0.4}{\includegraphics{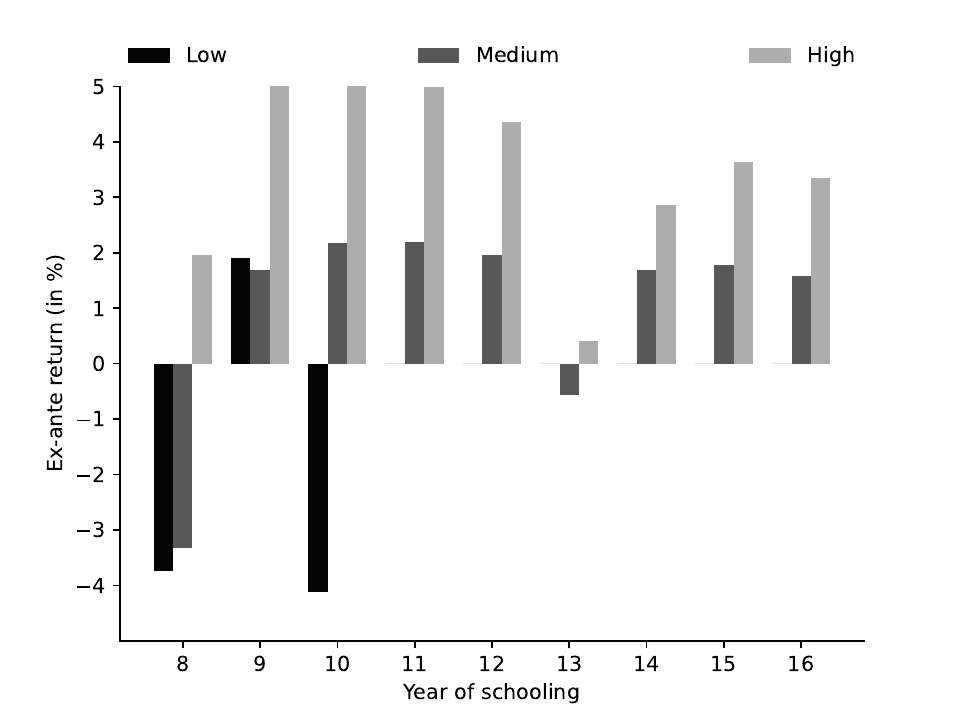}}}\hspace{0.3cm}
\subfloat[Vocational Schooling]{\scalebox{0.4}{\includegraphics{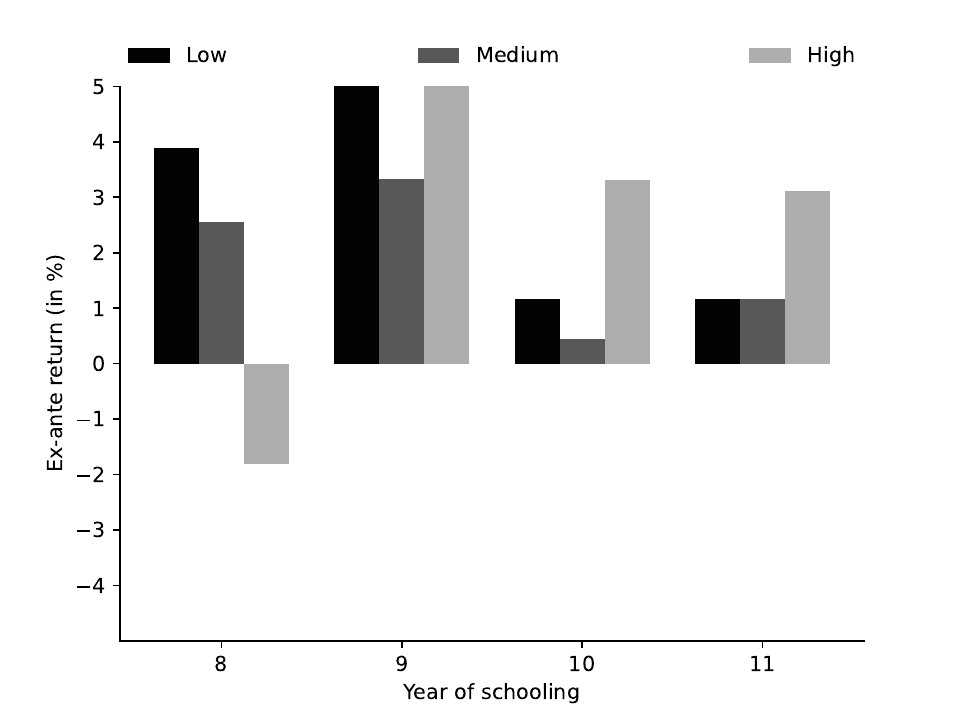}}}
\begin{center}
\begin{minipage}[t]{\columnwidth}
\emph{\scriptsize{}Note:}{\scriptsize{} The figure is based on samples of 5,000 simulated schooling careers for each ability group based on the estimated model under different specifications of transitory shocks. Each panel contains average ex-ante returns for each year-track-ability cell for alternative model specifications. Panel A shows results based on the baseline model, while panel B removes shocks to productivity (i.e., no wage risk) and panel C further also removes taste shocks. In each panel, the left figure shows how ex-ante returns to academic schooling develop over time for each group while the right figure shows the same for the vocational track. Each bar shows the average ex-ante return to a particular year of schooling for the subset of the respective ability group that has reached the relevant transition in our model. For instance, the bar for the high-ability group in the academic panel in year 11 shows the average ex-ante return of the 11th year for those that have had an uninterrupted academic schooling career until the 10th year. We compute the ex-ante return as defined in Equation (\ref{Calculation ex-ante return}). Whenever there are only a few people of a particular ability group that reaches a particular transition we do omit this group from the calculation.}{\scriptsize\par}
\end{minipage}
\end{center}
\end{figure}

\begin{figure}[h!]\centering
\caption{Change in Final Schooling without Public Sector Job Opportunites.}\label{Final Schooling Sectoral Choice}
\scalebox{0.4}{\includegraphics{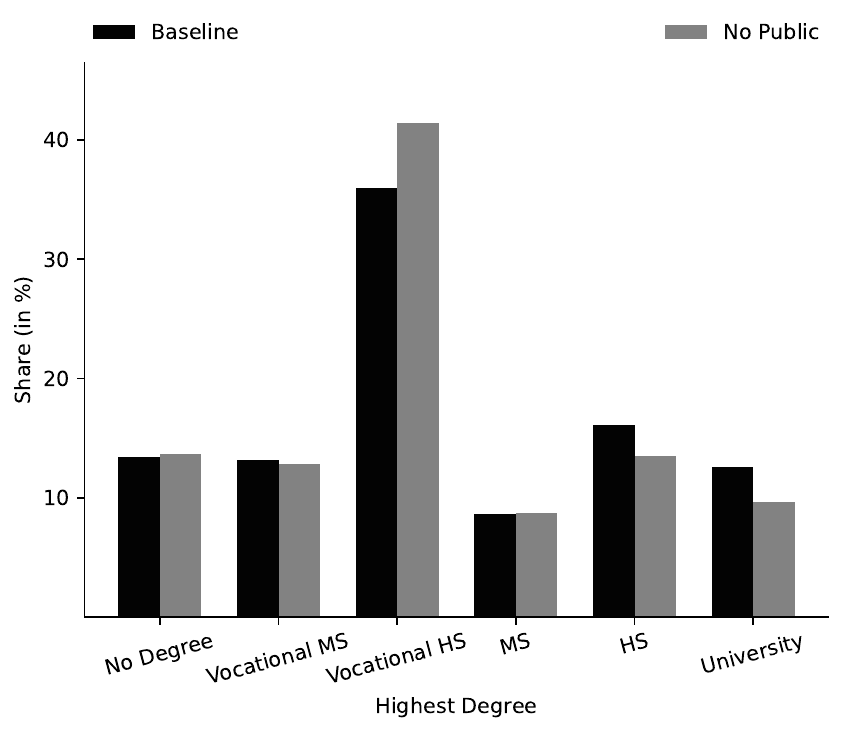}}
\begin{center}
\begin{minipage}[t]{\columnwidth}
\emph{\scriptsize{}Note:}{\scriptsize{} The figure is based on two samples of 100,000 simulated schooling careers under alternative policies.
The first bar shows the distribution of final schooling degrees in the baseline model.
The second bar shows the distribution of final schooling in an alternative model without a public sector.
Throughout, we keep the random realizations of the productivity and taste shocks $\bm{\epsilon_t}$ fixed, and we are thus able to compare the schooling decisions of the same individual under the two different regimes. Finally, we illustrate the distributions of final years of schooling under each policy simulation.
The labels on the x-axis refer to different degrees. The abbreviation MS refers to a middle school degree. The abbreviation HS refers to a high school degree. Both of these can be obtained in an academic and a vocational school.
}{\scriptsize\par}
\end{minipage}    
\end{center} 
\end{figure}

\begin{figure}[h!]\centering
\caption{Change in Final Schooling with Reduction of Public Sector Jobs.}\label{Final Schooling Sectoral Choice Ability}
\subfloat[Overall]{\scalebox{0.4}{\includegraphics{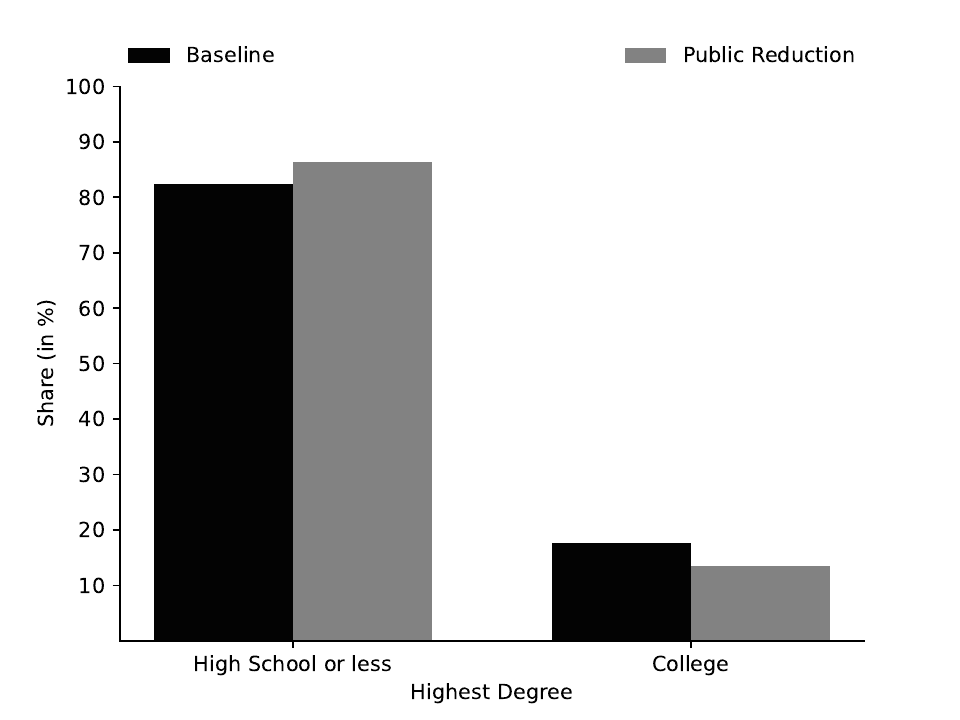}}}\hspace{0.3cm}
\subfloat[High Ability]{\scalebox{0.4}{\includegraphics{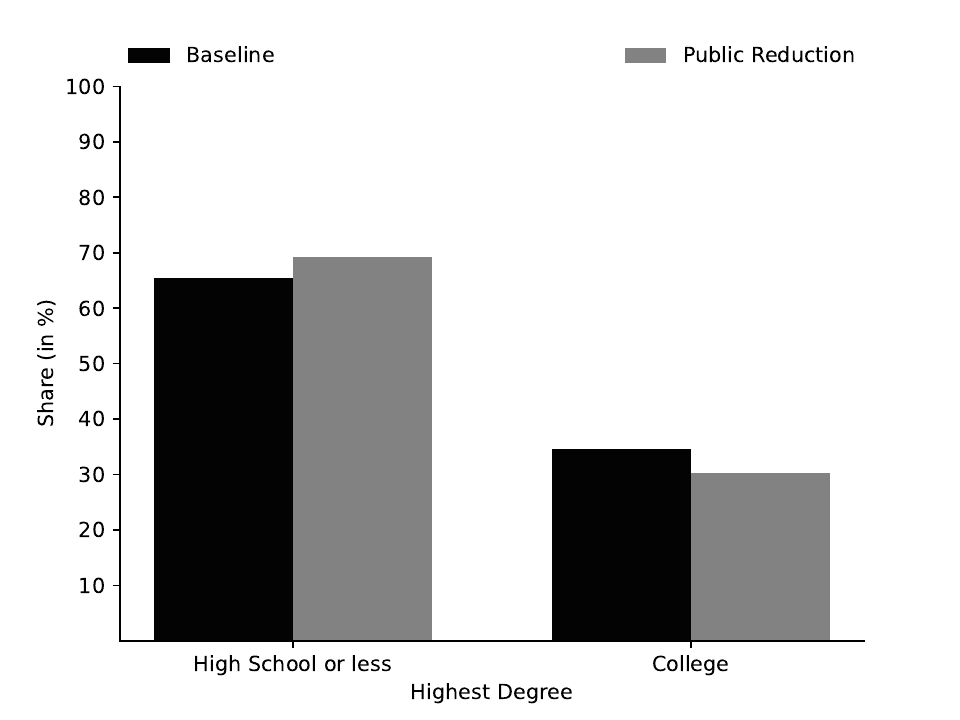}}}\\
\subfloat[Medium Ability]{\scalebox{0.4}{\includegraphics{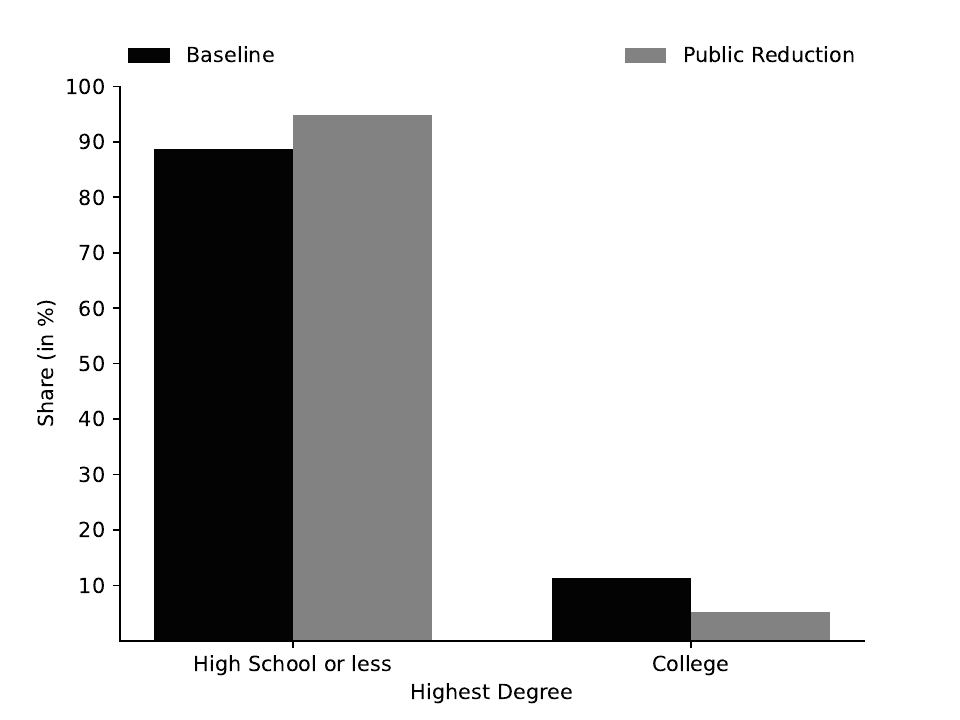}}}\hspace{0.3cm}
\subfloat[Low Ability]{\scalebox{0.4}{\includegraphics{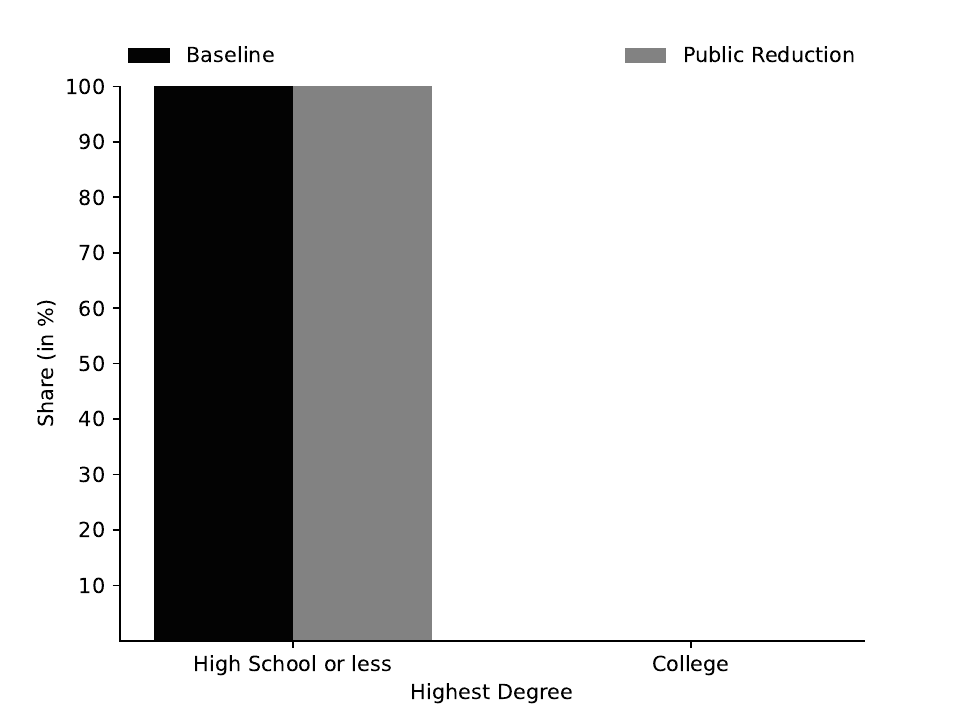}}}\begin{center}
\begin{minipage}[t]{\columnwidth}
\emph{\scriptsize{}Note:}{\scriptsize{} The figure is based on two samples of 100,000 simulated schooling careers under alternative policies. Using the point estimates, we first simulate the baseline model. Next, we rerun the simulation but reduce the baseline nonpecuniary utility associated with the public sector by a half. Throughout, we keep the random realizations of the productivity and taste shocks $\bm{\epsilon_t}$ fixed, and we are thus able to compare the schooling decisions of the same individual under the two different regimes. Finally, we illustrate the distributions of final schooling under each model. 
'College' represents all individuals with at least 16 years of academic schooling and 'High School or Less' represents all individuals with less than sixteen years of academic schooling.

}{\scriptsize\par}
\end{minipage}    
\end{center} 
\end{figure}

\FloatBarrier

\begin{figure}[h!]\centering
\caption{The Impacts of Compulsory High School Enrollment Policy (`Reform 10').}\label{Increase to 10 years of compulsory schooling}
\subfloat[Overall]{\scalebox{0.45}{\includegraphics{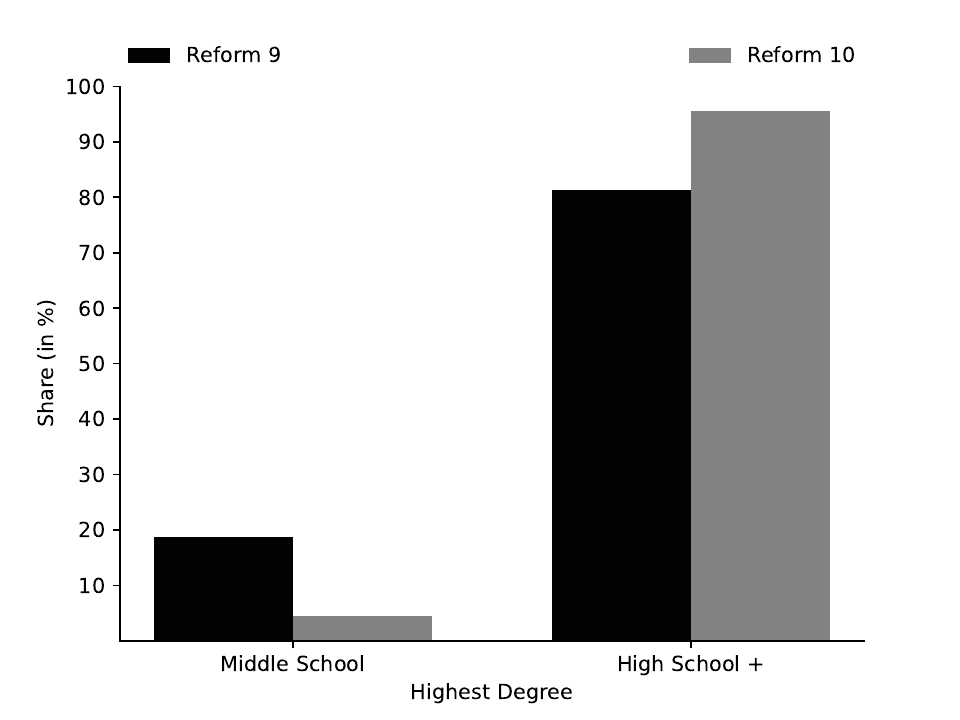}}}\hspace{0.3cm}
\subfloat[High Ability]{\scalebox{0.45}{\includegraphics{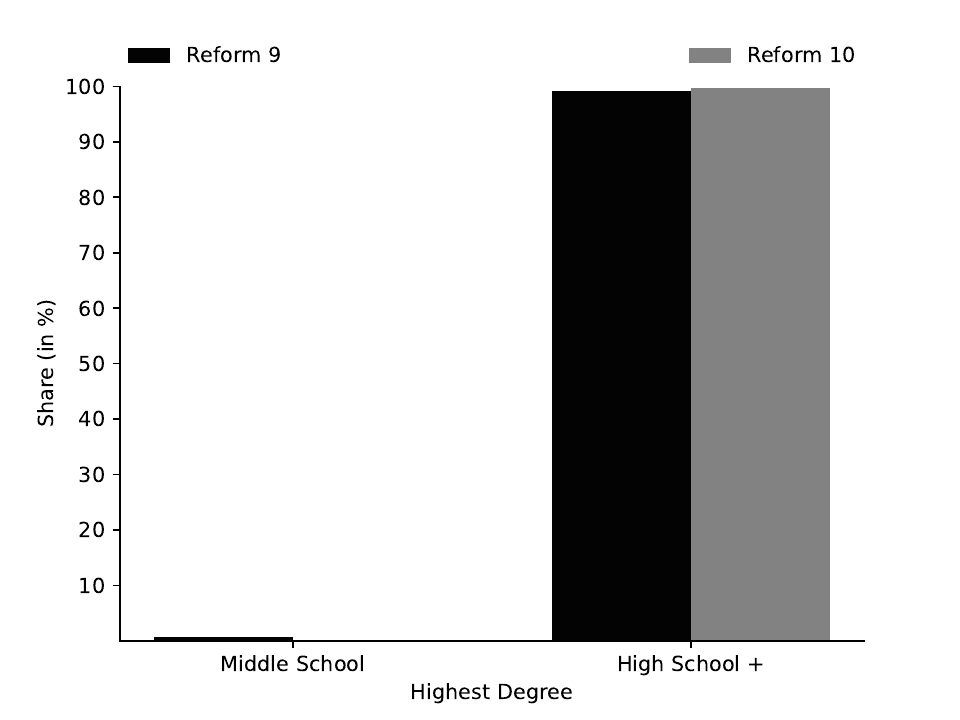}}}\\
\subfloat[Medium Ability]{\scalebox{0.45}{\includegraphics{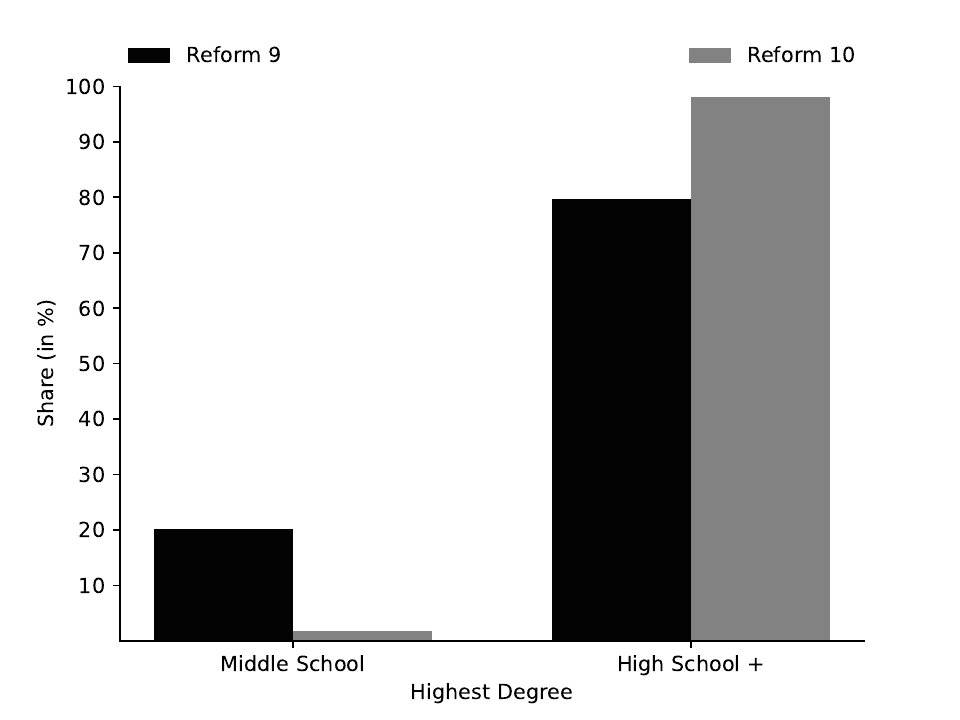}}}\hspace{0.3cm}
\subfloat[Low Ability]{\scalebox{0.45}{\includegraphics{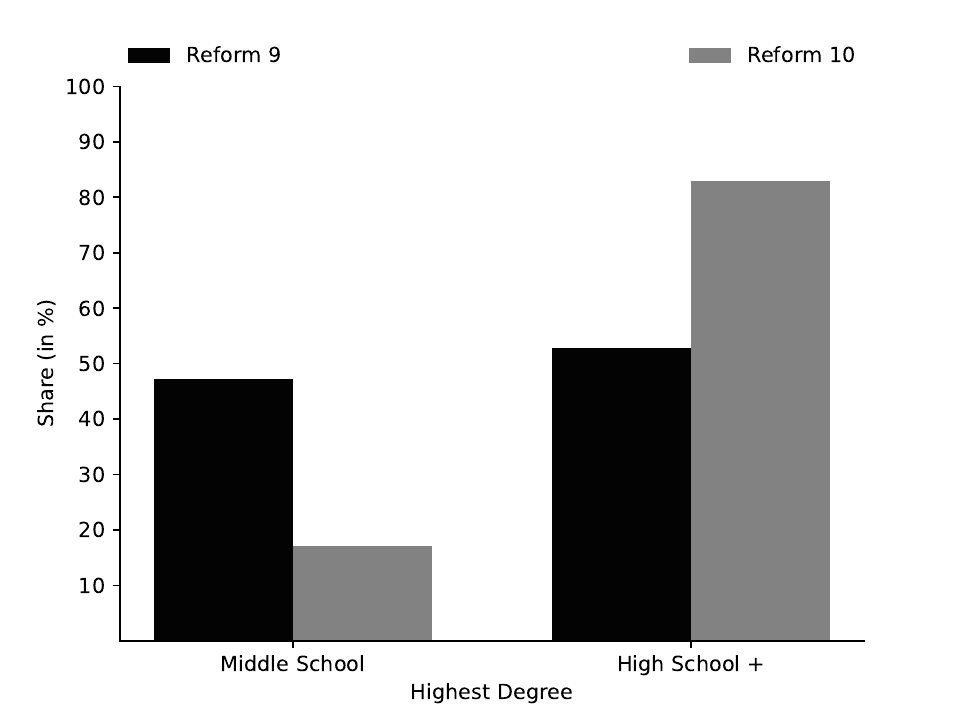}}}
\begin{center}
\begin{minipage}[t]{\columnwidth}
\emph{\scriptsize{}Note:}{\scriptsize{} The figure is based on two samples of 100,000 simulated schooling careers under alternative policies. Using the point estimates, we first simulate the model with the nine years of compulsory schooling (`Reform 9'). Next, we rerun the simulation but impose ten years of compulsory schooling to illustrate the compulsory high school enrollment policy (`Reform 10'). Throughout, we keep the random realizations of the productivity and taste shocks $\bm{\epsilon_t}$ fixed, and we are thus able to compare the schooling decisions of the same individual under the two different regimes. Finally, we illustrate the distributions of final schooling under each model. `Middle School' represents all individuals between 9--12 years of academic schooling and 9--11 years of vocational schooling. `High School +' represents all individuals with 12 or more years of academic schooling and with 11 or more years of vocational schooling.
}{\scriptsize\par}
\end{minipage}    
\end{center} 
\end{figure}

\FloatBarrier

\begin{figure}[h!]\centering
\caption{The Impacts of Tuition Fees Policy.}\label{Tuition}
\subfloat[Overall]{\scalebox{0.45}{\includegraphics{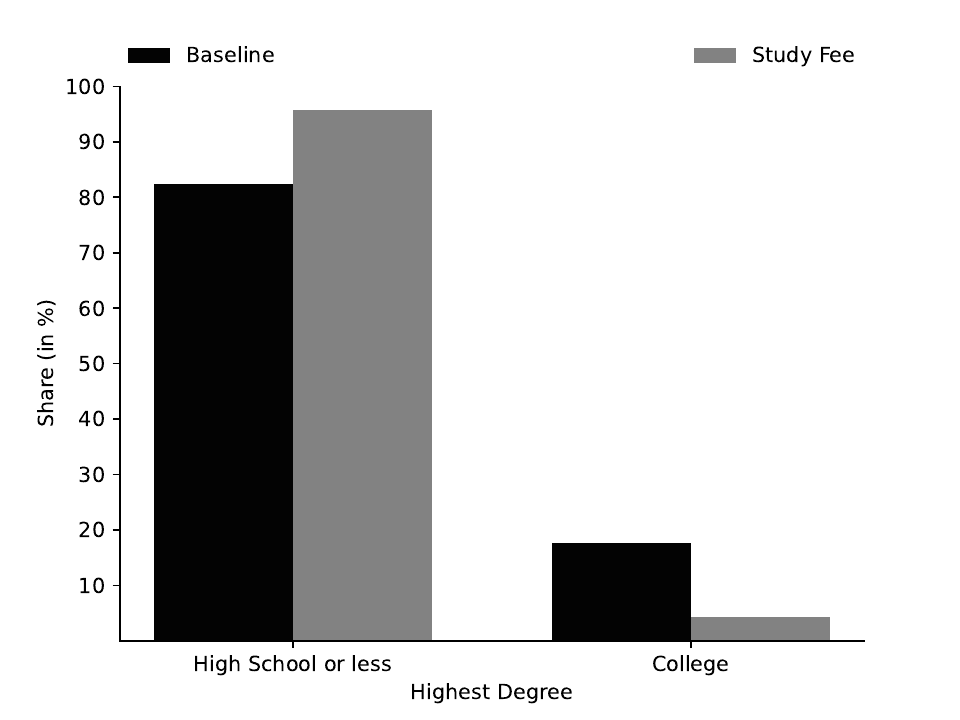}}}\hspace{0.3cm}
\subfloat[High Ability]{\scalebox{0.45}{\includegraphics{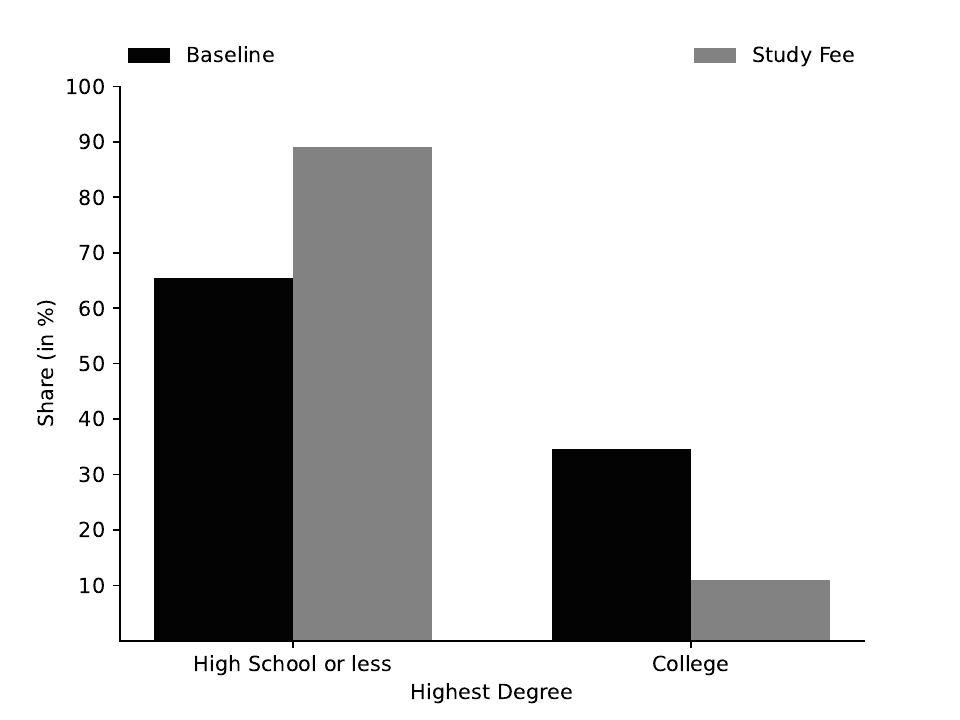}}}\\
\subfloat[Medium Ability]{\scalebox{0.45}{\includegraphics{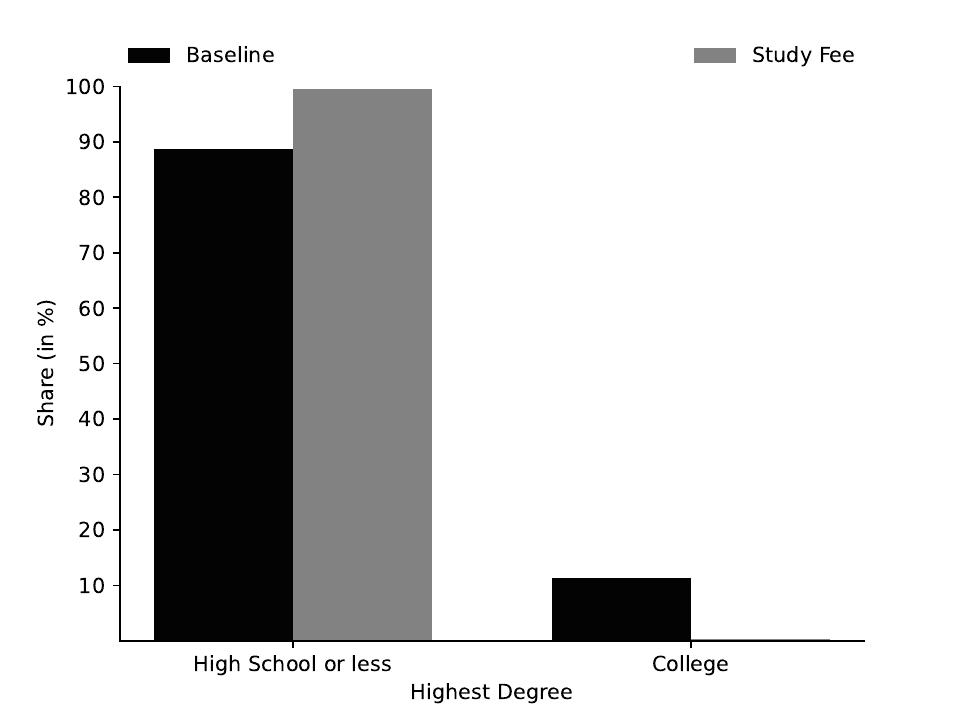}}}\hspace{0.3cm}
\subfloat[Low Ability]{\scalebox{0.45}{\includegraphics{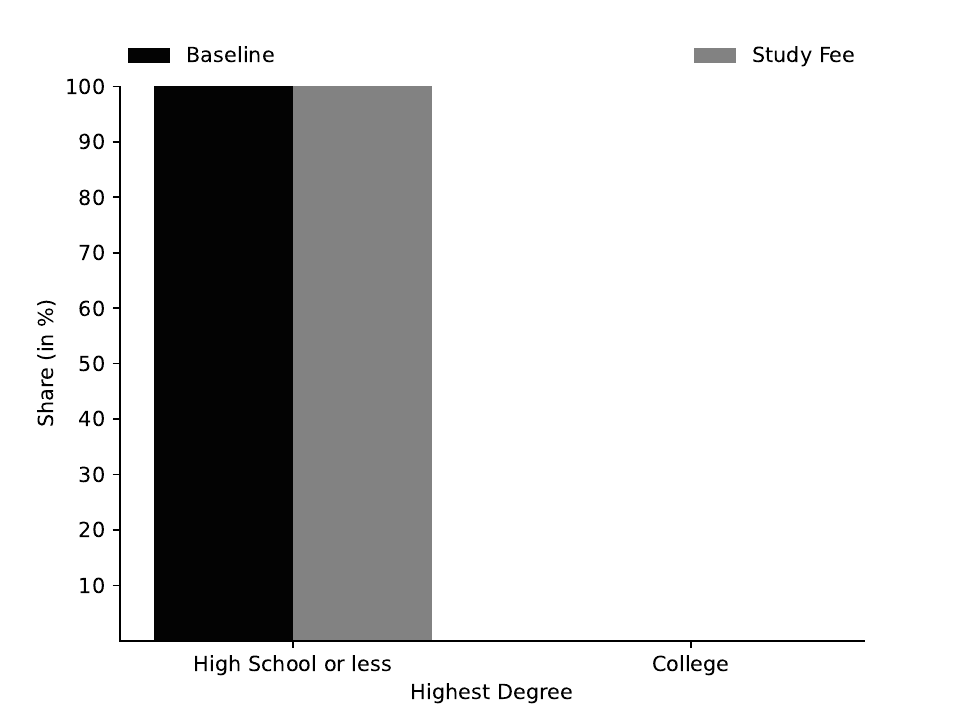}}}
\begin{center}
\begin{minipage}[t]{\columnwidth}
\emph{\scriptsize{}Note:}{\scriptsize{}  The figure is based on two samples of 100,000 simulated schooling careers under alternative policies. Using the point estimates, we first simulate the baseline model. Next, we rerun the simulation but impose a college tuition fees of 50,000 Norwegian Kroner (2018). Throughout, we keep the random realizations of the productivity and taste shocks $\bm{\epsilon_t}$ fixed, and we are thus able to compare the schooling decisions of the same individual under the two different regimes. Finally, we illustrate the distributions of final schooling under each model.
`College' represents all individuals with at least sixteen years of academic schooling and `High School or Less' represents all individuals with less than sixteen years of academic schooling.

}{\scriptsize\par}
\end{minipage}    
\end{center} 
\end{figure}

\begin{figure}[h!]\centering
\caption{Option Value Contributions -- The Role of Shocks to Productivity and Tastes.}\label{Option value uncertainty}
  \subfloat[Academic Schooling at 11th Year (Medium)]{\scalebox{0.45}{\includegraphics{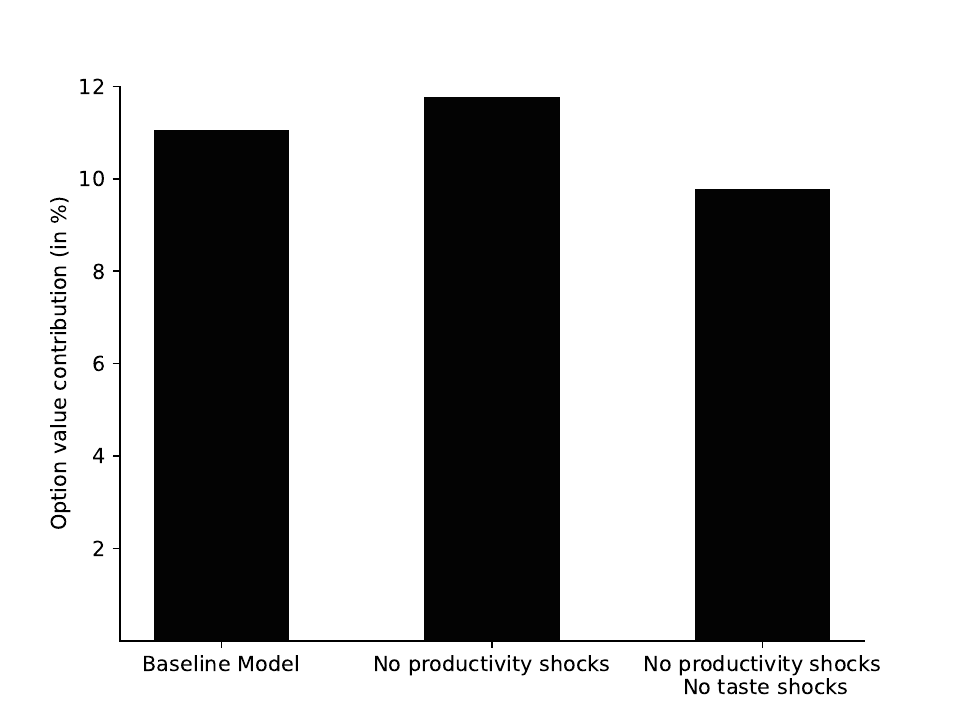}}}\hspace{0.3cm}
  \subfloat[Vocational Schooling at 8th Year  (Low)]{\scalebox{0.45}{\includegraphics{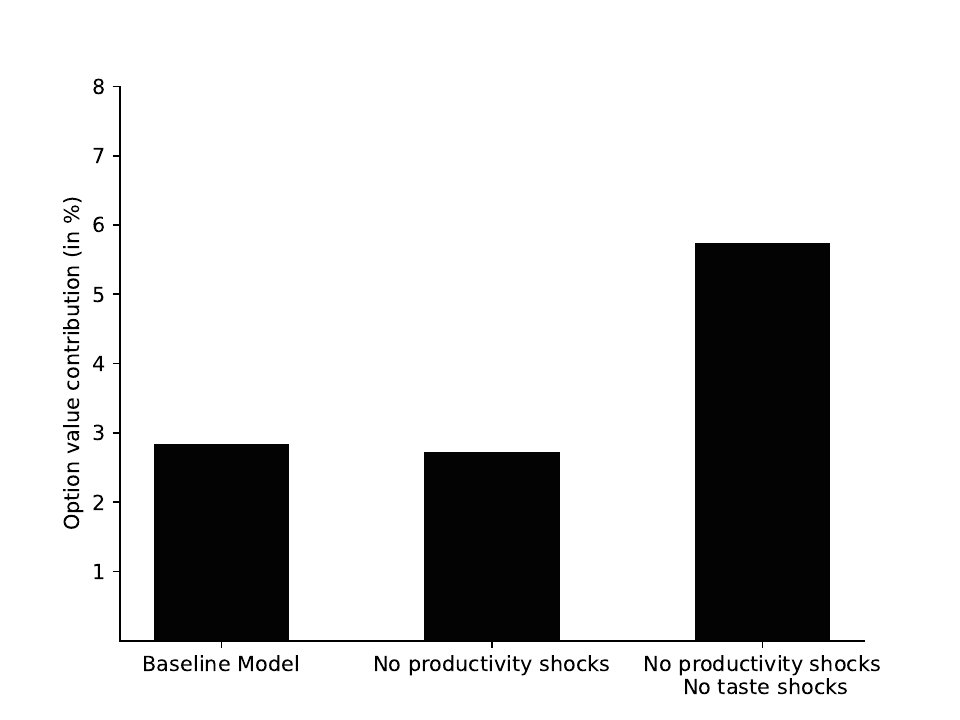}}}
\begin{center}
\begin{minipage}[t]{\columnwidth}
\emph{\scriptsize{}Note:}{\scriptsize{} The figure is based on samples of 5,000 simulated schooling careers for each ability group based on alternative model specifications. The left panel shows the different option value contributions for medium-ability individuals for the 11th year of academic schooling, while the right panel shows the option value contributions for low-ability individuals for the 8th year of vocational schooling. The option value contribution is defined in Equation (\ref{Calculation option value contribution}). The bars correspond to different model specification;  the first bar corresponds to the estimated model, the medium bar corresponds to an adapted version of the estimated model where productivity shocks (i.e., wage risk) is turned off and the final bar to a model where both productivity and taste shocks are turned off. Whenever there are only a few people of a particular ability group that reaches a particular transition we do omit this group from the calculation.}{\scriptsize\par}
\end{minipage}    

\end{center}
\end{figure}

\begin{figure}[h!]\centering
\caption{Option Value Contributions -- The Role of Sectoral Opportunities.}\label{Option value sector}
  \subfloat[Academic Schooling at 8th Year (Medium)]{\scalebox{0.45}{\includegraphics{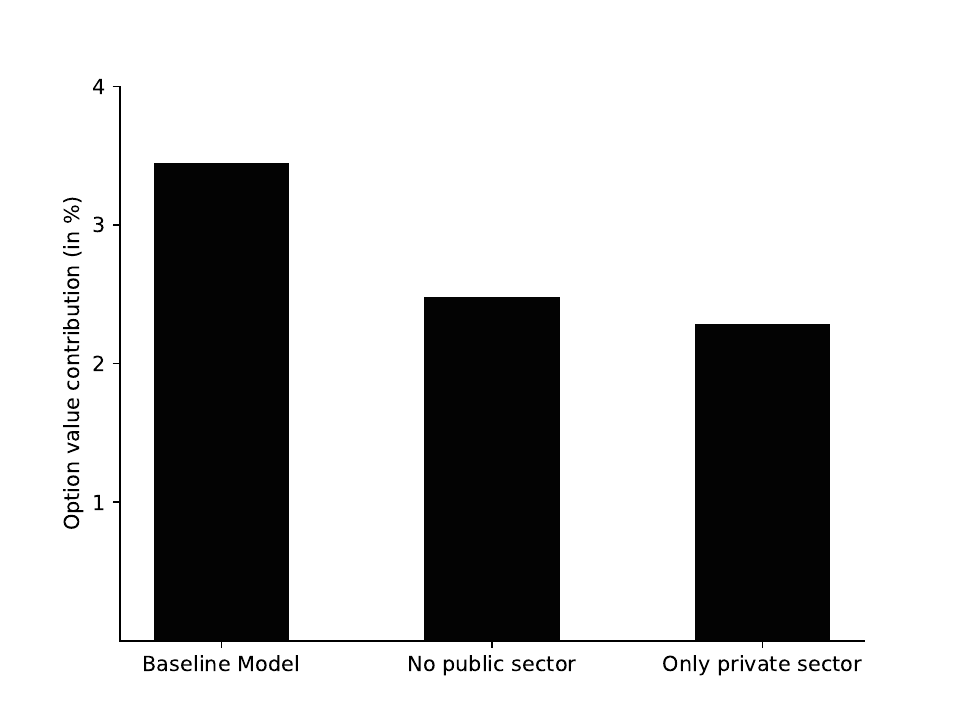}}}\hspace{0.3cm}
  \subfloat[Vocational Schooling at 8th Year  (Low)]{\scalebox{0.45}{\includegraphics{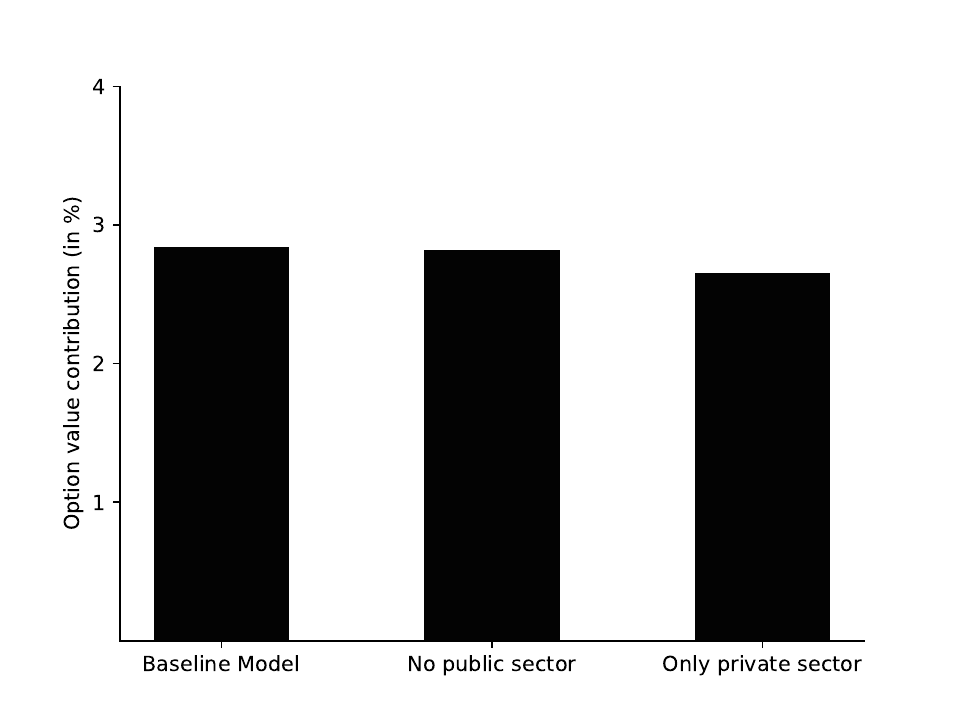}}}
\begin{center}
\begin{minipage}[t]{\columnwidth}
\emph{\scriptsize{}Note:}{\scriptsize{} The figure is based on samples of 5,000 simulated schooling careers for each ability group based on alternative model specifications. The left panel shows the different option value contributions for medium-ability individuals for the 8th year of academic schooling, while the right panel shows the option value contributions for low-ability individuals for the 8th year of vocational schooling. The option value contribution is defined in Equation (\ref{Calculation option value contribution}). The bars correspond to different model specification;  the first bar corresponds to the estimated model, the medium bar corresponds to an adapted version of the estimated model where individuals can only work as self employed and in the private sector and the final bar to a model where agents can only work in the private sector. Whenever there are only a few people of a particular ability group that reaches a particular transition we do omit this group from the calculation.}{\scriptsize\par}
\end{minipage}    
\end{center}
\end{figure}

\setcounter{section}{0}
\setcounter{equation}{0}
\setcounter{table}{0}
\setcounter{figure}{0}
\renewcommand{\thesection}{B}
\renewcommand{\theequation}{B.\arabic{equation}}
\renewcommand{\thetable}{B.\arabic{table}}
\renewcommand{\thefigure}{B.\arabic{figure}}

\end{document}